%% file: main.tex
\documentclass[twocolumn]{aastex63}
\usepackage{amsmath,amstext}
\usepackage[T1]{fontenc}
\usepackage{apjfonts} 
\usepackage[figure,figure*]{hypcap}
\usepackage{commath, mathrsfs}
\usepackage{color, upgreek}

\usepackage{mhchem}
\input{newcommands}
   \newcommand{\mph}{\dot{M}_{\rm ph}}
   \newcommand{\nondimts}{\tilde{t}_{\rm s}}
	\renewcommand{\mum}{{\rm \,\upmu m}}
	
	\newcommand{\km}{{\,\rm km}}
	\renewcommand{\Msun}{{\,M_\odot}}
	\newcommand{\solidmass}{M_{\rm s}}
	\renewcommand{\Mearth}{{\, M_\oplus}}
	
\newcommand{\mycomment}[1]{\textcolor{blue}{{\bf RN:} #1}}

\newcommand{\figdir}{.}
\newcommand{\revision}[1]{{#1}}

\shorttitle{Debris disk photoevaporation}
\shortauthors{Nakatani et al.}

\accepted{May 11, 2021}
\begin{document}

\title{Photoevaporation of Grain-Depleted Protoplanetary Disks around Intermediate-Mass Stars: Investigating Possibility of Gas-Rich Debris Disks as Protoplanetary Remnants}
\author[0000-0002-1803-0203]{Riouhei Nakatani}
\affiliation{RIKEN Cluster for Pioneering Research, 2-1 Hirosawa, Wako-shi, Saitama 351-0198, Japan}
\email{ryohei.nakatani@riken.jp}
\author[0000-0000-0000-0000]{Hiroshi Kobayashi}
\affiliation{Department of Physics, Nagoya University, Furo-cho, Chikusa-ku, Nagoya, Aichi 464-8602, Japan}
\author[0000-0003-2309-8963]{Rolf Kuiper}
\affiliation{Institute of Astronomy and Astrophysics, University of T\"ubingen, Auf der Morgenstelle 10, D-72076 T\"ubingen, Germany}
\author[0000-0002-7058-7682]{Hideko Nomura}
\affiliation{National Astronomical Observatory Japan (NAOJ), Osawa 2-21-1, Mitaka, Tokyo 181-8588, Japan}
\author[0000-0003-3283-6884]{Yuri Aikawa}
\affiliation{Department of Astronomy, School of Science, The University of Tokyo, 7-3-1 Hongo, Bunkyo, Tokyo 113-0033, Japan}

\begin{abstract}
Debris disks are classically considered to be gas-less systems, but recent (sub)millimeter observations have detected tens of those with rich gas content. The origin of the gas component remains unclear; namely, it can be protoplanetary remnants and/or secondary products from large bodies. In order to be protoplanetary in origin, the gas component of the parental protoplanetary disk is required to survive for $\gtrsim10\Myr$. However, previous models predict $\lesssim 10\Myr$ lifetimes because of efficient photoevaporation at the late stage of disk evolution.
We investigate photoevaporation of gas-rich, optically-thin disks around intermediate-mass stars at a late stage of the disk evolution. The evolved system is modeled as those \deleted{where radiation force is sufficiently strong to continuously blow out} 
devoid of small grains ($\lesssim 4 \mum$).
\deleted{, which are an essential component for driving photoevaporation via photoelectric heating induced by stellar far-ultraviolet (FUV). }
We find that the grain depletion reduces photoelectric heating, so that far-ultraviolet photoevaporation is not excited.
Extreme-ultraviolet (EUV) photoevaporation is dominant and yields a mass-loss rate of the order of $1\e{-11}(\LEUV/10^{38}\sec^{-1})^{1/2}\Msun\yr^{-1}$, 
where $\LEUV$ is the EUV emission rate of the host star. 
The estimated gas-disk lifetimes are $\sim 100 (M_{\rm disk}/10^{-3}\Msun)(\LEUV/10^{38}\sec^{-1})^{1/2}\Myr$ and depend on the ``initial'' disk mass at the point small grains have been depleted in the system. 
We show that the gas component can survive for a much longer time around A-type stars than lower-mass (F-, G-, K-type) stars \added{\revision{owing to their atypical low EUV (and X-ray) luminosities}}. This trend is consistent with the higher frequency of gas-rich debris disks around A-type stars, implying the possibility of the gas component being protoplanetary remnants.

\end{abstract}


\section{Introduction}
    A protostar-disk system forms through gravitational collapse of the clouds, and it evolves to a planetary system at the end. 
    The initially small interstellar grains, whose sizes are typically $0.005\mum \lesssim a \lesssim 0.1\mum$, collect to grow and form larger solid bodies, such as meteoroids, asteroids, planetesimals, planetary cores, etc., within the disks. 
    Observationally, small grains are traced by infrared (IR) and (sub)millimeter emission from young systems. 
    \replaced{Near-, mid-, and far-IR observations have shown that excessive IR emission from (hot) dust systematically decreases with stellar age}
    {\revision{Near- and mid-IR observations have shown that excessive IR emission from hot, inner dust systematically decreases with stellar age}}
    and disappears on a timescale of $\lesssim 10\Myr$ \citep[e.g.,][]{2001_Haisch, 2007_Meyer, 2007_Hernandez, 2009_Mamajek, 2014_Ribas, 2015_Ribas}, 
    \replaced{which gives a typical protoplanetary disk (PPD) lifetime. }
    {\revision{which gives a typical lifetime of the inner dust component in protoplanetary disks (PPDs).}}
    Similarly, mass accretion rates onto the host stars have been observed to decrease with stellar age and 
    drop below observable levels within $\sim 10\Myr$
    \citep[e.g.,][]{2005_Calveta, 2010_Fedele, 2010_Sicilia-Aguilar}. The observations give an estimate for the dispersal timescale of the (inner) gas component of PPDs. 
    
    \deleted{
    Some young systems show lack of near-IR excess emission while exhibit mid-/far-IR excess emission comparable to (or higher than) classical PPDs \citep[e.g.,][]{1990_Skrutskie, 2010_Muzerolle}. These objects, which are termed ``transitional'' disks, are associated with circumstellar disks with large cavities at $\sim 10$--$100\au$ \citep{2006_McCabe,2007_Najita, 2009_Salyk,2010_Sicilia-Aguilar, 2010_Muzerolle, 2011_Andrews, 2011_Espaillat,2014_Keane}. The fraction of transitional disks to the total population increases with age from a few percent at $\sim 1$--$2\Myr$ to $\sim 10$--$20\%$ at $\gtrsim 3\Myr$ \citep{2010_Muzerolle}. Therefore, transitional disks are considered to be evolved objects 
    where the inner materials have been dispersed. The small frequency of transitional disks implies a relatively short timescale of the transitional phase \citep[$\sim 0.1\Myr$;][]{1990_Skrutskie, 2010_Muzerolle}.
    }
    
    After the disk dispersal on a timescale of $\sim 10\Myr$, the system enters the debris phase where the ratio of the IR-excess luminosity to the total luminosity is $\lesssim 10^{-2}$ \citep{2000_Lagrange,2015_Wyatt}, i.e., it has turned into an optically-thin object to the stellar radiation. 
    Total dust mass measured with submillimeter observations is typically $\lesssim 1\, M_\oplus$ for debris disks, while PPDs have $>1\,M_\oplus$ \citep{2008_Wyatt}. The excessive infrared emission is observable over 
    a longer time than the dispersal time of PPDs. Old debris disks ($\gg 10\Myr$) are therefore considered to sustain an amount of small grains observable in both scattered light and thermal emission by grinding larger solids by collisions. Classically, debris disks were classified as the post-protoplanetary systems devoid of gas, but recent observations have detected tens of those with rich gas content, such as \ion{C}{1}, \ion{O}{1}, and CO, especially for $\lesssim 50\Myr$-old systems \citep[e.g.,][]{2013_Kospal,2014_Dent,2016_White,2017_Matra,2017_Hughes,2017_Marshall,2017_Higuchi,2018_Hughes, 2019_Higuchib, 2019_Higuchia}. 
    \replaced{The origin of the gas component has been unclear, yet it is possible to be both of protoplanetary remnants and secondary products.}
    {
    The origin of the gas component has been unclear. Two hypotheses have been proposed; one is the secondary-origin scenario that explains the origin of the gas to be secondary products from planetesimals. The other is the primordial-origin scenario that explains the gas to be  protoplanetary remnants having survived regardless of disk dispersal.
    Note that the two scenarios are not necessarily incompatible. 
    }
    
    \added{\revision{
    In the secondary-origin scenario, CO is produced from volatile-rich solids. The molecule is photodissociated into C and O and then is redistributed according to viscous evolution \citep{2017_Kral}. The accumulation of C is essential for shielding CO from photodissociating photons to extend the lifetime longer than the viscous timescale; otherwise, it is only $\sim 100\yr$ \citep{2019_Kral, 2019_Moor, 2020_Marino}. 
    Many CO line emission studies have been conducted to investigate the origin of the gas \citep[e.g.,][]{2013_Moor,2017_Moor, 2019_Moor, 2019_Kral, 2019_Hales}. Recent studies have found that the shielded disk model well agrees with the observed gas properties in debris disks \citep{2019_Moor, 2019_Kral, 2019_Hales, 2020_Marino}, yet these models are still under development.
    Regarding the primordial-origin scenario, on the other hand, only a few studies have investigated its likelihood/unlikelihood in detail. It is necessary to study disk-dispersal processes especially at the later epoch of the disk evolution for testing the plausibility. 
    }}
    
    As for theoretical disk dispersal processes, 
    viscous accretion \citep[e.g.,][]{1973_ShakuraSunyaev,1974_LindenbellPringle}, magnetohydrodynamics (MHD) winds \citep[e.g.,][]{2009_SuzukiInutsuka,2013_Bai_a,2013_Bai_b},
    and photoevaporation \citep[e.g.,][]{1994_Shu,1994_Hollenbach} have major effects on the evolution to evolved disks; namely transition and debris disks. Mass loss due to the individual process as well as the interplay between them
    have been studied with detailed modeling in the literature recently \citep[e.g.,][]{2013_Bai,2013_Bai_b,2015_Gressel,2013_Simon,2013_Simon_a,2015_Simon,2016_Bai,2016_Bai_a,2018_Wang,2020_Gressel}. 
    \cite{2020_Kunitomo} follow secular disk evolution with 1D~hydrodynamics simulations incorporating all of these processes. They show that the dominant dispersal process is MHD winds while the surface density is high, and it is replaced by photoevaporation at a later stage when the disk mass decreases to $\sim 0.01\Msun$ at an age of $\sim 1\Myr$. Photoevaporation is likely a major dispersal process for the gas component at later stages of the disk evolution including transitional/young debris phases. 
    
    Photoevaporation is driven by stellar far-ultraviolet \citep[FUV; $6\eV \lesssim h\nu \leq 13.6\eV$; e.g.,][]{1996_YorkeWelz, 1997_Richling, 2009_Gorti, 2017_Wang, 2018_Nakatani, 2018_Nakatanib}, extreme-ultraviolet \citep[EUV; $13.6\leq h\nu \lesssim 100\eV$; e.g.,][]{1994_Hollenbach, 2004_Font, 2004_Alexander}, and X-ray \citep[$h\nu \gtrsim 100\eV$; e.g.,][]{2008_Ercolano,2009_Ercolano, 2010_Owen,2011_Owen_a,2012_Owen}. 
    FUV and X-ray generally attenuate at a larger column density \deleted{($\sim 10^{21}\cm^{-2}$, assuming the interstellar composition)} compared to EUV, \replaced{($\sim 10^{19}$--$10^{20}\cm^{-2}$)}{assuming the interstellar composition}.
    It indicates that FUV and X-ray can heat the deep, high-density interior of PPDs. 
    Consequently, FUV- and X-ray-driven photoevaporation yield mass-loss rates of $\dot{M} \sim 10^{-8}\Msun\yr^{-1}$ for solar-type stars \citep[e.g.,][]{2009_GortiHollenbach, 2010_Owen}, which is orders of magnitude larger than EUV photoevaporation rates \citep[$\sim 10^{-10}\Msun\yr^{-1}$; e.g.,][for a review]{1994_Hollenbach, 2014_Alexander}. 
    
    Mass-loss rates of $\dot{M}\sim 10^{-8}\Msun$ are so high that 
    \replaced{the gas component of PPDs entirely disperse on a timescale of $\lesssim 1\Myr$}
    {\revision{the gas component of PPDs with a mass of $<0.01 \Msun$, which has been reduced by mass-loss due to MHD winds, would disperse in less than $1\Myr$ at the final clearing stage}}. 
    In this case, photoevaporation models are incompatible with the scenario of gas-rich debris being protoplanetary remnants. However, since small grains are an essential component for the thermochemical structure of the disk, \added{\revision{FUV}} photoevaporation rates can vary with the grain growth and disk evolution. \cite{2015_Gorti} investigate the impact of dust growth on FUV photoevaporation with 1D~two-component simulations. 
    It is shown that increasing the average grain size results in reducing FUV photoevaporation rates  because of less efficient photoelectric heating.  
    \deleted{Remove this part? This part may lead readers to think the grains are depleted only at the truncated radii. This is not what we assume in this study.\cite{2019_Owen} propose radiation force of low-mass stars to be a mechanism for depleting small grains at pressure traps in disks truncated at inner radii. The grain depletion occurs on a timescale as short as $\sim 0.1\Myr$ and leaves a gas-rich, optically-thin disk. It is shown that such disks can observationally appear as gas-rich debris disks. }
    
    In optically-thin disks around luminous sources, grains are continuously subject to strong radiation forces. 
\deleted{
    The magnitude of the radiation force is often measured with respect to the gravity of the host star as
\eq{
\splitting{
    \beta & \equiv 
    \frac{3L_*Q_{\rm pr}}{16\pi c a \rho_{\rm b} GM_*}
    =  4.1
    \braket{\frac{L_*}{20\,L_\odot}} \braket{\frac{Q_{\rm pr}}{1}} \\
    &\times \braket{\frac{\rho_{\rm b}}{1.4\gram\cm^{-3}}}^{-1} \braket{\frac{a}{1\mum} }^{-1} \braket{\frac{M_*}{2\Msun}}^{-1},
     }  \label{eq:betarad} 
}
where $L_*$ is the stellar luminosity, 
$\rho_{\rm b}$ is the bulk density of grains,
$a$ is the grain size, 
$c$ is the speed of light,
and $Q_{\rm pr}$ is the transfer efficiency from radiation to momentum \citep[e.g.,][]{1979_Burns, 2006_Krivov}. 
If the radiation force is much stronger than the gravity ($\beta \gg 1$), 
grains would be blown out even in gas-rich systems.
The condition ($\beta > 1$) sets the minimum size of grains above which grains would remain in the gas-rich disks regardless of the radiation force,
\eq{
\splitting{
    a_{\rm min, rem}
    &\equiv
    4.1 \mum
    \braket{\frac{L_*}{20\,L_\odot}} \braket{\frac{Q_{\rm pr}}{1}} \\
    &\times \braket{\frac{\rho_{\rm b}}{1.4\gram\cm^{-3}}}^{-1} \braket{\frac{M_*}{2\Msun}}^{-1}.
    }\label{eq:maxs}
}
    Hence, small grains with $a < a_{\rm min, rem}$  
    }
    Since small grains have a large opacity, they are preferentially susceptible to the effects of the radiation force. This effect works to sustain a large average size of grains despite collisional shuttering of large bodies in the disk. Thus, if gas disk is present in such optically-thin disks, the mass-loss rates are expected significantly smaller than those of primordial PPDs. The dispersal timescale would be extended to much longer than $\sim 1\Myr$ in this case. 
    \added{\revision{
    On the other hand, the reduced disk opacity allows FUV photons to heat a higher-density region. Besides, dust settling leads to a higher ratio of dust to gas there, which can increase the specific heating rate. Therefore, it is also possible that the dispersal timescale of the optically-thin disks would be actually shortened.}}
    \deleted{However, it has been unclear to what extent the gas dispersal timescale can be extended quantitatively.}

    In this study, we investigate photoevaporation of gas-rich, optically-thin disks around A-type stars with two-dimensional (2D) axisymmetric radiation hydrodynamics simulations. 
    Small grains are assumed to be continuously removed from the disk by the strong radiation force \citep{2019_Owen}. 
    The main purpose of this study is to quantify the extension/reduction of the gas dispersal timescale 
    for gas-rich, optically-thin disks around intermediate-mass stars. 
    We compare the derived dispersal timescale with the system ages of gas-rich debris disks to examine if a photoevaporation model can be compatible with the primordial-origin scenario.
    This paper is organized as follows. In \secref{sec:methods}, we present the methods of our radiation hydrodynamics simulations. The results are shown in \secref{sec:results}, and we give discussions in \secref{sec:discussion}. Summary and conclusion are presented in \secref{sec:conclusion}. 

\section{Modeling Evolved Disks} \label{sec:methods}
	We perform radiation hydrodynamics simulations of gas-rich, optically-thin (at visual wavelengths), aged disks irradiated by the stellar FUV, EUV, and X-ray, assuming that small grains are continuously depleted by the effect of radiation forces despite collisional shuttering of large bodies at the midplane. We define such an optically-thin disk as ``evolved disk''. We consider an A-type star for the central radiation source in our fiducial model. We also consider the cases where the central source is a solar-type pre-main-sequence star in \secref{sec:solartype} for comparison. 
	\replaced{We note that our interest is in the dispersal timescale of the given evolved objects, while it is out of scope in this paper when PPDs become optically thin. }
	{\revision{In the present study, we are interested in mass-loss rates of such optically-thin disks to investigate plausibility of primordial-origin scenarios. This is motivated by the consideration that if the dispersal timescale of the evolved disks is shorter than the lifetime of debris disks, it makes the primordial-origin scenarios unlikely.
	Therefore, we start the simulations with an already optically thin disk, and the evolution toward this stage is out of the scope in this study.}}
	
	We use a modified version of the publicly available hydrodynamics simulation code, PLUTO \citep{2007_Mignone}, where we have implemented a variety of physics such as UV/X-ray photoheating, photochemical reactions, and multispecies chemical network \citep{2018_Nakatani, 2018_Nakatanib}. 
	Our methods largely follow \cite{2018_Nakatani} and \cite{2018_Nakatanib} (hereafter Papers~I and II, respectively) while the thermochemistry model is suitably updated for the evolved disks. 
	In this section, we describe the updates with a brief review of our methods in Papers~I and II.

	\subsection{Initial Configuration of Disks}   
        
	We consider an axisymmetric and midplane symmetric stratified disk surrounding a central star with a mass of $M_*$. The disk is initially isothermal in the vertical direction and is in hydrostatic equilibrium on the poloidal plane.  
	The density distribution of the hydrostatic disk is derived as 
        \eq{
                \nh = n_{\rm m}(R) 
                \exp \left[ -\frac{z^2}{2H^2} \right] , \label{eq:inidenstr}
        }
	where $(R,z)$ are the radial distance and the vertical height in cylindrical coordinates, respectively;
	$\nh$ is number density of hydrogen nuclei;
	$n_{\rm m} (R)$ is that at the midplane (i.e., the boundary condition at $z= 0$);
	and $H$ is the pressure scale height defined as the ratio of 
	the local, isothermal sound speed $\cs$ to the orbital frequency $\Omega \equiv \sqrt{GM_*/R^3}$, where $M_*$ is the stellar mass. 
    The initial isothermal sound speed is assumed to depend only on $R$. 
	The corresponding surface density is obtained by integrating \eqnref{eq:inidenstr} with respect to $z$,
	\eq{
        		\Sigma (R) =  \sqrt{2\pi} H \rho_{\rm m} (R) , 
	}
	where $\rho_{\rm m}$ is mass density on the midplane 
	and relates to $n_{\rm m}$ in terms of mean gas particle mass per hydrogen nucleus, $m$, 
	as $\rho_{\rm m} = m  n_{\rm m}$.

	We give the initial radial profiles of surface density and temperature as 
	\gathering{
		\Sigma = \Sigma_0 \braket{\frac{R}{R_0}}^{-1} 
		\label{eq:sigma}	\\	
		T = T_0  \braket{\frac{R}{R_0} }^{-1/2}, 
	}
	respectively, 
	where $R_0$ is a scale radius, 
	$\Sigma_0$ and $T_0$ are reference values at $R_0$.
	For the scale radius $R_0$, 
	we use the gravitational radius of EUV-driven flows, 
	whose sound speed is typically $10\kms$, 
	\eq{
		R_{\rm g} \equiv \frac{GM_*}{\braket{10\kms}^2} \approx 8.9\au \braket{\frac{M_*}{\Msun}}.
	}
	The total disk mass is 
	\eq{
		M_{\rm disk} = \int_{R_{\rm min} }^{R_{\rm max}} 2\pi R \Sigma \, \dd R
		= 2\pi R_0 (R_{\rm max} - R_{\rm min})\Sigma_0. 
		\label{eq:diskmass}
	}
	with an inner truncated radius $R_{\rm min}$ and an outer radial extent $R_{\rm max}$.
	The disk mass is a given parameter, and thus the surface density is set to $\Sigma\approx M_{\rm disk}/ (2\pi R_{\rm max} R)$.
	In the present study, an evolved PPD is assumed to be at ages of $\gtrsim 3 $--$6\Myr$. 
	Such disks can be truncated at several tens of au \citep{2016_Suzuki,2020_Kunitomo}, and thus we use $R_{\rm min} = 25\au$ in our fiducial model. 
	\deleted{We also run simulations with non-truncated (full) disks for comparison.}
	The outer radius is set to $R_{\rm max} = 20\,R_{\rm g}$.
	Old PPDs likely retain an amount of gas with $\sim 10^{-2}$--$10^{-1}\Msun$ even at $\sim 10\Myr$ regardless of continued mass loss via accretion, photoevaporation, and magnetic-driven winds \citep{2020_Kunitomo}. 
	Such high disk mass has also been observationally suggested for aged disks ($\sim 6\Myr$) around Herbig Ae/Be stars \citep{2008_Panic, 2017_Fedele, 2019_Booth, 2019_Miley}. 
	In our model, the initial disk mass ranges in 
	\replaced{$M_{\rm disk} \approx 10^{-3}$--$ 10^{-2} \Msun$.}
	{\revision{$M_{\rm disk} \approx 10^{-6}$--$ 10^{-2} \Msun$ depending on the assumed dust-to-gas mass ratio and the total amount of solids (\secref{sec:medium}).}}
	\deleted{The scale temperature is set to be $T_0 = 34\Kelvin$, which yields $T \approx 100\Kelvin $ at $1\au$ \citep[e.g.,][]{1987_KenyonHartmann}. }
	As will be shown in \secref{sec:results}, resulting mass-loss rates are primarily set by the EUV emission rate of the star and hardly depend on the disk mass or geometry. 
	The results of our simulations are also not sensitive to $T_0$. 
	\added{\revision{We give the initial abundances by computing thermochemistry and radiative transfer for $5\Myr$ with hydrodynamics disabled. The temperature is also set by this procedure, and it results in yielding an initial transient after the hydrodynamics simulations start. Nevertheless, the disk settles into a quasi-steady structure on a timescale of the vertical crossing time, which is well within the computational time ($> \text{a few}\times10^3\yr$). Therefore, this procedure does not significantly affect the resulting mass-loss rates. 
	}}

	\subsection{Stellar Properties and Disk Medium}	\label{sec:medium}

        \begin{table}[htp]
                \caption{Parameters for fiducial models}
                \centering
                \begin{tabular}{lcc}   
                & Strong-EUV model & Weak-EUV model \\
                \hline \hline
                [{\it Stellar parameters}]                                                       \\
                Spectral type                   & \multicolumn{2}{c}{A-type}\\
                Mass $(M_*)$     				& \multicolumn{2}{c}{$ 2  \Msun$ }                           \\
                Radius $(R_*)$                 	& \multicolumn{2}{c}{$1.88\Rsun$} \\
                Bolometric luminosity  $(L_*)$  &  \multicolumn{2}{c}{ $19\Lsun$}\\
                FUV luminosity $(\LFUV)$        & $3.0\e{33}\unit{erg}{}\unit{s}{-1}$ &  $6.7\e{33}\unit{erg}{}\unit{s}{-1}$ \\
                EUV emission rate $(\LEUV)$     & $5.3\e{40}\sec^{-1}$ & $4.4\e{38}\sec^{-1}$      \\      
                X-ray luminosity $(L_{\rm X})$	& \multicolumn{2}{c}{$1\e{29} \erg \sec^{-1}$}  	\\	
                \hline
                [{\it Disk parameters}]                                                       \\
                Configuration					&	

                \multicolumn{2}{c}{
				\begin{tabular}[t]{c}
                Truncated at $R_{\rm min} = 25\au$
				\deleted{,\\ or full, smooth disk}
				\end{tabular}}	
                                                    \\
                Outer extent $(R_{\rm max})$	&		\multicolumn{2}{c}{$20 \, R_{\rm g}$}	\\	
				Disk mass $(M_{\rm disk})$  	&    \multicolumn{2}{c}{
				\replaced{$10^{-3}$--$10^{-2}\Msun$}{\revision{$10^{-6}$--$10^{-2}\Msun$}}} \\
                \hline
                [{\it Gas/dust properties}]                                                     \\
                Gas species                                     & 
                \multicolumn{2}{c}{
				\begin{tabular}[t]{c}
					H, \ce{H+}, \ce{H2}, \ce{H2+}, \added{\revision{\ce{H-}}}, 
					CO, \\
					O, \added{\revision{\ce{C}}}, \ce{C+}, 
					\added{\revision{\ce{He}, \ce{He+}, \ce{H3+},}} \\
					\added{\revision{CH, \ce{CH+}, \ce{CH2}, \ce{CH2+}, \ce{CH3+},}}\\
					\added{\revision{\ce{CO+}, \ce{HCO+}, OH, \ce{OH+}, \ce{H2O},}}\\
					\added{\revision{\ce{H2O+}, \ce{H3O+}, \ce{O+}, \ce{O2},}} \ce{e-}
				\end{tabular}
				}
                      \\
                Carbon abundance $(\abn{C})$&                  \multicolumn{2}{c}{ $ 0.927\e{-4} $ }                                       \\
                Oxygen abundance $(\abn{O})$&                   \multicolumn{2}{c}{$ 3.568\e{-4} $ }                                        \\      
                Dust-to-gas mass ratio $(\dgratio)$	&       
                \multicolumn{2}{c}{
                \replaced{$ 0.01 $}{\revision{$ 0.001\text{--}0.1$}} 
                }
                \\
                Min. grain size ($a_{\rm max}$)		&	\multicolumn{2}{c}{\replaced{$10\mum$}{$4\mum$}}\\
                Max. grain size ($a_{\rm min}$)		&		
                \multicolumn{2}{c}{\replaced{$1\cm$--$100\km$}{\revision{$1\cm$}}}	\\
                \revision{Dust mass $(\solidmass)$}	&	\multicolumn{2}{c}{\revision{0.1--$10\Mearth$}}\\
                \hline
                \end{tabular}
                
                \label{tab:fiducialmodel}
        \end{table}

	The central star is a young A-type star with $M_* = 2 \Msun$ and $R_* = 1.9 \Rsun$ in our fiducial model. Since we consider an aged system, we assume that the UV radiation originates mostly from photospheric emission rather than driven by magnetic activity or accretion. 
	In this study, we use two UV luminosity models; (i) the UV spectrum is given by the blackbody radiation with an effective temperature of $\approx 9000\Kelvin$. The corresponding FUV luminosity and EUV emission rate are $L_{\rm FUV} = 3\e{33}\erg\sec^{-1}$ and $\Phi_{\rm EUV} = 5\e{40}\sec^{-1}$, respectively. (ii) We take into account the stellar atmosphere's absorption, which can reduce $\Phi_{\rm EUV}$ by orders of magnitude 
	\citep[e.g.,][]{1968_Spitzer}. We use $\LFUV$ and $\LEUV$ derived from the model spectrum of a $\approx10^4\Kelvin$ star \citep{2013_Husser, 2018_Fossati}. The corresponding FUV luminosity and EUV emission rate are $\LFUV = 6.7\e{33}\erg\sec^{-1}$ and $\LEUV = 4.4\e{38}\sec^{-1}$. 
	Hereafter, we refer to the models of (i) and (ii) as strong- and weak-EUV models, respectively (cf.~\tref{tab:fiducialmodel}). Note that EUV emission rates are very poorly known for young stars, and therefore it is essential to treat it as a parameter. 
	In the present study, we use the EUV emission rate of the strong-EUV model as an approximate high-end of $\LEUV$ for young A-type stars. The EUV emission rate is sensitive to the photospheric temperature, and it can be even as low as $\LEUV\sim 10^{34}\sec^{-1}$ for a $2\Msun$ star \citep{2021_Kunitomo}.
	As for X-ray luminosity, since A-type stars are lack of convective zones, it is typically weak compared to young low-mass stars \citep{2007_Schroder}. We adopt $L_{\rm X} = 1\e{29}\erg\sec^{-1}$. 
	\added{\revision{We also run additional simulations with a high-end X-ray luminosity ($L_{\rm X} = 10^{30}\erg \sec^{-1}$) to compare the results in \secref{sec:heatingprocesses}.}}
	In \secref{sec:solartype}, we also consider a pre-main-sequence solar-type star for the central radiation source 
	to compare the results with those of the A-type star model. 
	
	\replaced
	{The disk medium consists of dust and gas with a dust-to-gas-mass ratio of 0.01. 	
	The gas includes H, \ce{H+}, \ce{H2}, O, \ce{C+}, CO, and \ce{e-} as chemical components. }
	{\revision{
	The disk medium consists of dust and gas. 
	We treat the total amount of the solids from, $\solidmass$, 
	and the dust-to-gas-mass ratio, $\dgratio$,
	as free parameters ranging in $0.1\Mearth \leq \solidmass \leq 10\Mearth$
	and $0.001 \leq \dgratio \leq 0.1$, respectively. 
	The solid mass $\solidmass$ here accounts for the mass of $\sim \mum$- to cm-sized dust, as will be introduced below.
	We choose $\dgratio$ so that dust's drag force on gas is negligible, and that the total disk mass is less than $\sim 10^{-2}\Msun$; evolved disks are supposed to be lower in mass than (young) protoplanetary disks.  
	The total disk mass is computed by 
	\[
        M_{\rm disk} \approx 3\e{-4} \braket{\frac{\dgratio}{0.01}}^{-1} 
        \braket{\frac{\solidmass}{1\Mearth}}\Msun
	\]
	The dust-to-gas mass ratio is spatially uniform and is fixed throughout our simulations. Such a complete dynamical coupling between dust and gas might be an unrealistic assumption for large grains that have smaller Stokes numbers compared to the turbulent viscous $\alpha$; these grains would have settled into the midplane in evolved systems. Nevertheless, this assumption of complete coupling between gas and dust does not influence our conclusions owing to the limited impacts of large grains on the thermal structure of the disk, as will be shown in Sections~\ref{sec:results} and \ref{sec:discussion}.
	
	We solve nonequilibrium chemistry including advection
	for these 27 chemical components:
	H, \ce{H+}, \ce{H2}, \ce{H2+}, \ce{H-}, O, C, \ce{C+}, CO, \ce{e-},
	\ce{He}, \ce{He+}, \ce{H3+}, CH, \ce{CH+}, \ce{CH2}, \ce{CH2+}, \ce{CH3+}, \ce{CO+}, 
	\ce{HCO+}, OH, \ce{OH+}, \ce{H2O}, \ce{H2O+}, \ce{H3O+}, \ce{O+}, 
	and \ce{O2}.
	In our previous thermochemistry model of Papers~I and II, we have assumed that 
	the carbon atoms produced by CO photodissociation are immediately ionized
	to yield the ionization front identical to the dissociation front \citep{2000_RichlingYorke},
	and the rates of CO formation initiated by \ce{C+ + H2 -> CH2 + \gamma} 
	have been given following \citep{1997_NelsonLanger, 2005_NomuraMillar}. 
	These methods have allowed us to follow the evolution of the \ce{C+} and CO abundances
	without explicitly including atomic carbon in the chemical network. 
	In this study, we use an updated version of the previous chemical network 
	to accurately derive the atomic carbon abundance near the ionization front, 
	which is essential to calculate the heating rates there (see also \secref{sec:coolingheating}). 
	The chemical network includes the chemical reactions listed in \appref{app:chemicalnetwork}.
	The chemical abundances are updated at each time step in the simulations.}}
	The gas-phase elemental abundances of carbon and oxygen are 
	$\abn{C} = 0.927\e{-4}$ and $\abn{O} = 3.568\e{-4}$, respectively \citep{1994_Pollack,2000_Omukai}.

    In contrast to Papers~I and II where an ISM grain model is adopted, we update the grain model suitably for an evolved disk in this study. 
    Grains are comprised by mixture of water ice, silicate, and organics with a bulk density of 
    $\rho_{\rm b} = 1.4\gram \cm^{-3}$ \citep{2014_Kobayashi}.
    We assume that small grains are swept out from the evolved disks if the radiation force is stronger than the gravity.
    The magnitude of the radiation force with respect to the stellar gravity is measured by
\eq{
\splitting{
    \beta & \equiv 
    \frac{3L_*Q_{\rm pr}}{16\pi c a \rho_{\rm b} GM_*}
    =  4.1
    \braket{\frac{L_*}{20\,L_\odot}} \braket{\frac{Q_{\rm pr}}{1}} \\
    &\times \braket{\frac{\rho_{\rm b}}{1.4\gram\cm^{-3}}}^{-1} \braket{\frac{a}{1\mum} }^{-1} \braket{\frac{M_*}{2\Msun}}^{-1},
     }  \label{eq:betarad} 
}
where $L_*$ is the stellar luminosity, 
$\rho_{\rm b}$ is the bulk density of grains,
$a$ is the grain size, 
$c$ is the speed of light,
and $Q_{\rm pr}$ is the transfer efficiency from radiation to momentum \citep[e.g.,][]{1979_Burns, 2006_Krivov}. 
The condition ($\beta > 1$) 
reduces to 
\eq{
\splitting{
    a_{\rm min, rem}
    &\equiv
    4.1 \mum
    \braket{\frac{L_*}{20\,L_\odot}} \braket{\frac{Q_{\rm pr}}{1}} \\
    &\times \braket{\frac{\rho_{\rm b}}{1.4\gram\cm^{-3}}}^{-1} \braket{\frac{M_*}{2\Msun}}^{-1}.
    }\label{eq:maxs}
}
    Hence, the minimum grain size $a_{\rm min}$ is set to $a_{\rm min} = a_{\rm min, rem}$. 
    The fiducial stellar parameters yield $a_{\rm min, rem} \approx 4\mum$. 
    Technically, collisional shuttering of large bodies plus turbulent diffusion can resupply small grains $a < a_{\rm min, rem}$ into the upper layers of the evolved disk to increase the abundance.
    \replaced{Efficient collisions would make the evolved disks fail to sustain a grain-depleted state.}
    {\revision{
    The produced grains can also be trapped within the gas disk by drag force if gas density is sufficiently high. However, we have found that the outward-drift timescale is much shorter than the production timescale of small grains from a planetesimal disk due to collisional cascade \citep{2007_Wyatta, 2010_KobayashiTanaka}. The drift timescale is also much shorter than the dispersal time of the evolved disk, which will be derived by our simulations in the later sections (\secref{sec:lifetime}). Both of the timescales roughly scale with the disk mass, and thus we can assume our evolved disks to remain in a grain-depleted state regardless of the above effects. 
    }}
    In \secref{sec:blowout}, we present quantitative discussions in more detail. 
    \deleted{to validate our assumption on $a_{\rm min} $.} 
    Also, we provide $Q_{\rm pr}$ and $\beta$ as a function of grain size for A-type and solar-type stars in \secref{sec:stellardependence}. 

	\replaced{We range the maximum grain size $a_{\rm max}$ in $10^4 \mum \leq a_{\rm max} \leq 10^{11} \mum$ as a model parameter.} 
	{\revision{The maximum grain size is fixed at $a_{\rm max} = 1\cm$.}}
	A power-law size distribution is adopted with an exponent of $-3.5$. 
	The adopted parameters are summarized in \tref{tab:fiducialmodel}.

\deleted{
    \subsection{Minimum Dust Sizes in the Evolved Disk Model}   \label{sec:dustsize}
	In optically-thin layers of PPDs, grains are subject to radiation forces especially around luminous sources like A-type stars. The radiative acceleration exerting on a spherical compact grain is expressed as 
\[
\splitting{
    a_{\rm rad} &=	\frac{3L_*Q_{\rm pr}}{16\pi r^2 c s \rho_{\rm b} } \\
    & \approx  0.0054
    \braket{\frac{L_*}{20\,L_\odot}} \braket{\frac{Q_{\rm pr}}{1}} \braket{\frac{r}{30\au}}^{-2} \\
	&\times \braket{\frac{\rho_{\rm b}}{1.4\gram\cm^{-3}}}^{-1} \braket{\frac{s}{1\mum} }^{-1}
				\cm \sec^{-2},
	} 
\]
where $s$ is the grain size and $Q_{\rm pr}$ is the transfer efficiency from radiation to momentum \citep[e.g.,][]{1979_Burns, 2006_Krivov}. 
The transfer efficiency ranges in $0\leq Q_{\rm pr} \leq 2$ depending on dust size and peak wavelengths of stellar radiation $\lambda_{\rm peak}$ \citep{2008_Kobayashi}. 
Geometrical optics approximation is valid for $s \gg \lambda_{\rm peak}$, and therefore $Q_{\rm pr}$ is nearly unity. It decreases with $s$ from unity for grains smaller than $\lambda_{\rm peak}$. We discuss $Q_{\rm pr}$ of our model in \secref{sec:blowout}.

Accelerated grains reach the escape velocity on a timescale of 
\[
    \splitting{
    t_{\rm esc} & \equiv \frac{\sqrt{2GM_*/r}}{a_{\rm rad}}
    = \frac{16\pi r^{3/2} c s \rho_{\rm b} \sqrt{2GM_*}}{3L_* Q_{\rm pr}}\\
       &\approx 6.4 \braket{\frac{L_*}{20\,L_\odot}}^{-1} \braket{\frac{Q_{\rm pr}}{1}}^{-1} \braket{\frac{r}{30\au}}^{3/2}\\
                & \times \braket{\frac{\rho_{\rm b}}{1.4\gram\cm^{-3}}} \braket{\frac{s}{1\mum} } \braket{\frac{M_*}{2\Msun}}^{1/2} \yr. 
    }
\]
We define a nondimensional escape time as 
\eq{
\splitting{
    \beta & \equiv \frac{a_{\rm rad}}{GM_*/r^2}=  4.1
    \braket{\frac{L_*}{20\,L_\odot}} \braket{\frac{Q_{\rm pr}}{1}} \\
    &\times \braket{\frac{\rho_{\rm b}}{1.4\gram\cm^{-3}}}^{-1} \braket{\frac{s}{1\mum} }^{-1} \braket{\frac{M_*}{2\Msun}}^{-1},
     }  \label{eq:betarad} 
}
where $\Omega\equiv \sqrt{GM_*/r^3}$ is the Keplerian orbital frequency. Note that the nondimensional escape time is the ratio of gravity to radiation force and is independent of $r$. 
For $\beta > 1$, radiation force is stronger than the star's gravity. 
Such grains are radially accelerated into the outer direction. 
The condition, $\beta > 1$, corresponds to
\eq{
\splitting{
    s  & < s_{\rm esc,max} \equiv
    4.1 \mum
    \braket{\frac{L_*}{20\,L_\odot}} \braket{\frac{Q_{\rm pr}}{1}} \\
    &\times \braket{\frac{\rho_{\rm b}}{1.4\gram\cm^{-3}}}^{-1} \braket{\frac{M_*}{2\Msun}}^{-1}, 
    }\label{eq:maxs}
}
for $s \gtrsim \lambda_{\rm peak}$, where $Q_{\rm pr} \sim 1$. 
Assuming that grains with $s < s_{\rm sec,max}$ have been completely swept out from the system by radiation pressure, we set the minimum grain size of dust to $10\mum$ in our fiducial model. Technically, $Q_{\rm pr}$ depends on grain size, and thus $\beta = 1$ can give multiple roots in general (see also \secref{sec:blowout} and \fref{fig:Qpr}). There is also a lower limit to $s $ below which grains can remain in the disk regardless of radiation pressure because of reduced $Q_{\rm pr}$ in this case. In addition, aerodynamical friction between dust and gas can decelerate the radial velocity induced by the radial pressure and can extend the escaping time. We discuss these effects on $s_{\rm esc, max}$ and the validity of the adopted minimum grain size in detail in \secref{sec:blowout}. 
}

    \subsection{Hydrodynamics}     \label{sec:basicequations}
    We perform simulations in 2D spherical polar coordinates $(r, \theta)$.
	The code solves the temporal evolution of gas density $\rho$,
	three dimensional velocity $\vec{v} = (v_r,~ v_\theta, ~ v_\phi)$,
	total gas energy density,
	and chemical abundances $\{y_i\}$.
    The basic equations are
        \gathering{
                \frac{\partial \rho}{\partial t} + \nabla \cdot \rho \vec{v}                                             =  0 ,                  \\
                \frac{\partial \rho v_r}{\partial t} + \nabla \cdot \left( \rho v_r \vec{v} \right)              =  -\frac{\partial P}{\partial r}
                                -\rho \frac{GM_*}{r^2} + \rho \frac{v_\theta^2 + v_\phi^2}{r}                   ,                       \\
                \frac{\partial \rho v_\theta}{\partial t} + \nabla \cdot \left( \rho v_\theta \vec{v} \right)    = - \frac{1}{r}\frac{\partial P}{\partial \theta }
                                - \rho \frac{v_\theta v_r}{r} + \frac{\rho v_\phi^2}{r} \cot \theta                     ,                       \\
                \frac{\partial \rho v_\phi}{\partial t} + \nabla ^l \cdot \left( \rho v_\phi \vec{v} \right)     = 0     ,       \label{eq:euler_phi}            \\              
                \frac{\partial E}{\partial t} + \nabla \cdot \left(H \vec{v} \right)                                     = - \rho v_r \frac{ GM_* }{r^2} +\rho \left( \Gamma -\Lambda    \right),                        \\
                \text{and} \quad
                \frac{\partial \nh y_i }{\partial t} + \nabla \cdot \left( \nh y_i \vec{v} \right)               = \nh R_i       .                       \label{eq:chemevoeq}
        }
        Here $P$ denotes the gas pressure;
        $E$ and $H$ are the total gas energy density and enthalpy density, 
        \[
        \begin{gathered}
        E =  \frac{1}{2} \rho v^2 + \frac{P}{\gamma-1}\\
        H =  E + P, 
        \end{gathered}
        \]
        respectively, where $\gamma$ is specific heat ratio;
        $\Gamma$ is the total heating rate per unit mass (specific heating rate),
        and $\Lambda$ is the total cooling rate per unit mass (specific cooling rate);
        and $R_i$ is the total reaction rate for the corresponding chemical species $i$.
	The azimuthal component of Euler equation (\eqnref{eq:euler_phi}) is written in the angular momentum conserving form \citep[see][for detail]{2018_Nakatani}. 
	The disk self-gravity is negligible in our simulations. 
	The sound crossing time is much shorter than the viscous timescale in this study. 
	Besides, mass loss is dominated by photoevaporation for evolved disks. 
	Therefore, we do not incorporate viscous friction in the conservation equations for energy and angular momentum. 
    Further detailed information are described in Papers~I and II.

    \subsection{Thermochemistry}        \label{sec:coolingheating}
	In Papers~I and II, we have developed a multispecies chemical network
	that includes relevant collisional reactions 
	and photochemical reactions such as H/\ce{H2} photoionization,
	\ce{H2} photodissociation \citep{1996_DraineBertoldi}, 
	and CO photodissociation \citep{1996_Lee}. 	
	For photoheating processes, 
	we implement EUV- and X-ray-induced ionization heating 
	\citep{1996_Maloney, 2000_Wilms, 2004_Gorti}
	and grain photoelectric heating caused by FUV absorption
	\citep{1994_BakesTielens}. 
	We also incorporate cooling sources: 
	radiative recombination cooling of \HII{} \citep{1978_Spitzer}, 
	Ly${\rm \alpha}$ cooling of \HI{} \citep{1997_Anninos}, 
	fine-structure line cooling of \OI{} and \CII{} \citep{1989_HollenbachMcKee,1989_Osterbrockbook,2006_SantoroShull},
	molecular rovibrational line cooling of \ce{H2} and CO \citep{1998_GalliPalla,2010_Omukai}, 
	and dust-gas collisional heat transfer \citep{1996_YorkeWelz}. 
	\added{\revision{In this study, 
	we additionally incorporate \ce{H2} photodissociation heating \citep{1979_HollenbachMcKee, 1996_DraineBertoldi}, 
	heating by \ce{H2} pumping \citep{2006_Rollig}, 
	\CI{} ionization heating \citep{1987_Black, 2004_Jonkheid, 2012_UMIST},
	chemical heating/cooling \citep{1979_HollenbachMcKee, 2000_Omukai}.
	We describe the implementation of these processes in \appref{app:newheatcool}.
	}}

	The UV and X-ray fluxes are computed in a frequency-dependent manner with ray-tracing radiative transfer. 
	\added{\revision{We ignore the diffuse EUV based on the results of \citet{2013_Tanaka} and use case~B recombination (see Paper I and II for more detailed discussions).}}
	We consistently determine dust temperatures by solving 2D radiative transfer for both of the direct stellar irradiation and diffuse thermal emission of dust with a hybrid scheme \citep{2010_Kuiper, 2013_Kuiper, 2020_Kuiper}. We use the two-temperature approach of the radiative transfer. 
    Dust opacity is consistently given according to the adopted dust properties. 
	\deleted{Note that opacity is dominated by grains with $a \lesssim 10^{4} \mum$.} 

    
    \deleted{
    In contrast to Papers~I and II, where an ISM grain model is adopted, we update it suitably for an evolved disk in this study. 
    Grains are comprised by mixture of water ice, silicate, and organics with a bulk density of $\rho _{\rm b} = 1.4\gram \cm^{-3}$.
	The maximum size of grains, $a_{\rm max}$, is a parameter, which ranges in $10^4 \mum \leq a_{\rm max} \leq 10^{11} \mum$. (Recall that the minimum grain size is $a_{\rm min} = 10\mum$.)
	A power-law size distribution is adopted with an exponent of $-3.5$. 
	Note that opacity is dominantly contributed by grains with $1\mum \leq a \lesssim 10^{4} \mum$. 
    }
        
	In Papers~I and II, we give the dust extinction factor for FUV 
	as $\exp\braket{-C \Av}$, where $C (\simeq 2\texttt{--}3)$ is a constant and $\Av$ is the visual extinction.
    Since the dust model is lack of small grains in this study, 
	FUV extinction is approximately equal to extinction at visual wavelengths. 
	Therefore, we adopt $C = 1$
	for all of the photochemical reactions and heating processes. 
	The visual extinction relates to column density $\col{H}$ as 
	\[
		\Av = \Sigma_{\rm V} \col{H},
	\]
	where $\Sigma_{\rm V}$ is visual extinction per hydrogen column density. ISM grains typically yield $\Sigma_{\rm V} \approx 5 \e{-22} \unit{mag}{}\cm^2$ with $a_{\rm min} \approx 0.005 \mum$ and $a_{\rm max} \approx 0.1\texttt{--} 1\mum$. 
	The magnitude of $\Sigma_{\rm V}$ is determined by the total surface area of grains at wavelengths longer than grain size, and thus it can be approximately proportional to $(a_{\rm min} a_{\rm max})^{-1/2}$ with the assumed power-law index for $a_{\rm max} \gg a_{\rm min}$. 
	In this study, we scale $\Sigma_{\rm V}$ with $a_{\rm min}$ and $a_{\rm max}$ as
		\eq{
			\Sigma_{\rm V} = 5\e{-22} \braket{\frac{a_{\rm min}}{5{\rm \,nm}}}^{-1/2}
			\braket{\frac{a_{\rm max}}{1\mum}}^{-1/2}
			\unit{mag}{}\cm^2.	
			\label{eq:sigmaV}
		}		
		This scaling of $\Sigma_{\rm V}$ can overall give a good approximation to actual values of visual extinction; the deviation is within a factor of three. 
	With the adopted $a_{\rm min}$ and $a_{\rm max}$, $\Sigma_{\rm V}$ ranges in $10^{-7} \lesssim (\Sigma_{\rm V}/5\e{-22}\unit{mag}{}\cm^2) \lesssim 2\e{-4}$. 
		Papers~I and II use the photoelectric heating rate presented by \cite{1994_BakesTielens} (hereafter BT94), where
		carbonaceous grains are assumed with $ a_{\rm min} = 3\e{-4}\mum$ and $a_{\rm max} = 0.01\mum$. 
		BT94 shows that small grains ($\lesssim 15$\,\AA) are responsible for a half of the total heating rate. \cite{2001_WeingartnerDraine} incorporate larger grains up to $a_{\rm max} = 1\mum$ and also found that small grains dominate heating. 
        Thus, modifications are necessary to photoelectric heating rates for our evolved disk model where small grains are absent.
		Photoelectric yield generally decreases as the grain size increases, and it reaches laboratory values of bulk solids for sufficiently large-sized grains. 
		Our large grain sizes would result in a small photoelectric heating efficiency. Large grain sizes also significantly reduces the total cross section of grains to FUV photons depending on $a_{\rm min}$ and $a_{\rm max}$. These two effects make the photoelectric heating rates smaller by orders of magnitude compared to those for ISM grains. 
		Photoelectric heating is therefore presumably ineffective 
		to heat the gas in evolved disks, as opposed to young disks that supposedly have abundant small grains and polycyclic aromatic carbons (PAHs). 
		Nevertheless, we incorporate it in our thermochemistry model,
		employing an approximate modification to the photoelectric heating rate of BT94 
		so that it is computed consistently with \eqnref{eq:sigmaV} 
		(see \appref{app:modificationFUVheating} for more detail). 		
		More detailed modeling of photoelectric heating for evolved disks is out of the scope in the present work. This issue will be addressed in future work (Nakatani \& Tazaki in prep). 

        As well as photoelectric heating, dust-gas collisional heat transfer rates and grain-catalyzed \ce{H2} formation rates need to be modified suitably for the evolved disk model in the present study. 
        We give specific dust-gas collisional heating/cooling rates 
        for the adopted dust size distribution ($\dd n_{\rm d}/\dd a\propto a^{-3.5}$) by
		\eq{
		\splitting{
		    \Lambda_{\rm dust} &= 
			- 4 \pi \cs \,
			k (T - T_{\rm dust})
			\frac{\nh}{\rho}			
		    \int _{a_{\rm min}} ^{a _{\rm max}} a^2 \frac{\dd n_{\rm d}}{\dd a} \,\dd a \\
		    & \approx 
		    - 4 \pi \cs \,k (T - T_{\rm dust}) \frac{\nh}{\rho_{\rm b}}	
            \frac{\rho_{\rm dust}}{\rho}
            \frac{3}{4\pi \sqrt{a_{\rm min}a_{\rm max}}},
		    }
		    \label{eq:dust-gascollisional}
		}
		where $\rho_{\rm dust}$ is dust density and 
		\replaced{$\rho_{\rm dust} = 0.01 \rho$}
		{\revision{$\rho_{\rm dust} = \dgratio \,\rho$ is the density of dust grains smaller than $1\cm$.}} 
        We use grain-catalyzed \ce{H2} formation rates scaled by grain sizes in the same manner as $\Sigma_{\rm V}$ in the chemical network.

        \subsection{Numerical Configuration}
        The computational domain extends as
        $r = [0.1, 20]\times R_{\rm g}$        
		and $\theta = [0, ~\pi/2] {\rm \, rad}$.
        We use a large outer radial extent to avoid the spurious reflection of outflows 
        at the boundary \citep{2018_Nakatani}. 
        The disk is axisymmetry around the rotational axis $(\theta = 0)$ and is symmetric with respect to the midplane $(\theta = \pi /2)$.
        The computational domain is logarithmically spaced with 128 cells in the radial direction. 
        The meridional direction is divided into two domains at $\theta = 1\rad$; each of them is uniformly spaced with 80 cells. 
        We use an approximately two-times higher spacial resolution for $1\leq \theta \leq \pi/2$ than for $0\leq \theta \leq \pi/2$ to resolve the small scale height of PPDs and the launching points of photoevaporative flows (so-called base).

\section{Results} \label{sec:results}
Results of the radiation hydrodynamics simulations are presented in this section. 
We describe the physical structures of photoevaporating evolved disks in \secref{sec:structure}. 
In \secref{sec:evarates}, measured photoevaporation rates are given and are compared with those for primordial PPDs, where an ISM grain model is assumed. 
We estimate lifetimes for the gas components of evolved disks using the measured photoevaporation rates in \secref{sec:lifetime}. 
We also run simulations with evolved disks around a solar-type central star and compare the results with those of the fiducial A-type star cases in \secref{sec:solartype}.

\subsection{Overview of Physical Structures}	\label{sec:structure}

\begin{figure*}[htbp]
\begin{center}
\includegraphics[clip, width = \linewidth]{\figdir/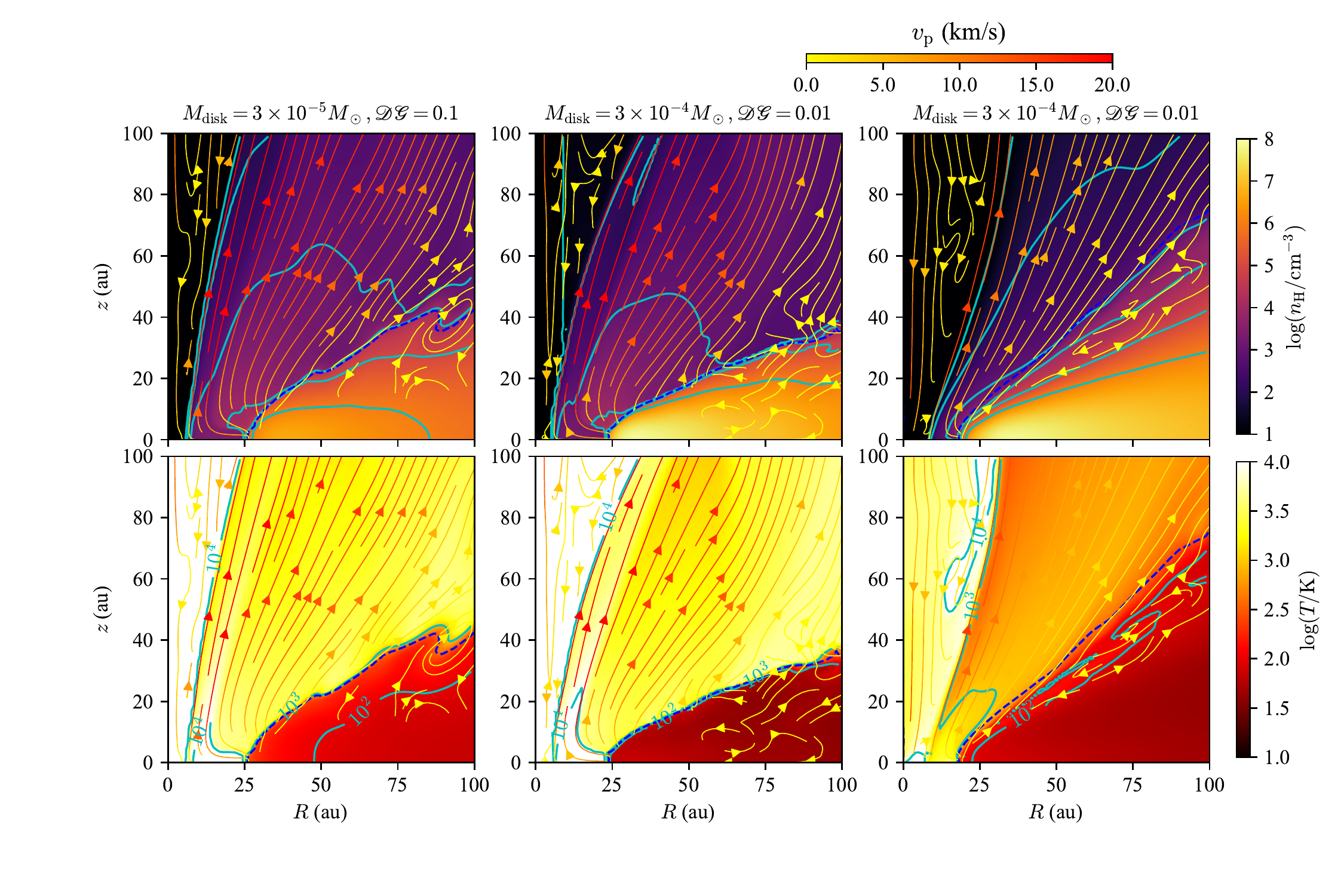}
\caption{
		Snapshots of the strong-EUV model with 
		\replaced{$M_{\rm disk} = 10^{-3}\Msun$ and $a_{\rm max} = 1\cm\text{ (left column)}, 10{\rm \, m} \text{ (middle column)}, 100\km \text{ (right column)}$. }
		{\revision{$M_{\rm disk} = 3\e{-5}\Msun$ and $\dgratio = 0.1$ (left panels), $\dgratio = 0.01$ (middle panels). The right panels show the snapshots of the weak-EUV model with $M_{\rm disk} = 3\e{-4}\Msun$ and $\dgratio = 0.01$. The total solid mass is $\sim 1\Mearth$ for each case.}}
		The top and bottom panels show the density and temperature structures, respectively. 
		The cyan contours indicate $\nh = 10^3$, $10^4$, $10^5$, $10^6$, $10^7$, and $10^8\cm^{-3}$ in the upper panels,
		and $T = 10$, 100, 1000, $10^4\Kelvin$ in the lower panels. 
		\added{\revision{The blue dashed lines indicate the boundary at which $\eta = 0$ (\eqnref{eq:eta})}}. 
		The arrows show poloidal velocity fields, and the colors represent the magnitude.
		The systems have reached a quasi-steady state up to this point. 
		}
\label{fig:snapshots}
\end{center}
\end{figure*}
\fref{fig:snapshots} shows the density and temperature structures of photoevaporating evolved disks   
\replaced{for the strong-EUV model with $a_{\rm max} = 1\cm, 10 {\rm \,m}, 100\km$, which corresponds to $\Sigma_{\rm V} \approx 10^{-25}, 10^{-27}, 10^{-29} \unit{mag}{}\cm^{2}$, respectively. }
{\revision{for various $M_{\rm disk}$ and $\dgratio$ in the strong- and weak-EUV models.}}
Photoevaporative flows are excited from the surfaces of the cold disks (dark-orange regions in the temperature maps of \fref{fig:snapshots}).
For the strong-EUV runs, the flows have velocities of $\sim 10\kms$ and go outward after being excited at the base.
The flow regions are fully photoionized via \ce{H + $\gamma$ -> H+ + e-} by the stellar EUV ($13.6\eV \leq h\nu\lesssim 100\eV$) to form an \ce{H+}-dominated region (so-called \ion{H}{2} region).
The typical temperature is of the order of $10^3$--$10^4\Kelvin$ there. 
Density is $\sim 10^4$--$10^5\cm^{-3}$ at the launching base of the photoevaporative flows
and decreases by the expansion with increasing distance from the disk surface.
The launching base is largely identical to the H-ionization front.
Regarding the weak-EUV runs, 
we observe partially ionized photoevaporative flows at a slow velocity of $< 10\kms$. The flow density and temperatures are smaller by an order of magnitude and by a factor, respectively, compared to the corresponding strong-EUV runs. 
These results show that photoevaporative flows are excited by the stellar EUV for the evolved disks, and a higher $\LEUV$ yield a stronger photoevaporation (and thereby higher mass-loss rates; see \secref{sec:evarates}).

\begin{figure*}[htbp]
\begin{center}
\includegraphics[clip, width = \linewidth/2-0.1cm]{\figdir/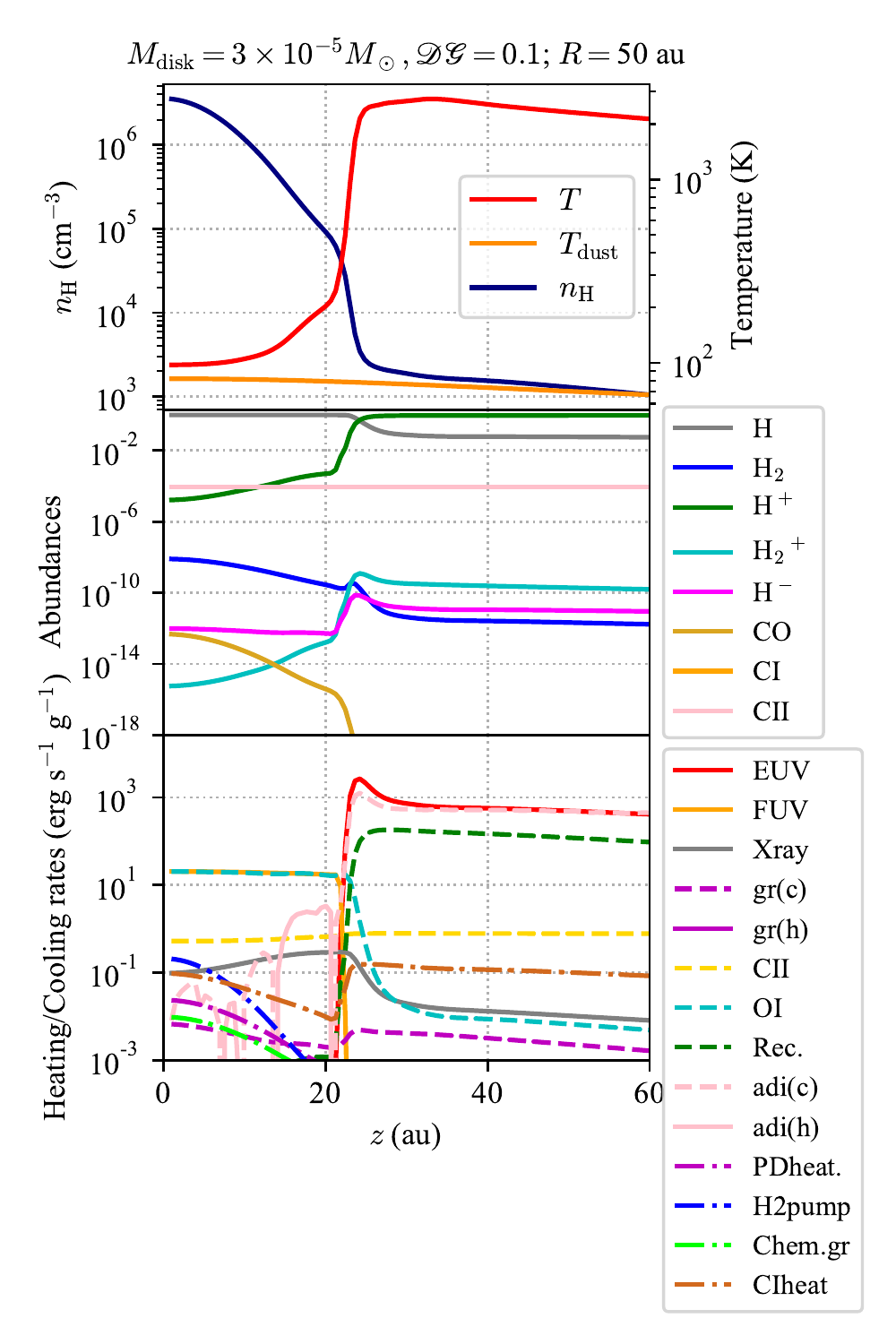}
\includegraphics[clip, width = \linewidth/2-0.1cm]{\figdir/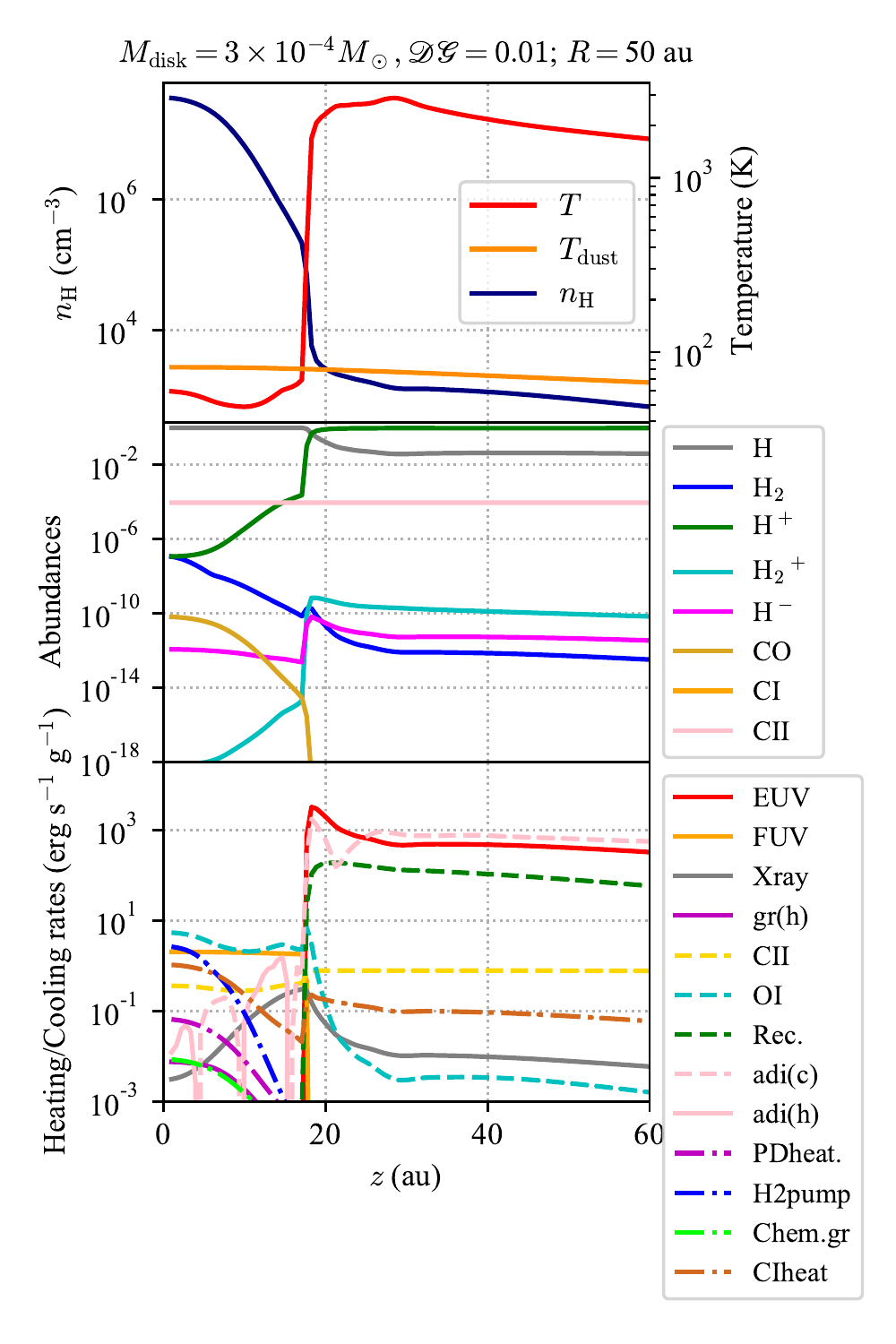}
\caption{Vertical profiles of physical quantities at $R = 50\au$ 
for the strong-EUV runs with 
\replaced{$a_{\rm max} = 1\cm$ (left) and $a_{\rm max} = 100\km$ (right).}
{\revision{$M_{\rm disk} = 3\e{-5}\Msun$ and $\dgratio = 0.1$ (left), and with $M_{\rm disk} = 3\e{-4}\Msun$ and $\dgratio = 0.01$ (right).}}
The top panels show density $\nh$ and temperatures of gas and dust. 
The middle panels show abundances of major hydrogen-/\revision{carbon-}bearing species,
and the bottom panels show specific heating/cooling rates. 
The solid and dash dotted lines of the bottom panels indicate heating, while dashed lines indicate cooling. In the legend, ``FUV'' means photoelectric heating, ``CII'' and ``OI'' are line cooling of each species, ``Rec'' is radiative recombination cooling of hydrogen, \revision{``PD heat'' is photodissociation heating, ``H2pump'' is \ce{H2} pumping, ``Chem.gr'' is chemical heating by \ce{H2} formation on dust, and ``CIheat'' is photoheating associated with \ion{C}{1} ionization}. 
``gr(c/h)'' and ``adi(c/h)'' in the legend stand for cooling/heating due to dust-gas collisional heat transfer and due to adiabatic expansion/compression, respectively. \revision{We do not show the rates of the other heating/cooling incorporated in the simulations because they are below the lower frames.} 
}
\label{fig:coolingheating}
\end{center}
\end{figure*}
\fref{fig:coolingheating} shows the vertical profiles of various physical quantities at $R = 50\au$ 
\replaced{for the strong-EUV runs with $a_{\rm max} = 1\cm$ and $100\km$.}
{\revision{for the runs shown in the left and middle panels of \fref{fig:snapshots}.}} 
Thermalization of the ejected electrons due to EUV photoionization results in the high temperatures of the gas in the \ion{H}{2} regions. 
The EUV heating largely balances with adiabatic cooling associated with the gas expansion. 
This trend is common in all runs. \deleted{with various $a_{\rm max}$.}
For the weak-EUV cases, 
the EUV heating rates are an order of magnitude smaller than the strong-EUV cases. The resulting temperatures of the \ion{H}{2} region is correspondingly smaller compared to the strong-EUV runs (\fref{fig:snapshots}). 

\deleted{Thermalizing electrons are ejected from solids in cases of FUV photoelectric heating. The heating efficiency scales with the absorption cross-section. As $a_{\rm max}$ increases, the total grain surface area is effectively reduced. FUV photoelectric heating accordingly weakens as $a_{\rm max}$ increases. 
On the other hand, thermalizing electrons are mainly ejected from gas species in cases of photoheating associated with X-ray ionization, and therefore the heating rate does not depend on the grain sizes. 
Consequently, the dominant heating source is FUV photoelectric heating for $a_{\rm max } \lesssim 1{\rm \,m}$, while it is replaced by X-ray heating for $a_{\rm max} \gtrsim 10{\rm \,m}$. 
Temperatures reaches $\sim 100\Kelvin$ in either case, but it is not sufficiently high to drive photoevaporation. The temperature affects the geometry of the disks via the scale height. Since FUV attenuates at a larger column $(\sim \Sigma_{\rm V} ^{-1})$ than X-ray ($\sim 10^{21} \cm^{-2}$) in the evolved disks, it can heat in a deeper interior of the disks than X-ray.
FUV heating is sufficiently strong to increase the scale height for $a_{\rm max } \lesssim 1{\rm \, m}$. It is evident in \fref{fig:snapshots} where the disk has a relatively inflated structure compared to the cases with $a_{\rm max} = 10\mum$.}

As for the neutral (H- or \ce{H2}-dominated) region, dust-gas heat transfer is negligible compared to line cooling of metals in the evolved disks, as opposed to young disks where the total surface area of grains is orders of magnitude larger owing to abundant small grains \citep{2018_Nakatani, 2018_Nakatanib}. Hence, dust and gas thermally decouple. 

\deleted{Regarding \ce{H2} abundance, the strong FUV radiation almost completely photodissociates \ce{H2} in $R\lesssim50\au$. The geometrical aspect ratio of \ce{H2} disks is $\approx0.1$--$0.15$ in $R\gtrsim 50\au$ for $a_{\rm max} = 1\cm$ and $\lesssim 0.05$ for $a_{\rm max} = 100\km$. A smaller total surface area of grains makes \ce{H2} production less efficient. and thus the total \ce{H2} mass is smaller with increasing $a_{\rm max}$.}
\replaced{In the H- or \ce{H2}-dominated (neutral) regions, either of FUV or X-ray works as a dominant heating source.}
{
\revision{
\subsubsection{Chemical Abundances of \ce{H2} and CO}
The neutral region is overall dominated by atomic hydrogen for low-mass disks ($M_{\rm disk} \lesssim 10^{-3} \Msun$) in a quasi-steady state. This is in contrast to young protoplanetary disks, where grain-catalyzed \ce{H2} production is efficient owing to a larger total surface area of grains. This \ce{H2} formation process is significantly reduced in the evolved disks. Instead, the \ce{H-}-process, \ce{H- + e -> {\gamma} + H-} followed by \ce{H- + H -> H2 + e}, dominates \ce{H2} formation. The \ce{H2} abundance is essentially determined by the \ce{H-} process, \ce{H2} photodissociation, and \ce{H-} photo-detachment, \ce{H- + {\gamma} -> H + e}. The photochemical reactions are fast with the intense FUV radiation of intermediate-mass stars and destroy \ce{H2} and \ce{H-} quickly. The resulting \ce{H2} abundance is no larger than $\lesssim 10^{-6}\text{--}10^{-5}$ even in the midplane. 

Ionized carbon is the most abundant among carbon-bearing species. The ion is produced through photoionization of atomic carbon preceded by photodissociation of CO. Atomic carbon is almost completely destroyed, and CO exists at very small abundances ($\lesssim 10^{-8}$). The molecule forms mainly via \ce{H + CO+ -> CO + H+}, where \ce{CO+} is supplied from \ce{H + O -> OH} followed by \ce{OH + C+ -> CO+ + H}. The total CO mass $M_{\rm CO}$ is of the order of $10^{-8}\Mearth$ for $M_{\rm disk} \sim 10^{-4}\Msun$ and $\lesssim 10^{-11}\Mearth$ for $M_{\rm disk} \sim 10^{-5}\Msun$. (See also \secref{sec:debris} for comparison of the resulting $M_{\rm CO}$ with observations towards gas-rich debris disks.)

\begin{figure*}
    \centering
    \includegraphics[clip, width = \linewidth]{\figdir/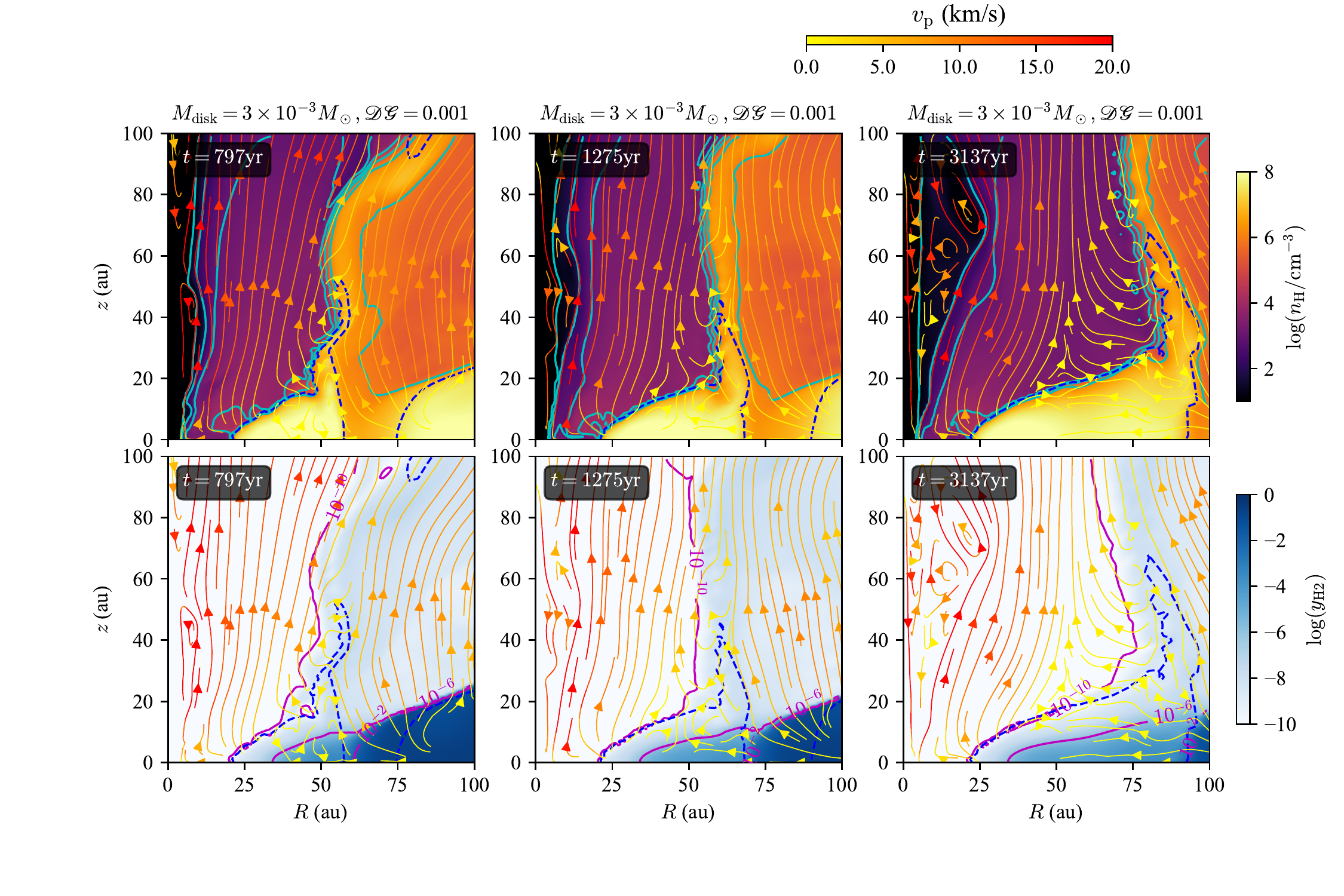}
    \caption{Same as \fref{fig:snapshots} but we show the sequence of the evolution for $M_{\rm disk} = 3\e{-3}\Msun$ and $\dgratio = 0.01$ from left to right. The bottom panels show the \ce{H2} abundances instead of $T$. The magenta contours indicate $\abn{\ce{H2}} = 10^{-10}, 10^{-6}, 10^{-2}$. }
    \label{fig:snapshotsH2}
\end{figure*}
For $M_{\rm disk} \gtrsim 10^{-3}\Msun$, since density is high ($\gtrsim 10^8\text{--}10^{9}\cm^{-3}$), disks have the \ce{H2}-rich layer around the midplane that survives for more than thousands of years. The \ce{H2} abundance is of the order of $\sim 10^{-2}\text{--}10^{-1}$ in the disk, implying that atomic hydrogen is still (moderately) dominant. The \ce{H2}-rim gradually moves outwards, as the \ce{H2}-disk is eroded from its surface by photodissociation (\fref{fig:snapshotsH2}). Simultaneously, chemically-neutral photoevaporative flows are excited from the \ce{H2}-rim by heating due to \ce{H2} pumping. 
When the head of the \ce{H2}-rim is at $R = R_{\rm rim}$, the dissociation timescale within the disk is approximately given by 
\eq{
\splitting{t_{\rm dis} (R) 
                        & =  \braket{I_0 f_{\rm shield} }^{-1}\\
                        & \approx 5\e{2}\yr 
                        \braket{\frac{\LFUV}{10^{34}\erg\sec^{-1}}}^{-1}
                        \braket{\frac{H/R_{\rm rim}}{0.1}}^{-3/4}
                        \braket{\frac{\abn{\ce{H2}}}{0.1}}^{3/4}
                        \\  &   \times
                        \braket{\frac{R_{\rm rim}}{30\au}}^{2}
                        \braket{\frac{\Sigma(R_{\rm rim})}{0.1\gram\cm^{-2}}}^{3/4}
                        \braket{1- \frac{R_{\rm rim}}{R}}^{3/4},
}
\label{eq:tdis}
}
where $f_{\rm shield} = (N_{\rm \ce{H2}}/10^{14}\cm^{-2})^{-3/4}$ is the self-shielding factor \citep{1996_DraineBertoldi}, and $I_0$ is an unattenuated photodissociation rate given by $I_0= 4\e{-11} \FFUV \sec^{-1}$ with a normalized FUV flux of $\FFUV$ (see \appref{app:modificationFUVheating} for the definition). In the derivation of $N_{\rm \ce{H2}}$, we have assumed constant $H/R$ and $N_{\rm H2} \approx \abn{\ce{H2}} N_{\rm H}$. The former is well justified from the resulting disk structure, and the latter is valid for an order-of-magnitude estimation here. 
The dissociation front sweeps the \ce{H2}-disk on timescales of $t_{\rm dis}$, dispersing the disk gas. However, not all of the gas is lost if $\Sigma/\dot{\Sigma} > t_{\rm dis}$. This is the case for $10^{-3}\Msun \lesssim M_{\rm disk} \lesssim 10^{-2} \Msun$. In these simulations, the dissociation front expands to $R \approx 100\au$ by $t \sim 3000\yr$. About 70\% of the initial mass is lost in the swept region, while the rest forms an H-disk. The H-disk is not capable of reproducing an \ce{H2}-rich region because photodissociation is quite efficient with unattenuated dissociating radiation.

Photodissociation, photo-detachment, and \ce{H-} process determine \ce{H2} abundances ($\lesssim10^{-3}$) in the swept region and in the photoevaporative flows. Replenishment of \ce{H2} molecules by advection is also important near the launching base. An exchange reaction, \ce{H2 + O -> OH + H}, along with photodissociation is effective as a destruction process there owing to fairly high temperatures ($\approx 10^3\Kelvin$). If the density is high ($\gtrsim 10^8\cm^{-3}$), \ce{H2O+ + e -> O + H2} contributes to production of the molecules.  

Similarly to \ce{H2}, a CO-rich layer is present within the disk. 
The molecule is protected by self-shielding and mutual-shielding of \ce{H2}. Atomic carbon and ionized carbon are largely absent in the CO disk. The CO dissociation front also moves outwards and is generally ahead of the \ce{H2} dissociation front. 

It is noteworthy that our results for $M_{\rm disk} \gtrsim 10^{-3}\Msun$ are more or less influenced by the initially given \ce{H2} amount. We have set the initial abundances by simply computing thermochemistry for $5\Myr$ with hydrodynamics disabled, but in practice the abundances would also be affected by the history of grain growth, stellar/disk evolution, etc. Further follow-up studies are necessary to address this issue by following a long-term evolution of disk chemistry. 
Nevertheless, it is intriguing that our evolved disks show a significant $M_{\rm disk}$-dependence in photoevaporation. This is in contrast to optically-thick, primordial disks; since photoevaporation is a ``surface process'', it is essentially determined by stellar luminosities and does not directly depend on $M_{\rm disk}$ or midplane density \citep[e.g.,][]{2019_Wolfer}. 
Our results demonstrate that density dependence of (photo)chemical processes can result in a certain $M_{\rm disk}$-dependence of photoevaporation for inner truncated disks (see also \secref{sec:evarates} for discussion on mass-loss rates).

\subsubsection{Heating Processes}   \label{sec:heatingprocesses}
Grain photoelectric effect dominates heating for disks with low gas density ($\nh \lesssim 10^7\cm^{-3}$), i.e. low-mass disks ($M_{\rm disk} \approx 10^{-6} \text{--}10^{-4}\Msun$). This is the case when $\dgratio = 0.1$ with $\solidmass \lesssim 10\Mearth$ and $\dgratio = 0.01$ with $\solidmass \lesssim 1 \Mearth$. For $M_{\rm disk} \approx 3\e{-4}\Msun$, \ce{H2} pumping and \ion{C}{1} ionization heating are comparable. They are slightly stronger than photoelectric heating in the midplane at $R \lesssim 50\au$ with $\dgratio = 0.01$ 
and dominate photoelectric heating with $\dgratio = 0.001$. 
Nevertheless, the resulting temperature is too low ($\lesssim 10^2\Kelvin$) to excite photoevaporative flows from the neutral region for these low-mass disks. 
 
For higher-mass disks ($M_{\rm disk} \gtrsim 10^{-3}\Msun$), the gas temperature can get $\approx 10^3\Kelvin$ at the rim of the \ce{H2} disks by \ce{H2} pumping, and it results in H-rich photoevaporative flows. Ionization heating of \ion{C}{1} also contributes to heating, but the effect is limited compared to \ce{H2} pumping. Photoelectric heating is even more weaker and is negligible. 
For $10^{-3}\Msun \lesssim M_{\rm disk} \lesssim 10^{-2} \Msun$, the \ce{H2} dissociation front is distant from the H ionization front. EUV-driven, ionized photoevaporative flows are continuously excited from the H-rich disk's rim at around $30\au$ as in \fref{fig:snapshots}. Pumping-driven, neutral photoevaporative flows are excited from the \ce{H2} dissociation front that moves outward. 
 Since EUV does not thermochemically affect the neutral disk region, the above trends are the same between the strong- and weak-EUV models for a given set of $M_{\rm disk}$ and $\dgratio$.

X-ray is found to be ineffective for driving photoevaporation.
The specfic X-ray heating rate is of the order of $\sim 1 \erg \gram^{-1}\sec^{-1}$ at $30\au$ with the fiducial X-ray luminosity of $L_{\rm X} = 10^{29}\erg \sec^{-1}$. 
It is comparable to or less than the FUV heating. 
The gas internal energy is reduced by adiabatic expansion and \ion{O}{1} cooling in the X-ray-heated region. 
The specific adiabatic cooling rate is approximately  $\cs^3/r \approx 2 (\cs/1\kms)^3  (r/30\au)^{-1}$ and is comparable with or greater than the specific X-ray heating rate. 
Therefore, the resulting temperatures do not go higher than $\sim 100\Kelvin$ in $> 30\au$. 

Although it is a valid assumption to use $L_{\rm X} \approx 10^{29}\erg\sec^{-1}$ for intermediate-mass stars, some Herbig Ae/Be stars are known to have X-ray luminosities up to $L_{\rm X} \approx 10^{30}\erg\sec^{-1}$ \citep{2008_Hamidouche, 2009_Hubrig}.
We have run additional simulations with $L_{\rm X} = 1\e{30}\erg\sec^{-1}$ to see if photoevaporative flows are excited with a high-end X-ray luminosity for intermediate-mass stars. 
The heating rate is $\sim 10 \erg \gram^{-1} \sec^{-1}$ at $30\au$, which is ten times higher than that with the fiducial $L_{\rm X}$. The resulting temperatures are increased from those of the fiducial case by a factor of a few at most, but X-ray-heated gas is still found to hardly drive photoevaporative flows. 
Hence, we conclude that the effect of X-ray is limited for photoevaporation of the evolved disks hosted by intermediate-mass stars.
}}

\subsubsection{Brief Summary of Physical Structures}
\replaced{
To summarize, the effects of varying $a_{\rm max}$ appear only in the geometry of the disk through heating the gas by photoelectric effects. 
The physical structures and trends described above are common among all of the disk configurations (either of a truncated or full disk; $M_{\rm disk} = 10^{-3}$--$10^{-2}\Msun$) considered in our model. }
{\revision{Photoevaporation of the evolved disks is largely divided into two classes depending on disk mass: lower-mass class ($M_{\rm disk} \lesssim 10^{-3}\Msun$) and higher-mass class ($M_{\rm disk}\gtrsim 10^{-3}\Msun$). In lower-mass cases, only ionized photoevaporative flows are excited. Heating processes in the neutral region, such as photoelectric heating and \ce{H2} pumping, are insignificant. The abundances of \ce{H2} and CO are quite low; atomic hydrogen and ionized carbon are the most dominant species among H-bearing and C-bearing species, respectively. For higher-mass disks, on the other hand, there are an \ce{H2}/CO disk in the outer region. Ionized photoevaporative flows are excited in the same manner as the lower-mass disks, but high-mass disks also have neutral photoevaporative flows excited at the rim of the \ce{H2}. Heating from \ce{H2} pumping is responsible for the neutral flow. }}

A caveat is that our model does not take into account independent motion between gas and large grains. In practice, large dust can settle onto the midplane within a short time compared to evolutionary timescales of PPDs. Upper layers of evolved disks would be devoid of grains by this effect regardless of turbulent diffusion, \added{\revision{because the production timescale of small grains is much longer than the outward-drift timescale (\secref{sec:blowout})}}. In this case, photoelectric heating is even reduced in the upper layers, and therefore
we may have overestimated the temperatures of upper layers. Nevertheless, large grains hardly contribute to photoelectric heating, and thus 
incorporating spatially inhomogeneous dust-to-gas mass ratio would not significantly change the flow structures and mass-loss rates. 

\subsection{Photoevaporation Rates} \label{sec:evarates}
The outgoing photoevaporative flows yield mass loss from the system. The mass-loss rates (photoevaporation rates) are an essential quantity that characterizes the lifetimes of the gas component in the evolved disks. 
We measure the photoevaporation rates by
\gathering{
	\dot{M}_{\rm ph} (r_{\rm S}) = \int_{S,\eta > 0} \dd\vec{S} \cdot \vec{v} \rho
					= 4\pi r_{\rm S} ^2 \int_{\eta > 0} \dd \theta \,  v_r \rho  \sin \theta \\
	\eta \equiv \frac{1}{2} v^2 + \frac{\gamma\cs^2}{\gamma - 1} - \frac{GM_*}{r}, \label{eq:eta}  
}
where $S$ is a spherical surface with a radius of $r_{\rm S}$. 
The outgoing mass flux is summed up only where $\eta > 0$. 
We adopt this measurement technique to distinguish a bound gas from an unbound gas in an approximate manner. \added{\revision{(Refer to \fref{fig:snapshots} to see where the $\eta = 0$ boundary is located.)}}

The measured $\mph$ reaches quasi-steady after $\sim 500 \yr$ for all of the runs
\added{\revision{except for those with $M_{\rm disk} \approx 3\e{-4} \Msun$, 
where the \ce{H2} dissociation front sweeps the disk until $t \sim 2000\yr$.
The mass-loss rates reach quasi-steady afterwards. 
We thus take the average over $t \gtrsim 2000\yr$ for these runs.}}
Photoevaporative flows within $\lesssim 50\au$ account for the bulk of the total mass-loss rates.
Therefore, 
we set $r_{\rm S} = 200 \au$. Note that using a fairly smaller $r_{\rm S}$ than the outer computational boundary allows avoiding contamination from the spurious reflection to the mass-loss rates \citep{2018_Nakatani}.

\begin{figure*}[htbp]
\begin{center}
\includegraphics[clip, width = \linewidth ]{\figdir/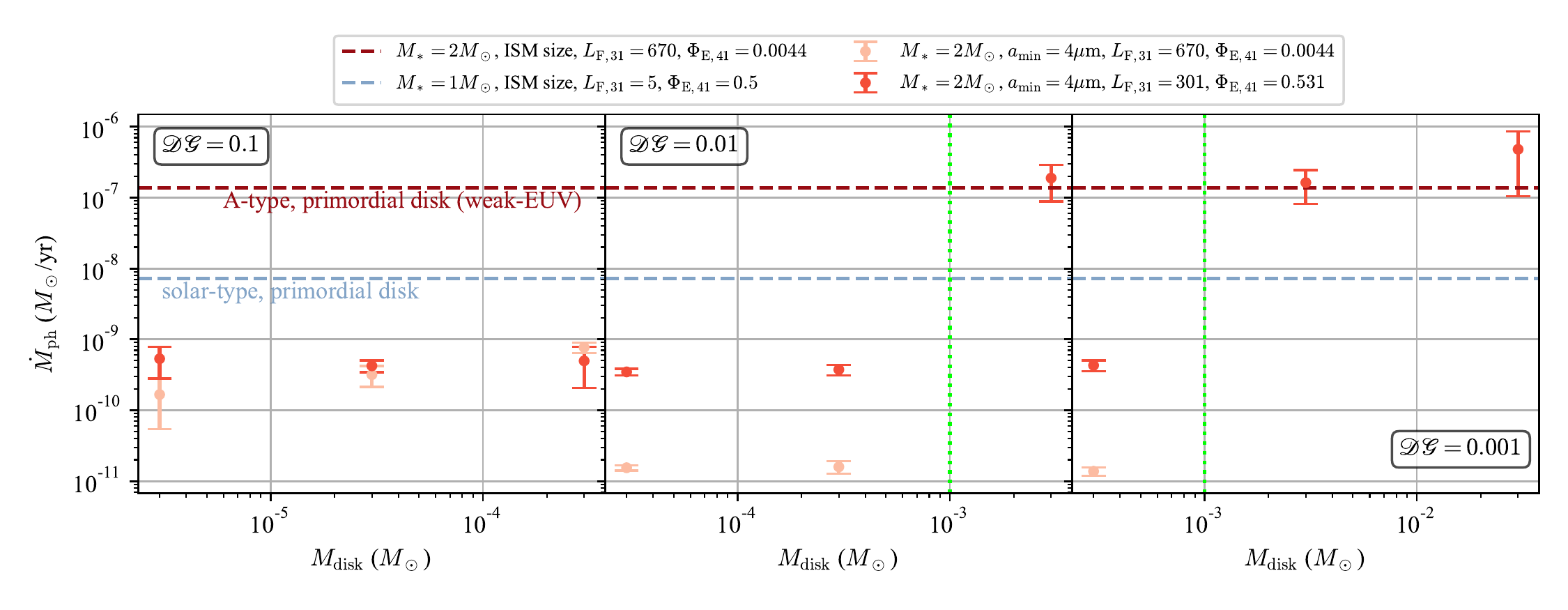}
\caption{
Time-averaged total photoevaporation rates, $\mph(< r_{\rm S} = 200\au)$\deleted{, with various $a_{\rm max}$}. 
The red and pink markers show $\mph$ of the strong- and weak-EUV models, respectively. The horizontal red dashed line indicates $\mph$ of the primordial disks for reference. 
The mass-loss rates are compared with those of the evolved disks around a solar-type star represented by the blue markers (cf.~\secref{sec:solartype}). The horizontal blue dashed line shows $\mph$ of the primordial disk. 
In the legend, $L_{\rm F,31}$ and $\Phi_{\rm E,41}$ are defined as $L_{\rm F,31}=\LFUV/(10^{31}\erg\sec^{-1})$ and $\Phi_{\rm E,41} = \LEUV/(10^{41}\sec^{-1})$. 
Photoevaporation rates are quasi-steady, i.e. they slightly fluctuate from the mean;
the error bars of the markers represent the root-mean-square of the fluctuation. 
\added{\revision{The vertical dotted green line indicates $M_{\rm disk} = 10^{-3}\Msun$ for reference.}}
}
\label{fig:evaporationrate}
\end{center}
\end{figure*}
The time-averaged total photoevaporation rates of the strong- and weak-EUV models are shown in \fref{fig:evaporationrate}. \deleted{for the inner-truncated disks at $25\au$.} 
\replaced{The mass-loss rates decrease by $\sim 40\%$ varying $a_{\rm max}$ from $1\cm$ to $1{\rm m}$. In this range,
the scale height of the neutral region is higher with decreasing $a_{\rm max}$ by the effects of photoelectric heating (\secref{sec:structure}). 
The geometrical cross-section of the disk is correspondingly larger to the stellar EUV, and it yields a slightly higher total ionization rate and thereby $\mph$. Although an overestimated photoelectric heating might have yielded a large geometrical cross-section of the evolved disks as discussed in \secref{sec:structure}, the effect is insignificant in terms of $\mph$. 

For $a_{\rm max} \gtrsim 10{\rm\, m}$, the dominant heating source is X-ray heating, which is not affected by $a_{\rm max}$, in the neutral regions. It results in the same geometrical structure of the disks over a wide range of $a_{\rm max}$ for $a_{\rm max} \gtrsim 10{\rm\, m}$. Hence, the photoevaporation rates are largely constant in this range. 
The same trend holds for other disk configurations and disk masses, and the mass-loss rates are essentially determined by $\LEUV$ (cf.\tref{tab:fiducialmodel}; see also \secref{sec:solartype}).  }
{\revision{For $M_{\rm disk} \lesssim 10^{-3}\Msun$, the mass-loss rates are $\approx 1\e{-11}\Msun \yr^{-1}$ with $\dgratio \leq 0.01$ in the weak-EUV model (pink dots). They are $\approx 3\e{-10} \Msun \yr^{-1}$ with any $\dgratio$ in the strong-EUV model (red dots) and with $\dgratio = 0.1$ in the weak-EUV model. The EUV emission rates essentially determine the mass-loss rates for these low-mass disks, but photoelectric heating is also efficient to heat the gas in cases of $\dgratio = 0.1$. Therefore, the mass-loss rates in the weak-EUV models with $\dgratio$ are high for the given $\LEUV$. 

For $M_{\rm disk} \gtrsim 10^{-3}\Msun$, neutral photoevaporative flows are excited by \ce{H2} pumping and have orders of magnitude higher density compared to EUV-driven ionized flows. The mass-loss rates are correspondingly higher ($\sim 10^{-7}\Msun \yr^{-1}$) than those of $M_{\rm disk} \lesssim 10^{-3}\Msun$ disks. 
However, we expect that a high mass-loss rate at the level of $10^{-7}\Msun \yr^{-1}$ would be sustained only before the whole disk is swept by the \ce{H2} photodissociation front on timescales of $t_{\rm dis}$. Afterward, if the swept region leaves an H-disk behind as observed in our simulations with $10^{-3} \Msun \lesssim M_{\rm disk}\lesssim 10^{-2}\Msun$, the mass-loss rates would be determined by EUV and reduce to $\approx 10^{-11}\mbox{--}10^{-10}\Msun \yr^{-1}$ depending on $\LEUV$. 
This means that although the values of $\dot{M}_{\rm ph}$ shown in \fref{fig:evaporationrate} are almost steady over the computational time ($\approx 3000\mbox{--}4000\yr$), they will be significantly reduced on a longer timescale (cf. \eqnref{eq:tdis}) even for $M_{\rm disk} \gtrsim 10^{-3}\Msun$. 
A longer-term computation ($\gtrsim 10^4\yr$) is necessary to understand whether the H-rich disk is left behind and to get, if any, the remained mass. Such long-term simulations may require simplifications in the thermochemistry model to reduce the computational cost. We will work on this issue in future studies. 
}}

In order to compare the mass-loss rates of the evolved disks with those of primordial disks, where grains are ISM-like, we additionally run simulations with the same stellar parameters but using an ISM size distribution. 
The resulting mass-loss rate is of the order of $10^{-7}\Msun\yr^{-1}$. The mass-loss rate is shown by the red horizontal line in \fref{fig:evaporationrate}. The figure directly shows that depletion of small grains results in reducing mass-loss rates by a few orders of magnitude even with the same stellar parameters. 

\deleted{
We assume a uniform gas-dust ratio within the disk in the present work, and therefore the abundance of small dust grains contributing to the FUV photoelectric heating declines for larger grain sizes. In practice, the decoupling of gas and dust, e.g. sedimentation of dust grains and ejection due to radiation pressure, could also reduce the dust abundances near the photoevaporation base. The derived photoevaporation rates in our model are thus applicable to 
disks with different $M_{\rm disk}$, allowing us to estimate dispersal times of the gas component for various-mass systems.
}

\subsection{Lifetime of the Gas Component in Evolved Disks}	\label{sec:lifetime}
\added{\revision{For $\dgratio \leq 0.01$}}, 
the measured photoevaporation rates are $ \approx 3$--$4\e{-10}\Msun\yr^{-1}$ for the strong-EUV model ($\LEUV = 5\e{40}\sec^{-1}$) and $ \approx 1$--$2\e{-11}\Msun\yr^{-1}$ for the weak-EUV model ($\LEUV = 4\e{38}\sec^{-1}$).
\added{\revision{There is little difference between the strong- and weak-EUV models with $\dgratio = 0.1$ (\fref{fig:evaporationrate}). 
We have obtained $\dot{M}_{\rm ph}$ of the order of $ \approx 10^{-7}\Msun \yr^{-1}$ for $M_{\rm disk} \gtrsim 10^{-3}\Msun$, but it is likely that the system has not yet reached a (quasi-)steady state within the computational time; $\dot{M}_{\rm ph}$ is expected to reduce to $\approx 10^{-11}\mbox{--}10^{-10}\Msun \yr^{-1}$ after the dissociation-front sweeping, as discussed in the previous section. 
Hence, we use the resulting $\dot{M}_{\rm ph}$ for $M_{\rm disk} \lesssim 10^{-3}\Msun$ in the following discussions.
}}

Since photoevaporation dominates mass loss in the later phase of disk dispersal \citep[e.g,][]{2001_Clarke, 2006_Alexander_b, 2010_Owen, 2015_Gorti, 2020_Kunitomo}, our results imply that the lifetime of the gas component is primarily set by the disk mass and the mass-loss rate in the evolved disks. 
We estimate the lifetime of the evolved disks around A-type stars as
\eq{
\splitting{
	T_{\rm life} & = \frac{M_{\rm disk}}{\mph} \\
				 & = 50\, \braket{\frac{M_{\rm disk}}{10^{-3}\Msun}} \braket{\frac{\mph}{2\e{-11}\Msun\yr^{-1}}}^{-1} \Myr. 
	}
	\label{eq:life}
}
The lifetime is proportional to the ``initial'' disk mass when the disk has reached a optically-thin grain-depleted state 
\replaced{lost the bulk of small grains ($\lesssim 4\mum$).}
{\revision{(or, possibly, to the remained mass after the dissociation-front sweeping).}}
This initial disk mass is determined in a complex manner influenced by various effects other than photoevaporation, such as accretion, magnetically driven winds, grain growth, radiation pressure, settling, and etc.
Here, we have used $M_{\rm disk} = 10^{-3} \Msun$ for a nominal value in \eqnref{eq:life}. Such high disk mass is plausible for evolved disks at ages of $\sim 10\Myr$ while being consistent with near-infrared observations of disk fractions, which have suggested $3$--$6\Myr$ for {\it inner} disk lifetimes \citep{2020_Kunitomo}. 

Our results propose that the dispersal timescale of the gas disk has been significantly extended after the system reaches the evolved state. The mass-loss rate is of the order of $10^{-7} \Msun\yr^{-1}$ while small grains are present in the upper layers of the disk, and it reduces to $\lesssim 10^{-11}\mbox{--}10^{-10}\Msun \yr^{-1}$ at the point when small grains has cleared out from the system. Depending on the disk mass at this point, the gas component of evolved disks can survive for $\gtrsim 50\Myr$ around intermediate-mass stars. It suggests that gas can remain in old systems over a timescale of $>10\Myr$ as protoplanetary remnants. We present further discussions on disk mass evolution in \secref{sec:discussion}.

\subsection{Grain-Depleted Disks around Solar-Type Stars}    \label{sec:solartype}
    We have discussed photoevaporation of the evolved disks around A-type stars in prior sections. In this section, we consider a pre-main-sequence solar-type star with $1\Msun$ and $2.6\Rsun$ for the radiation source to compare the resulting mass-loss rates and lifetimes of the evolved disks around different spectral types. We also run simulations with primordial disks to compare the results. 
    
	The stellar UV and X-ray luminosities of young low-mass stars are highly uncertain. In this study, we simply follow \cite{2008_GortiHollenbach, 2009_GortiHollenbach} and adopt $\LFUV = 5\e{31}\erg \sec^{-1}$, $\LEUV = 0.5\e{41} \sec^{-1}$, and $L_{\rm X} = 2.5\e{30} \erg \sec^{-1}$.
	The minimum grain size is set to $a_{\rm min} = 1\mum$ according to \eqnref{eq:maxs}. 
	In \tref{tab:model}, we compare the adopted parameters with the A-type star model. 
        \begin{table}[htp]
                \caption{Comparison of model parameters}
                \begin{center}
                \begin{tabular}{l c c}   
                &A-type (strong-EUV) 
                &solar-type   \\ \hline \hline
                $M_*$     				& $ 2  \Msun$ &      $1\Msun$                            \\
                $R_*$                 	& $1.88\Rsun$ 	&	$ 2.61 \Rsun$ \\
                 $L_*$  &   $19\Lsun$   &   $2.3 \Lsun$\\
                 $\LFUV$        & $3\e{33}\unit{erg}{}\unit{s}{-1}$  &$ 
                5\e{31}\unit{erg}{}\unit{s}{-1}$ \\
                 $\LEUV$     & $5\e{40}\sec^{-1}$  &$  5\e{40}\unit{s}{-1}$      \\ 
                 $L_{\rm X}$	& $1\e{29} \erg \sec^{-1}$ & $2.5\e{30} \erg \sec^{-1}$ 	\\
                 $a_{\rm min}$		&	$4\mum$ & 	$1\mum$ \\
                \hline
                \end{tabular}
                \end{center}
                \label{tab:model}
        \end{table}

In the solar-type pre-main-sequence star model, 
the resulting photoevaporation rates are $\mph \approx 3$--$6\e{-10}\Msun \yr^{-1}$. \replaced{(cf.~\fref{fig:evaporationrate})
for both of full and truncated disks, and again $\mph$ hardly depends on $a_{\rm max}$. }
{Photodissociation and pumping are insignificant with $\LFUV$ of the solar-type star. 
The disk is \ce{H2}-rich even with $M_{\rm disk}\lesssim 10^{-5}\Msun$, 
but \ce{H2} pumping hardly excites photoevaporative flows in contrast to the A-type disks. 
The mass-loss rates are determined by $\LEUV$ in this case. 
Thus, the mass-loss rates are largely independent of $M_{\rm disk}$ for solar-type disks 
and have similar $\mph$ to those of the strong-EUV model for A-type stars.
}

\deleted{The photoevaporation rates are almost the same as those of the strong-EUV model for A-type stars. 
Photoevaporation rates of evolved disks are essentially determined by the EUV emission rate of the host star; disk geometry, grain sizes, or the stellar mass does not strongly affect resulting total mass-loss rates.}

Generally, mass-loss rates are set by the EUV flux reaching the ionization front for the evolved disks. It is located at far above the midplane, and thus $\mph$ is largely independent of the disk mass. 
The insignificant dependence of mass-loss rates on $M_{\rm disk}$ has also been reported by \cite{2019_Wolfer} in the context of X-ray photoevaporation for PPDs. 
We have also confirmed the insignificant dependence of $\mph$ on $M_{\rm disk}$ or $R_{\rm min}$ by performing simulations with different $M_{\rm disk}$ and $R_{\rm min} (= 50\au)$. \deleted{for a disk truncated at $50\au$ with a mass of $M_{\rm disk} \approx 10^{-3}\Msun$ for the strong-EUV model. The resulting photoevaporation rates are $\approx2$--$ 4\e{-10}\Msun\yr^{-1}$.}

\revision{
Prior hydrodynamics models predict nearly an order of magnitude higher EUV photoevaporation rates for truncated disks than for full disks, owing to direct irradiation onto the inner rim. 
The underlying assumption is that in cases of full disks, the direct EUV field is completely attenuated at radii smaller than the gravitational radius, and the diffuse EUV field dominates in the outer region. On the other hand, if a cavity is present, the direct field can directly illuminate the inner rim of the truncated disk. 
However, we have not observed such complete attenuation of the direct EUV in our simulations, where we solve hydrodynamics and radiative transfer in a self-consistent manner while resolving the gravitational radius. The direct EUV is found to reach the outer region whether or not the disk has a cavity, because the disk has a flared structure. The mass-loss rates are thus almost independent of disk geometry. 
}

In \fref{fig:evaporationrate}, we also compare the photoevaporation rates of \deleted{the evolved disks (blue dots) and}
primordial disks (cyan dashed line) for the solar-type star model. The mass-loss rate of the primordial disk is of the order of $\sim 10^{-8}\Msun\yr^{-1}$ and is higher than those of the evolved disks by about an order of magnitude. The reduction of the mass-loss rates due to grain depletion is also significant for solar-type stars, yet it is relatively small compared to A-type star cases owing to the much smaller $\LFUV$.

\section{Discussions} 	\label{sec:discussion}
We have investigated photoevaporation of the evolved disks around A-type stars and have compared the results with those of the solar-type star cases. 
The evolved disk is modeled as a gas-rich optically-thin disk where grains smaller than $a_{\rm min, rem}$ are continuously removed from the system by the radiation pressure of the host star. FUV heating has been found insignificant to drive photoevaporation in the evolved disks, which extends the dispersal timescale of the gas component compared to that of PPDs with a primordial composition. 
The results are valid as long as the evolved systems sustain a grain-depleted state regardless of small grain production by fragments of large bodies. 
In \secref{sec:blowout}, we discuss whether the evolved disks can maintain the grain-depleted state by an order estimation. Then, we investigate the impacts of grain depletion on the stellar type dependence of photoevaporation rates and disk mass evolution in \secref{sec:stellardependence}. We apply our results to recently observed statistics of gas-rich debris disks in \secref{sec:debris}. We also examine the effects of heating processes that have not been implemented in this study on our simulation results in \secref{sec:otherheating}.

\subsection{Can Evolved Disks Maintain a Grain-Depleted State?} \label{sec:blowout}
\replaced{
\newcommand{\tdifver}{t_{\rm dif, \perp}}
We have set the minimum grain size in the evolved disks to $a_{\rm min, rem}$ defined by $\beta = 1$ (cf.~\eqnref{eq:maxs}). The adopted minimum grain size has been $a_{\rm min} = 4\mum$ for the A-type star model and $a_{\rm min} = 1\mum$ for the solar-type star model, and they are fixed throughout the simulations. However, in practice, small grains ($a<a_{\rm min, rem}$) are produced by collisional fragmentation of larger bodies near the midplane and are continuously (re)supplied in the vertical direction by turbulent diffusion.
An efficient (re)supply could even result in increasing mass-loss rates by inducing photoelectric heating \citep{2015_Gorti}. 

For $\beta > 1$, grains are accelerated in the radially outward direction by radiation force to be blown out on a timescale of
\[
t_{\rm blow} 
            \approx 0.5(\beta -1)^{-1}\Omega^{-1}.
\]
In gas-rich disks, grains are also susceptible to gas drag. The magnitude of dynamical coupling between gas and grains is characterized by a dimensionless stopping time
\[
\splitting{
    \nondimts  & 
    =	 \sqrt{\frac{\pi}{8}}\frac{\rho_{\rm b} a }{\rho \cs} \Omega
    = \frac{\pi}{2} \frac{\rho_{\rm b} a }{\Sigma \exp\braket{-z^2/2H^2}}\\
 &\approx 0.0022  \braket{\frac{\rho_{\rm b}}{1.4\gram \cm^{-3}}} \braket{\frac{a}{1\mum}}\\
 & \times \braket{\frac{\Sigma}{0.1\gram\cm^{-2}}}^{-1} \exp\braket{\frac{z^2}{2H^2}}.
 } 
\]
The nominal $\Sigma$ corresponds to a surface density of a disk with $\sim 10^{-3}\Msun$ at $30\au$ in our model. 
\replaced{
The turbulent diffusion can stir small grains up from the near-midplane region to a height of a few times the pressure scale height \citep{2002_TakeuchiLin}. 
Small grains with $\beta > 1$ are unbound and are blown out from the layers where $t_{\rm esc} \Omega \lesssim \nondimts$ on a timescale of $\sim (\beta -1)^{-1}\Omega^{-1}$. This timescale is much shorter than the collisional timescale of large bodies $t_{\rm col} \sim (\tau \Omega)^{-1}$, where $\tau$ is the vertical optical depth of grains, with a typical optical thickness for debris disks ($\tau \lesssim 10^{-2}$).}{ 
The gas drag is more important than the radiation pressure near the midplane where $\nondimts \ll 1$. We note that dust grains ejected via collisional fragmentation of large bodies, which have $\beta \sim 0$, get unbound orbits if $\beta > 0.5$ \citep[e.g.,][]{2006_Krivov}, but the gas drag damping converts them to bound orbits on the gas stopping time. These small grains may survive near the midplane in gas-rich disks and would be supplied to high altitudes by turbulent stirring. Nevertheless, the radiation pressure blows out the stirred-up dust grains, which are also pushed in the vertically upward direction for $\beta > 1$, since $t_{\rm blow} \ll  \nondimts \Omega^{-1}$ at a high altitude \citep[e.g. a few times the pressure scale height;][]{2002_TakeuchiLin}.  
}
Therefore, high-altitude small grains with $\beta > 1$ would be efficiently removed from the evolved disk before (re)supplied from the midplane. 

For $\beta < 1$, 
\replaced{the vertical component of the effective gravity (radiation force plus the stellar gravity) is directed downward.   
The produced grains can be stirred up to the height where the effective gravity equilibrates with the stirring-up force due to turbulent diffusion. 
The grains that have reached the equilibrium height would not be allowed to escape from the system vertically, and hence the motion of grains occurs on timescales of gas dynamics.
Radial motion of grains can be induced by angular momentum transfer with the ambient gas. The radial velocity is approximated to $v_{R,\rm d}\approx \beta \nondimts R\Omega$ for $\nondimts \ll 1$ \citep{2003_TakeuchiLin}. Hence, the escape time of small grains are likely much longer than $t_{\rm col}$ for the systems with $\beta < 1$. 
}
{radiation pressure is too weak to blow out dust grains stirred up to high altitudes by turbulent diffusion. Dust grains are removed by gas drag on a timescale of $\nondimts^{-1} \Omega^{-1} \beta^{-1}$ \citep{2001_TakeuchiArtymowicz}, which is similar to the stir-up timescale. Hence, gas-rich disks can retain high-altitude grains with $\beta < 1$.
}
Note that small grains ($\lesssim 1\mum$) contribute to increase the effective opacity of the disks and FUV photoevaporation rates at a large distance for solar-type stars \citep{2015_Gorti}. 

To summarize, the evolved disks likely maintain a grain-depleted state for $\beta > 1$ owing to an efficient vertical removal of small grains due to the strong radiation force. In contrast, for $\beta < 1$, removal of small grains appears to be inefficient, and it may lead to accumulation of small grains within the disks. In such cases, the evolved disks would fail to maintain a grain-depleted state. For a further detailed and very accurate discussion, an extensive numerical study is indispensable to solve the 2D advection-diffusion equation with taking into account radiative transfer and grain fragmentation/coagulation. It is out of scope in the present study and is left for future studies.
}   
{\revision{
\input{blowout}
}}

\deleted
{
Even for grains in a range of the blown-out size, the aerodynamical friction between gas and dust may affect to restrain the grains accelerated by the radiation pressure within the disk. It is also necessary to assess such possibilities to validate our assumptions on the minimum grain sizes for the A-type star model. 
The magnitude of the aerodynamical friction is characterized by the Stokes number,
\[
\splitting{
    {\rm St}  & = \Omega t_{\rm tstop}  
    =	 \sqrt{\frac{\pi}{8}}\frac{\rho_{\rm b} s }{\rho \cs} \Omega\\
    & \approx 0.64 \braket{\frac{\rho_{\rm b}}{1.4\gram \cm^{-3}}} \braket{\frac{s}{1\mum}}\braket{\frac{M_*}{2\Msun}}^{1/2}\\
 & \times \braket{\frac{\nh}{10^6\cm^{-3}}}^{-1} \braket{\frac{r}{30\au}}^{-3/2} \braket{\frac{\cs}{1\kms}}^{-1} . }
\]
The friction is strong enough to couple the motion of grain and gas for grains with ${\rm St} < 1$. 
This condition sets the lower limit of grain sizes that dynamically decouples from gas 
\eq{
\splitting{
    s & > s_{\rm stop, max} \equiv 1.56
    \braket{\frac{\rho_{\rm b}}{1.4\gram \cm^{-3}}}^{-1} \braket{\frac{M_*}{2\Msun}}^{-1/2}\\
    & \times \braket{\frac{\nh}{10^6\cm^{-3}}} \braket{\frac{r}{30\au}}^{3/2} \braket{\frac{\cs}{1\kms}} \mum . }
    \label{eq:mins}
}
Grains satisfying both of ${\rm St} > 1$ and $\beta > 1$ can escape from the optically-thin layers in a short time without being held back by gas. 
\footnote{Note that grain sizes satisfying both of ${\rm St} > 1$ and $\beta > 1$ necessarily fulfills $t_{\rm stop} > t_{\rm esc}$, i.e., the grains escape from the system before dynamically coupling with the gas.} 
\footnote{In cases of $t_{\rm stop} < t_{\rm esc}$, 
transferred radiation momentum is immediately transferred to the gas. If the transferred momentum is sufficiently large, it can drive an outward movement of the gas. \mycomment{this footnote will be removed soon. It is here at this moment just for my use.}}
Note that this is a sufficient condition; not only these grains are capable of escaping from the system (see below). 
For the existence of grains with ${\rm St} > 1$ and $\beta > 1$, 
$s_{\rm stop, max} < s_{\rm esc,max}$ is required. This necessity reduces to a restriction to gas density as
\eq{
\splitting{
        \nh & < n_{\rm H, crit}   \equiv  1.1\e{7}\braket{\frac{\cs}{1\kms}}^{-1} 
        \braket{\frac{Q_{\rm pr}}{1}}\\
        &\times \braket{\frac{L_*}{20\,L_\odot}}
        \braket{\frac{r}{30\au}}^{-3/2}\braket{\frac{M_*}{2\Msun}}^{-1/2} \cm^{-3}.   \label{eq:critdensity}
}
}
Hence, grains are able to escape by the effects of radiation pressure without being dragged by gas friction in the low-density layers where $\nh > n_{\rm crit}$. 
The critical gas density 
implicitly depends on $s$ and $\rho_{\rm b}$ in terms of $Q_{\rm pr}$. For $s < \lambda_{\rm peak}$, $n_{\rm H,crit}$ is small. Typical peak wavelength of stellar radiation is at $\sim 0.5\mum$ for A-type stars and at $\sim 1\mum$ for solar-type stars with the blackbody radiation.

We stress again that \eqnref{eq:maxs} is a necessary condition for escape driven by radiation pressure, while \eqnref{eq:critdensity} is a sufficient condition; that is, not only the grains satisfying \eqnref{eq:critdensity} but also small grains with ${\rm St} \ll 1$ are able to move outward to escape from the system under strong radiation. Acceleration due to radiation pressure yields nonzero radial velocity unless ${\rm St}  = 0$. Very small ${\rm St}$ affects to reduce the radiation pressure force. The radial distance of small grains accelerated by radiation pressure would increase at an approximate rate of 
\[
\splitting{
    v_{r, \rm drag} & \equiv r \Omega (\beta - 1) {\rm St} \\
    & \approx 1.04 \, \braket{\frac{\rho_{\rm b}}{1.4\gram \cm^{-3}}} \braket{\frac{s}{1\mum}}\\
 & \times   \braket{\frac{M_*}{2\Msun}}
  \braket{\frac{\nh}{10^6\cm^{-3}}}^{-1} 
  \braket{\frac{r}{30\au}}^{-2}\\ 
  & \times  \braket{\frac{\cs}{1\kms}}^{-1} (\beta - 1) \au \yr^{-1}.\\
    }
\]
The accelerated grains are blown out from a disk with a size of $R_{\rm disk}$ on a timescale of 
\eq{
\splitting{
   t_{\rm blow} & \equiv \frac{R_{\rm disk}}{v_{r,\rm drag}}\\
        & \approx 10^{-4}\,  \braket{\frac{\rho_{\rm b}}{1.4\gram \cm^{-3}}}^{-1} \braket{\frac{s}{1\mum}}^{-1}\\
 & \times   \braket{\frac{M_*}{2\Msun}}^{-1}
  \braket{\frac{\nh}{10^6\cm^{-3}}}
  \braket{\frac{r}{30\au}}^{2}\\ 
  & \times  \braket{\frac{\cs}{1\kms}} 
  \braket{\frac{R_{\rm disk}}{100\au}} (\beta - 1)^{-1}  \Myr.\\
  \label{eq:blow}
  }
}
All grains that have $\beta$ fairly larger than unity appear to be blown out within $\sim 1\Myr$ from the optically-thin layer of PPDs, even with ${\rm St} \ll 1$. 
Hence, radiation pressure is expected to blows out small grains with $s < s_{\rm esc, max}$ around A-type stars within a fairly short time regardless of the size (or Stokes number).

To summarize, the adopted minimum size of grains are well justified for the A-type star model. On the other hand, the blow-out effects of the radiation pressure is likely significantly overestimated in our solar-type star model in \secref{sec:solartype}. The solar-type star model corresponds to an extreme case where small grains have been depleted to an unrealistic extent. We conclude that EUV-driven photoevaporation is dominant in the evolved disks around A-type stars because of a reduced photoelectric heating, while for solar-type stars, small grains can remain in the disks so that FUV-driven photoevaporation is active.
This spectral-type dependence gives an interesting implication to the statistics of gas-rich debris disks, and we discuss it in the next section. 
}

\subsection{Disk Mass Evolution of Low- and Intermediate-Mass Stars}    \label{sec:stellardependence}


\begin{figure}
    \centering
    \includegraphics[clip, width = \linewidth]{\figdir/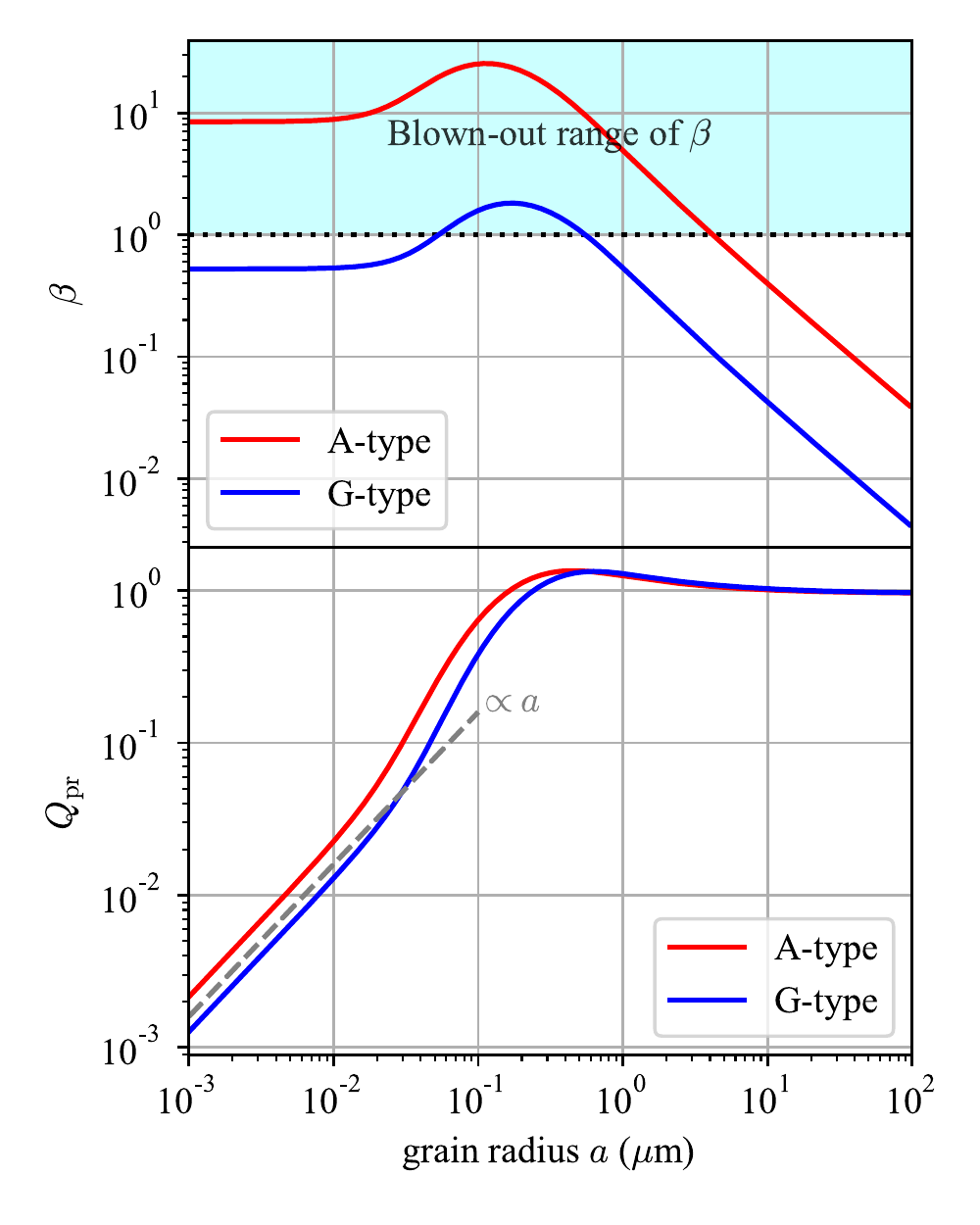}
    \caption{Magnitude ratio of radiation pressure to the host star gravity $\beta$ (top), and the momentum transfer efficiency $Q_{\rm pr}$ (bottom) for our models. The red and blue lines represent those for the A- and solar-type models, respectively. The cyan-shaded region shows $\beta > 1$, where radiation pressure dominates the gravity, leading to an efficient removal of small grains. Note that $Q_{\rm pr}$ scales as $\propto a$ for $a \ll \lambda_{\rm peak}$ as guided by the black dashed line.}
    \label{fig:Qpr}
\end{figure}
\fref{fig:Qpr} shows $\beta$ derived for icy grains composed of pure ice, astronomical silicate, and organics adopted in this study \citep{2014_Kobayashi}. 
The value of $\beta$ takes the maximum at $a$ comparable to the peak wavelength of the stellar radiation $\lambda_{\rm peak}$; it is constant for $a \ll \lambda$, where the grain size is in the Rayleigh regime and $Q_{\rm pr} \propto a$. In the Rayleigh regime, $\beta$ is approximated to
\begin{equation}
 \beta_{\rm R} \approx 0.52 \left(\frac{L_*}{L_\sun}\right)
			      \left(\frac{M_*}{M_\sun}\right)^{-1},
\end{equation}
and is $\beta_{\rm R} \approx 10$ in the A-type star model. 
Once the grains are depleted in such systems, they are likely to maintain a grain-depleted state regardless of large body shuttering, according to the discussions in \secref{sec:blowout}.
FUV-driven photoevaporation does not occur in the grain-depleted disks; otherwise, mass-loss rates can be as high as $\sim 10^{-7}\Msun\yr^{-1}$ (\secref{sec:evarates}). Instead of that, EUV yields a mass-loss rate of $\sim 10^{-11}$--$10^{-10}\Msun \yr^{-1}$ depending on the emission rate. 
Hence, the mass-loss rates of PPDs around A-type stars can significantly reduce during the evolution from the primordial state to a grain-depleted state, extending the gas disk lifetimes. 

On the other hand, $\beta_{\rm R}$ exceeds unity only in a small range of $a$ at around $ 0.1\mum$ for a solar-type star and is less than unity for $a\lesssim 0.1\mum$ (see the blue line in the upper panel of \fref{fig:Qpr}).
\replaced{
It indicates that the radiation force is too weak to efficiently remove the produced small grains, and the disks would fail to maintain a grain-depleted state. In this case, 
FUV photoevaporation rates can remain high through the entire evolutionary stages of the disks. In fact, 
\citet{2015_Gorti} show that gas disk disperses on a timescale of $\sim 1\Myr$ by FUV photoevaporation for solar-type stars, using a 1D~model where dust size evolution is incorporated with a collision-fragmentation scheme \citep{2011_Birnstiel}. 
We note that the radiation force effect is not taken into account in the 1D~model. 
}
{
\revision{
The ratio of production timescale to removal timescale (\eqnref{eq:provsblow}) is about ten times smaller than A-type systems that have similar disk properties.
The environments are much more suitable for small grains to remain in the disks compared to those of A-type systems. This means photoelectric heating could be effective for a longer epoch in solar-type systems than in A-type systems, and, if so, disk lifetime could be longer for A-type disks. 

In any case, photoelectric heating is suppressed once a disk gets optically-thin to FUV photons. This optically-thin (grain-depleted) state can be maintained over the gas disk lifetime, as discussed in the previous section. 
If the disk is fairly massive $M_{\rm disk} \gtrsim 10^{-3}\Msun$, there can be an \ce{H2}-rich region within the disk.  Photoevaporation is driven from the \ce{H2} rim by heating due to \ce{H2} pumping while the \ce{H2} dissociation front sweeps the \ce{H2} disk. Some of the gas is dispersed through this process but a part of the gas can be left behind forming an H-rich disk \fref{fig:snapshotsH2}. 
EUV- and X-ray-driven photoevaporation matter in these grain-depleted H-rich disks. The EUV and X-ray luminosities are atypically low for intermediate-mass stars compared to young low-mass stars. Photoevaporation rates are correspondingly low \citep[see also][]{2021_Komaki}, and thus disks around intermediate-mass stars can survive much longer than those around low-mass stars; our weak-EUV model ($\LEUV \sim 10^{38} \sec^{-1}$) predicts about ten times longer lifetimes. Note that the lifetime approximately scales as $\propto \LEUV^{-1/2}$. EUV emission rates of intermediate-mass stars can be as small as $10^{34} \sec^{-1}$ depending on atmospheric opacity \citep{2021_Kunitomo}.
}
}

\begin{figure}
    \centering
    \includegraphics[clip, width = \linewidth]{\figdir/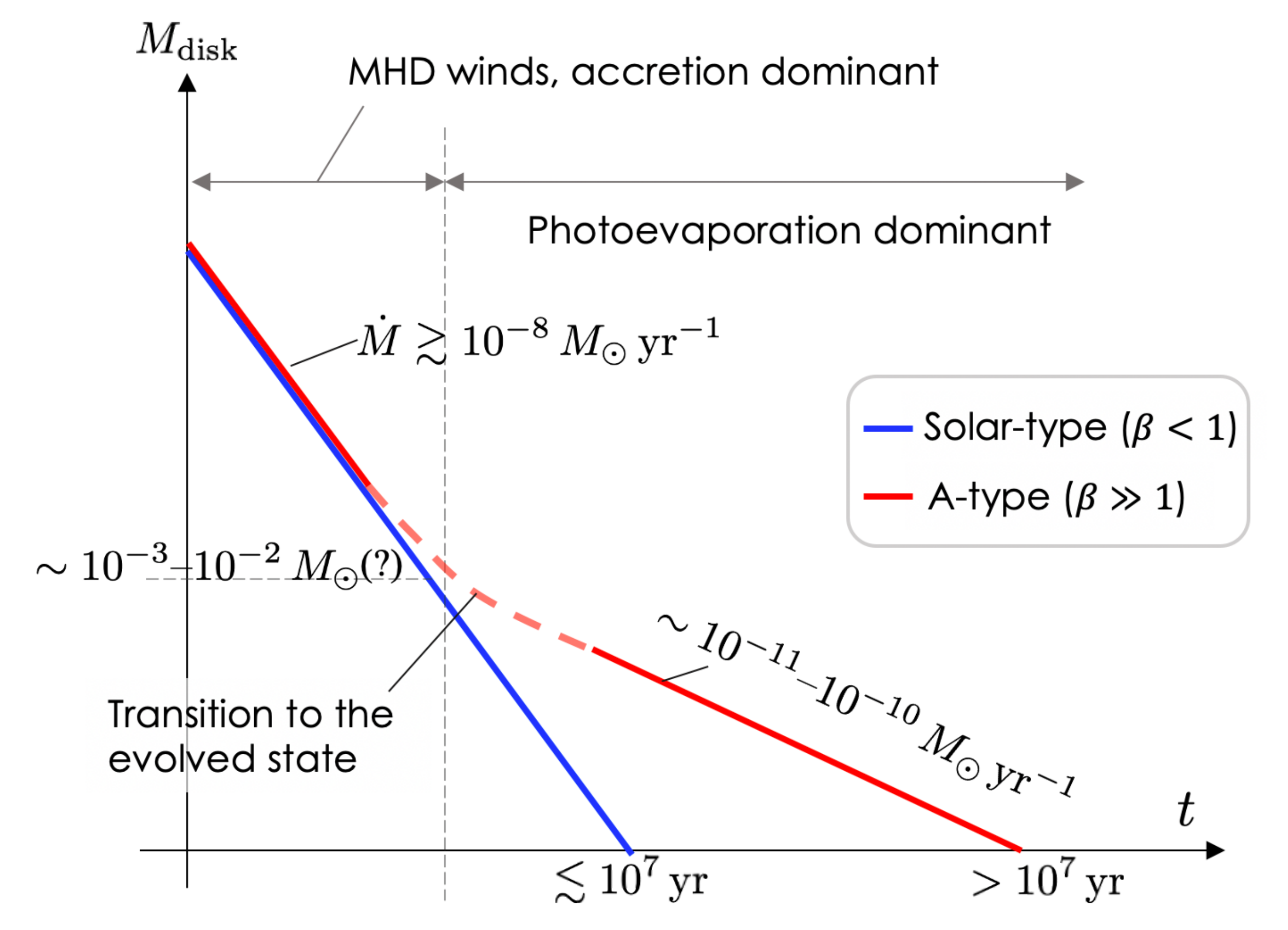}
    \caption{Schematic picture for our hypothesized evolutionary scenarios of disk mass for A-type (red; $\beta \gg 10$) and solar-type (blue; $\beta < 1$) central stars. Magnetic-driven winds and accretion dominate mass loss at the early phase, and photoevaporation dominates at the later phase \citep{2020_Kunitomo}. Photoevaporation rates are smaller with A-type radiation sources because of the grain depletion (and/or the stellar atmospheric absorption) than with solar-type sources. As a result, gas remains for a longer time.}
    \label{fig:schematic}
\end{figure}

We hypothesize different evolutionary scenarios of PPD mass as schematically depicted in \fref{fig:schematic}.
Mass loss is dominated by accretion and MHD-driven winds at the early stage \citep{2020_Kunitomo}. The energy source of MHD winds is the gravitational energy released by accretion process. Therefore, the mass-loss rate is relatively high (possibly, $\gtrsim 10^{-8}\mbox{--}10^{-7} \Msun \yr^{-1}$) in this epoch where the disk has a high surface density. \added{\revision{The winds take the mass particularly at the inner radii. It implies that the wind's mass-loss rate rapidly drops in the first $\sim 1\Myr$ while leaving the outer disk, assuming $M_{\rm disk} \sim 0.1\Msun$ for the initial disk mass. }}
The remained mass at outer radii are dominantly dispersed by photoevaporation.
\replaced{If the disk has reached a grain-depleted state by radiation force around the inter star ($\beta > 1$), the photoevaporation rate is significantly smaller than that around the solar-type systems where FUV-driven photoevaporation may remain effective to yield a mass-loss rate of $\sim 10^{-8}\Msun\yr^{-1}$. Even the cases where FUV photoevaporation is not effective around solar-type stars as well, EUV-driven photoevaporation rates of A-type stars can be an order of magnitude smaller than those of solar-type stars owing to the absorption by stellar atmosphere (\secref{sec:methods} and \fref{fig:evaporationrate}). Interestingly, gas disks around A-type stars have multiple factors to survive for a longer time than those around solar-type stars.}
{\revision{If the disk has reached or will reach a grain-depleted state with low mass ($\lesssim 10^{-3}\Msun$), then FUV-driven photoevaporation hardly contribute to mass loss. EUV- and/or X-ray-driven photoevaporation are the major contributors in this case. Atypically low EUV and X-ray luminosities of intermediate-mass stars owing to the absorption by stellar atmosphere and loss of convective zone (\secref{sec:methods} and \fref{fig:evaporationrate}) yields correspondingly smaller mass-loss rates, extending the gas disk lifetime of intermediate-mass stars.

In the hypothesized scenario, small grains are required to grow to larger bodies and/or to be lost from the disk before the dominant dispersal process switches to photoevaporation at a few $\Myr$; otherwise, all the remained gas would be lost less than another few $\Myr$ by efficient FUV photoevaporation.   
This implies that if our hypothesized scenario explains the evolution of long-lived disks hosted by intermediate-mass stars, the first few Myr sets a definite time-limit to dust growth (or planet formation) for such disks.

In this study, we have given an optically-thin, gas-rich disk to investigate the possibility of a primordial-origin scenario. 
One remaining issue is whether protoplanetary disks can achieve a grain-depleted state during the evolution (and, if so, how and when). 
This would be possible when disk dispersal timescale is longer than the outward-drift timescale or any other depletion timescales of small grains. The FUV-driven photoevaporation rates are $\sim 10^{-7}\Msun\yr^{-1}$ for intermediate-mass stars with assuming ISM-like grains. The corresponding dispersal time is $10^3 \yr (M_{\rm disk}/ 10^{-4}\Msun) (\dot{M}/10^{-7}\Msun\yr^{-1})^{-1}$. This is longer than the outward-drift timescale of nm-sized grains, which are responsible for photoelectric heating, in $z\gtrsim 3H$ at several tens of au for a $M_{\rm disk} \approx 10^{-4}\Msun$ disk. 
This appears that grains can be removed faster than disk disperses, but grain removal can let UV photons reach a deeper interior of the disk to drive dense photoevaporative flows \citep[see, e.g.,][for a 1D model]{2015_Gorti}. The outcome is decreasing disk dispersal time. On the other hand, photoelectric heating weakens as dust-to-gas mass ratios are reduced. It results in increasing the dispersal time. 
In this way, this remained issue is a very complex problem that is not answered rigidly only with the present work.
Further multidimensional numerical studies treating gas and dust separately are needed to address the issue and to investigate how likely the hypothesized mass evolution is.}}

\subsection{Gas-Rich Debris Disks}  \label{sec:debris}
Debris disks are the objects at the last stage of the stellar-system formation and are observed as systems at typical ages of $10$--$100\Myr$. 
They are optically thin ($\tau \lesssim 10^{-2}$) and have much weaker infrared emission than PPDs, which is associated with a little amount of small-sized grains resulting from effects of radiation forces, stellar winds, and Poynting-Robertson drag forces. 
The grains are considered to be constantly produced by grinding large bodies like planetesimal and planets rather than they are protoplanetary remnants. Debris disks are therefore frequently referred to as secondary disks. 

Classically, debris disks have been considered as disks devoid of molecular gas, but recent observations have detected rich gas content (e.g., CO, \ion{O}{1}, \ion{C}{1}, \ion{C}{2}) in tens of debris disks.
While models that explains production of the secondary grains are well established, the origins of the gas component remain unclear. The gas could be both of protoplanetary remnants and secondary products. 
Observations have revealed consistent spatial locations of dust and gas contents in asymmetric disks of $\beta$~Pic, HD181327, and Fomalhaut \citep{2014_Dent, 2016_Marino, 2017_Matra}. 
\replaced{It suggests that these gas and dust have common origins and have likely been produced by secondary processes.}
{\revision{The co-spatial dust and gas would be straightforwardly associated with secondary origin scenarios.}}
By contrast, most of axisymmetric disks show non-cospatial configurations between dust and gas \citep{2013_Kospal, 2016_White, 2017_Hughes, 2017_Marshall, 2017_Higuchi, 2018_Hughes}. 
\replaced{These debris disks may contradict the secondary-origin models and thus possibly have origins from protoplanetary gas.}
{\revision{These disks might appear contradicting the secondary-origin models according to typical CO lifetimes ($\sim 10^2\yr$), but recent models show that self-shielding and/or mutual shielding by \ion{C}{1} can extend CO lifetimes longer than viscous timescale \citep{2019_Kral, 2019_Moor, 2020_Marino}. The gas can viscously spread in systems, observed as a dislocated disk from the dust component.}}

Another intriguing statistics of gas-rich debris disks is a higher frequency around intermediate-mass (A- and B-type) stars than around lower-mass (F-, G-, K-, M-type) stars with CO, O, and \ion{C}{2} lines for $< 50\Myr$ systems \citep{2018_Hughes}. 
\replaced{Since the intermediate-mass stars have orders of magnitude stronger FUV radiation, which shortens CO gas lifetimes in debris disks, the secondary-origin models appear inconsistent to explain the statistics. On the other hand, it can be consistent with the primordial-origin scenario if protoplanetary gas tends to survive longer around intermediate-mass stars than around lower-mass stars. 
The primordial-origin scenario requires sufficiently low mass-loss rates to retain gas disks for $\gtrsim 10\Myr$ around A-type stars regardless of strong radiation. 
Our hypothesis on PPD mass evolution in \secref{sec:stellardependence} (also see \fref{fig:schematic}) simultaneously satisfies these requirements. Therefore, if the gas component of gas-rich debris disks is primordial-origin, the reduced photoevaporation rates due to a strong radiation force on small grains can be a good explanation for the higher frequency of gas-rich debris disks around AB-type stars. 
}
{\revision{To explain the statistics with secondary-origin scenarios, one may need a higher production rate of secondary C and CO for debris disks around intermediate-mass stars. 
Explaining the higher detection frequency with primordial-origin scenarios requires a longer dispersal time for primordial disks around intermediate-mass stars than around low-mass stars despite orders of magnitude stronger FUV radiation. It also requires an absolute dispersal time of $\gtrsim 10\Myr$ to be observed as debris disks. 
The statistics appears puzzling from the viewpoints of both primordial-origin and secondary-origin scenarios.
On the other hand, the statistics may provide clues to figure out the origins of the gas in debris disks.

Interestingly, our evolved disk model has the potential to provide a natural explanation to the statistics. Intermediate-mass stars have atypically low EUV/X-ray luminosities compared to low-mass stars. It is advantageous to yield a longer disk lifetime for intermediate stars than for low-mass stars, if photoevaporation dominates mass loss at the late stage. FUV luminosities are higher but need small grains to drive photoevaporation. If planetesimal formation proceeds at a rate where the dust-to-gas mass ratio can get very low, disks can survive for a long time as shown by this study. 
In addition, radiation force exerting on small grains is roughly 10 times higher around intermediate-mass stars and is stronger than the host star's gravity, i.e., $\beta \gg 1$. The removal timescale of small grains via the strong radiation is much shorter than the production timescale due to the collisional cascade (see \secref{sec:blowout}). This might indicate that the disks have environments where small grains are depleted relatively easily. This can result in a longer survival of gas disks selectively around intermediate-mass stars.
}}

Besides the radiation force, small grains are also depleted by grain growth in the neutral region, and the amount of small grains would approximately scale as $\propto (a_{\rm max}/0.1\mum)^{-1/2}$. For example, when grains have grown to $a_{\rm max} \sim 1\cm$, the number of small grains reduces by a factor of 100 from the ISM values. It results in reducing photoelectric heating efficiency as in the evolved disks. 
\cite{2018_Nakatani,2018_Nakatanib} show that reduced amount of grains from the ISM value (typically, by a factor of 100--1000) results in ceasing photoevaporative flows excited by photoelectric heating to extend the lifetimes of disks to $>10\Myr$. 

In conclusion, the observed high frequency of gas-rich debris disks around A-type stars are overall consistent with our reduced FUV photoevaporation model of the evolved disks. It proposes plausibility that the gas component of gas-rich debris disks is protoplanetary remnants. Also, the reduced FUV photoevaporation model may explain the observed rich gas content in the fairly aged disks ($\sim 6\Myr$) around Herbig Ae/Be stars \citep{2008_Panic, 2017_Fedele, 2019_Booth, 2019_Miley}. 
\added{\revision{Our reduced-photoevaporation model gives implications to gas accretion onto planets at a late stage of disk evolution ($\gtrsim 10\Myr$) and to formation of gas-rich, dust-poor disks. Such objects can be targeted by future ALMA and James Webb Space Telescope observations. Comparing our model with far-infrared observations and exoplanet statistics would also give intriguing links between disk-dispersal models and planet formation. We will work on this subject in future studies. }}

\subsection{Other Thermal Processes}    \label{sec:otherheating}
\deleted{We have incorporated FUV photoelectric heating and photoionization heating due to EUV and X-ray in our simulations. Prior works have suggested other heating processes that can be effective to PPDs. 
\cite{2017_Wang} demonstrate FUV pumping of \ce{H2} and its subsequent chemical effects to dominate heating with a strong Lyman-Werner (LW; $11.2 \eV \lesssim h\nu \lesssim 13.6\eV$) radiation of the order of $10^{31.5}\erg \sec^{-1}$ for PPDs around solar-type stars. 
Assuming blackbody spectrum with an effective temperature of $9000\Kelvin$, A-type stars have comparable LW luminosities, and therefore the heating process can be important. The heating efficiency depends on \ce{H2} abundance at low density layers in the neutral region, which is essentially determined by balance of photodissociation by LW photons and grain-catalyzed molecule formation. 
\footnote{Three-body reactions and \ce{H-} process also contribute to \ce{H2} formation in high- ($\nh \gtrsim 10^8\cm^{-3}$) and low-density layers ($\nh \lesssim 10^8\cm^{-3}$), respectively, but they are subdominant compared to grain-catalyzed formation with the ISM composition \citep{2000_Omukai}.  X-ray photoionization, cosmic ray, and OH formation are also destruction processes for \ce{H2}. Their destruction effects are partial; they do not turn molecular layers into fully atomic as photodissociation by LW photons. }
The efficiency of the latter is significantly lowered in the evolved disks because of a reduced total surface area of grains. 
This effect can reduce the \ce{H2} abundance and thereby heating by \ce{H2} pumping in the neutral layers. Therefore, it is likely that incorporating \ce{H2} pumping does not strongly change the results in the present study.

Using photodissociation region models, \cite{2020_Grassi} show effectiveness of chemical heating, which derives from energy release due to exothermic reactions. Resulting temperatures are of the order of $ 100\Kelvin$ and are not sufficiently high to drive photoevaporation from the neutral layers. These heating processes might increase the temperature of the neutral regions but would not contribution to mass loss. Again, including the heating processes would not differ our conclusions. 

Nevertheless, we note that these heating processes are essential to determine temperatures of neutral layers in PPDs and to make detailed comparisons of disk structures between theoretical models and observations. 
}

\added{\revision{In general, collisionally excited lines (CELs) such as [\ion{O}{2}] 3727, 3730\,\AA; [\ion{S}{2}] 6733, 6719\,\AA; and [\ion{N}{2}] 6585, 6550\,\AA can be important coolants in the \ion{H}{2} region. The specific cooling rate is roughly of the order of $\Lambda_{\rm CEL} \sim 1\text{--}10 \, n_{\rm e} \erg \cm^3 \gram^{-1} \sec^{-1} $ for $\nh \sim 10^2$--$10^5\cm^{-3}$ \citep{2011_Draine}. 
The adiabatic cooling rate is much higher than the CEL's cooling rate, as shown by \fref{fig:coolingheating} (see also Section~4.7 of \citet{2018_Nakatani}). 
Therefore, we expect that including the coolants would not significantly change the results of this study. 
Note that the net effect of coolants is to reduce the mass-loss rates. 
It means that if effective, CEL cooling could work in a way to make our conclusions firmer. 
}}

\added{\revision{
\subsection{Thermal Sweeping}
We have not observed a rapid inside-out dispersal of truncated disks \citep[so-called thermal sweeping][]{2013_Owen} except for the strong-EUV model with the initial disk mass of $M_{\rm disk} \approx 3\e{-6} \Msun$. The assumed disk mass is the lowest among the runs performed here. In the simulation, the inner hole expands from $25\au$ to $\sim 10^2\au$ on a timescale of $\sim 10^3 \yr $. 
Surface mass-loss timescale, which we define as $\Sigma/ \dot{\Sigma}$, is continuously $\lesssim 10^3\yr$ at the evaporating rim while expanding the hole size. Therefore, the apparent rapid dispersal can be simply interpreted as conventional photoevaporation of a very low mass disk; thermal sweeping could also be responsible for the rapid clearing, but it is in principle difficult to distinguish conventional photoevaporation from thermal sweeping in this case. Currently, the theory of thermal sweeping is under development. Both a high luminosity (mass-loss rate) and a low surface density (disk mass) are required for onset of the potential instability. \cite{2013_Owen} introduce an energy-limited clearing timescale, and it is qualitatively similar to the surface mass-loss timescale. This implies general difficulties to tell whether the disk disperses via thermal sweeping or via conventional photoevaporation. More detailed investigations are needed to understand how these compatible processes can be quantitatively discriminated, which is out of the scope of this study.
}}

\section{Summary \& Conclusions}   \label{sec:conclusion}
We have performed radiation hydrodynamics simulations of UV- and X-ray-irradiated evolved disks.  
The disks are modeled as gas-rich, optically-thin disks where small grains ($\lesssim 4 \mum$ for A-type star cases) are assumed continuously depleted by the blowing-out effect of radiation pressure. 
\replaced{The disk is modeled as either of a truncated disk or a full disk with $M_{\rm disk } = 10^{-3}$--$10^{-2}\Msun$. The maximum dust size has been ranged in $10^4\mum \leq a_{\rm max} \leq 10^{11} \mum$, and the corresponding visual extinction per hydrogen column density is $10^{-28}\cm^{-2}\lesssim \Sigma_{\rm V} \lesssim 10^{-25}\cm^{-2}$. }
{\revision{The minimum and maximum dust sizes are fixed at $4\mum$ and $1\cm$, respectively. The disk is modeled as a truncated disk at $25\au$ and has the total solid mass ($<1\cm$) of $\solidmass = 0.1\mbox{10}\Mearth$ and dust-to-gas mass ratios of $\dgratio = 0.001\mbox{--}0.1$. The corresponding total disk mass is $M_{\rm disk} \approx 10^{-6}\mbox{--}10^{-2}\Msun$. }}

The results show that grain photoelectric heating is inefficient to drive dense photoevaporative flows from the neutral layers in the evolved disks \added{\revision{with $\dgratio \lesssim 0.01$}}. It is in contrast to primordial disks with a rich content of small grains as in ISM. They yield mass-loss rates of $\sim 10^{-8}$--$10^{-7}\Msun\yr^{-1}$ by photoelectric heating. 
EUV photoevaporation dominates gas removal in these evolved disks, while X-ray effects are limited. The mass-loss rate scales with the EUV emission rate $\LEUV$ as $\dot{M} \approx1\e{-11} (\LEUV/10^{38}\sec^{-1})^{1/2}\Msun\yr^{-1}$.
\added{
\revision{With $\dgratio = 0.1$, on the other hand, photoelectric heating is efficient enough to yield $\dot {M} \approx 3\e{-10}\Msun\yr^{-1}$.  }
}
In general, the mass-loss rate does not strongly depend on \deleted{$a_{\rm max}$, }$M_{\rm disk}$ or disk geometry 
\added{for $M_{\rm disk}\lesssim 10^{-3}\Msun$}.

\added{\revision{
We have found that for $M_{\rm disk} \gtrsim 10^{-3}\Msun$, \ce{H2} pumping can drive neutral photoevaporative flows if an \ce{H2}-rich disk is initially given. The neutral photoevaporative flows are excited from the \ce{H2}-rim while the dissociation front sweeps the disk. About 70\% of the initial mass is lost by this process, but the rest forms a stable H-rich disk. The swept region never reproduces an \ce{H2}-rich environment because of inefficient \ce{H2} formation and strong photodissociation/photodetachment. EUV-driven photoevaporation dominates the mass loss of the remnant disk with the same $\dot{M}$ as above. 
We note that this \ce{H2} dissociation front sweeping requires an initially \ce{H2}-rich disk; if the disk is \ce{H2}-poor in the first place, the mass-loss process is the same as that of lower-mass disks ($M_{\rm disk} \lesssim 10^{-3}\Msun$).
The abundances of these species in the evolved disks would be determined by the history of disk evolution.
Further studies are necessary to derive a reasonable distribution of \ce{H2}. 
}}

The dispersal time of the gas component in evolved disks is set by $\LEUV$ and the ``initial'' disk mass at the point when the bulk of small grains have been lost from the disk. 
The estimated dispersal time of the gas component is roughly $\sim 100 M_3 \Phi_{38}^{1/2} \Myr$, where $M_3$ is the ``initial'' disk mass normalized by $10^{-3}\Msun$ and $\Phi_{38}\equiv (\LEUV/10^{38}\sec^{-1})^{1/2}$.
Hence, depletion of small grains is capable of extending the dispersal timescale of the gas disk to $> 10\Myr$ by ceasing dense FUV-driven photoevaporative flows from neutral layers \added{\revision{and by making \ce{H2} production inefficient. 
We note that generally, $\LEUV$ are not well known mainly because of significant interstellar extinction by atomic hydrogen. Our results show the importance of determining $\LEUV$ to understand the disk lifetimes for intermediate-mass stars.  
}}

\replaced{
With the radiation force of solar-type stars which is approximately an order of magnitude weaker than intermediate-mass stars, small grains are more likely to remain in the disk. Then, FUV photoevaporation is continuously active throughout the evolution and can maintain a mass-loss rate of $\approx 10^{-8}\Msun \yr^{-1}$. In contrast, we have estimated the radiation pressure of A-type stars sufficiently strong to deplete small grains continuously. The evolved disks around A-type stars likely maintain a grain-depleted state. The mass-loss rates are determined by $\LEUV$. The atmospheric absorption of A-type stars can reduce $\LEUV$ by orders of magnitude compared to young low-mass stars. This effect can also result in a significantly smaller photoevaporation rate around A-type stars than around solar-type stars.
These results indicate that gas component of PPDs around A-type stars can survive for a relatively long time. This trend is consistent with the higher frequency of gas-rich debris disks around AB-type stars than around FGK-type stars. Our results show the plausibility of the gas component in gas-rich debris disks being protoplanetary remnants.}
{\revision{
Intermediate-mass stars have atypically low EUV/X-ray luminosities compared to young low-mass stars, and thus EUV/X-ray photoevaporation rates are orders of magnitude smaller. 
The FUV luminosity is much higher, but abundant small grains and/or \ce{H2} molecules are needed to drive photoevaporation. Grain-depleted disks are devoid of these essential agents, and hence FUV photoevaporation is ineffective. 
}}
These effects can result in longer lifetimes of gas component in evolved disks around A-type stars than around low-mass stars. 
Interestingly, this trend is consistent with the higher frequency of gas-rich debris disks around AB-type stars than around FGK-type stars. It implies the plausibility of the gas component in gas-rich debris disks being protoplanetary remnants.

\acknowledgments %
We thank Masanobu Kunitomo, Ryo Tazaki, Sebasti\'an Marino, Aya Higuchi, and Hiroto Mitani for fruitful discussions and technical advises.
We are also grateful to the anonymous referee for insightful comments, which have helped improve the manuscript greatly. 
RN is supported by the Special Postdoctoral Researcher (SPDR) Program at RIKEN and by Grant-in-Aid for Research Activity Start-up (19K23469).
RK acknowledges financial support via the Emmy Noether Research Group on Accretion Flows and Feedback in Realistic Models of Massive Star Formation funded by the German Research Foundation (DFG) under grant no.~KU~2849/3-1 and KU~2849/3-2.


\bibliographystyle{aasjournal}
\bibliography{references}

\setcounter{table}{0}
\renewcommand{\thetable}{\Alph{section}\arabic{table}}
\begin{appendix}
\section{Heating and cooling processes}	\label{app:newheatcool}
	We take into account EUV- and X-ray-induced ionization heating \citep{1996_Maloney, 2000_Wilms, 2004_Gorti},
	FUV-induced grain photoelectric heating \citep{1994_BakesTielens},
	heating associated with \ce{H2} pumping \citep{1990_Burton, 2006_Rollig}, 
	\ce{H2} photodissociation heating \citep{1979_HollenbachMcKee, 1996_DraineBertoldi}, 
	chemical heating/cooling \citep{1979_HollenbachMcKee},
	and \CI{} ionization heating \citep{1987_Black, 2004_Jonkheid, 2012_UMIST}. 
	For cooling sources, we include radiative recombination cooling of \HII{} \citep{1978_Spitzer}, 
	Ly${\rm \alpha}$ cooling of \HI{} \citep{1997_Anninos}, 
	fine-structure line cooling of \OI{} 
	and \CII{} \citep{1989_HollenbachMcKee,1989_Osterbrockbook,2006_SantoroShull},
	molecular rovibrational line cooling of \ce{H2} and CO \citep{1998_GalliPalla,2010_Omukai}, 
	and dust-gas collisional heat transfer \citep{1996_YorkeWelz}. 
	We refer the readers to Papers~I and II for the implementation 
	of the heating/cooling processes
	except for \ce{H2} pumping, photodissociation heating, 
	and \CI{} photoionization heating;
	the implementation of these additional processes is shown in the following sections.
	In the present study, we modify the photoelectric heating rate (see \appref{app:modificationFUVheating})
	and dust-gas collisional heat transfer (\eqnref{eq:dust-gascollisional}) suitably for the evolved disk.

\subsection{Approximate Modification to Photoelectric Heating Rates}	\label{app:modificationFUVheating}
		The specific photoelectric heating rate of \cite{1994_BakesTielens} is given by
		\gathering{
		\Gamma_{\rm pe}=  10^{-24}\unit{erg}{}\unit{s}{-1}\,
		\epsilon_{\rm pe} \FFUV e^{-\Av} \frac{\nh}{\rho}, 
		\label{eq:photoeleheat} \\
		\splitting{
		\epsilon_{\rm pe} =  &
		\left[ \frac{4.87\e{-2}}{1+4\e{-3} \, \gamma_{\rm pe}^{~0.73}} \right.\\
		& \left.+ \frac{3.65\e{-2}(T/10^4\Kelvin)^{0.7}}{1+2\e{-4} ~ \gamma_{\rm pe}} \right],
		}
		\label{eq:photoelectricefficiency}
		}
		where $\epsilon_{\rm pe}$ is the photoelectric effect efficiency, 
		$\FFUV$ is a geometrically attenuated FUV flux ($\LFUV/4\pi r^2$) normalized by an average interstellar radiation field \citep[$\approx 1.6\e{-3}\erg \cm^{-2} \sec^{-1}$;][]{1968_Habing},
		and $\gamma_{\rm pe}$ is a parameter describing the ratio of photoionization rates 
		to grain recombination rates,
		$\gamma_{\rm pe} \equiv \FFUV \exp[-\Av] \braket{T/\Kelvin}^{1/2} \braket{n_{\rm e}/\cm^{-3}}^{-1}$.
		BT04 adopts $a_{\rm min} = 3\e{-4} \mum$ and $ a_{\rm max} = 0.01\mum$ and has found that smaller grains ($\lesssim 1.5\e{-3}\mum$) dominantly contribute to heating. Since our grains are larger than $1\mum$, corresponding $\epsilon_{\rm pe}$ is expected to be lower by orders of magnitude than those computed by \eqnref{eq:photoelectricefficiency}. 
		In order to incorporate these reducing effects approximately in this study, we simply multiply $\braket{a_{\rm min}/{0.005\mum}}^{-1/2} \braket{{a_{\rm max}}/{1\mum}}^{-1/2}$ to $\FFUV$ to be consistent with \eqnref{eq:sigmaV} \citep{2007_Nomura_II, 2015_Gorti}. 
		Technically, the coefficients of \eqnref{eq:photoelectricefficiency} can depend on dust size distribution. However, photoelectric heating is presumably expected to be ineffective in disks where small grains have been depleted. This rough approximation does not affect the results and conclusions in this study unless it underestimates the heating rates by several orders of magnitude. 

\added{\revision{
\subsection{Heating by pumping and photodissociation of molecular hydrogen} 
Hydrogen molecules are pumped to the excited electronic state by absorbing the Lyman-Werner photons. The excited \ce{H2} decays to the vibrational continuum of the ground electronic state at a probability of $\sim 10\%$, leading to photodissociation of the molecule. Approximately $\Delta E_{\rm diss}\sim 0.4\eV$ of energy is deposited to the gas per photodissociation \citep{1979_HollenbachMcKee}. 
The rest $90\%$ decays to the excited rovibrational levels of the ground electronic state, followed by further decay to the ground vibrational state either by collisions or spontaneous emission. The collisional de-excitation also results in heating the gas. 

We implement the heating rate of \ce{H2} photodissociation as 
\[
    \Gamma_{\rm diss} = \Delta E_{\rm diss} R_{\rm diss} n_{\ce{H2}}, 
\]  
where $R_{\rm diss}$ is a photodissociation coefficient \citep[][for a detailed implementation]{1996_DraineBertoldi, 2018_Nakatani}. 
As for \ce{H2} pumping, we largely follow \cite{2006_Rollig} and \citet{2020_Gressel}
and give the heating rate as
\[
\begin{gathered}
    \Gamma_{\rm pump} = { \Delta E_{\rm *}R_{\rm pump}} \braket{1 + \frac{n_{\rm cr}}{\nh}}^{-1} n_{\ce{H2}} \\
    n_{\rm cr} = \frac{A_{\rm eff} + R_{\rm eff, diss}^\prime}{\gamma_{\rm eff}} , 
\end{gathered}
\]
where $R_{\rm pump} = 9 R_{\rm diss}$ is the production rate of \ce{H2} at the vibrationally excited levels of the ground electronic state via FUV pumping, 
$\Delta E_{\rm  *} = 2.0\eV$ is the effective deposited energy by collisional de-excitation of pumped \ce{H2} \citep[][]{1990_Burton}, 
$n_{\rm cr}$ is a critical density defined by 
the effective spontaneous decay coefficient $A_{\rm eff} = 1.9\e{-6}\sec^{-1}$, the effective collisional de-excitation coefficient $\gamma_{\rm eff} = 5.4\e{-13}\sqrt{T/1\Kelvin}\cm^{3}\sec^{-1}$, 
and the effective photodissociation rates for vibrationally excited levels $R_{\rm eff, diss}^\prime \approx 7 R_{\rm diss}$ \citep{2006_Rollig, 2020_Gressel}. 
\footnote{With these definitions, $\Delta E_{\rm *} R_{\rm pump}$ yields $ \approx 1\e{-21}\erg\sec^{-1}$, which is almost the same as the original value ($0.94\e{-21}\erg\sec^{-1}$) in \citet{2006_Rollig}. For the estimation of $R_{\rm eff, diss}^\prime$, we have used the effective photodissociation rate $D_{\rm eff} = 4.7\e{-10}\sec^{-1}$ of \citet{2006_Rollig} \citep[see also Appendix~B of][]{2020_Gressel}.}

\subsection{Chemical Heating and Cooling}
Exothermic reactions also work as heating sources for the gas by collisionally converting the chemical energy to the kinetic energy. Similarly, endothermic reactions work as cooling sources by using the kinetic energy of the gas for the chemical reactions.
In the present study, we incorporate chemical heating/cooling due to \ce{H2} formation/dissociation and \ion{H}{1} ionization. 
The Formation reactions of \ce{H2} on grain surfaces (\ce{H + H ->C[grain] H2}) and via \ce{H-} process (\ce{H- + H -> H2 + e}), \ce{H2^+} process (\ce{H2^+ + H -> H2 + H+}), and three-body reactions (\ce{3H -> H2 + H}, \ce{2H + H2 -> 2H2}) are exothermic reactions, and the resulting \ce{H2} goes to the excited rovibrational levels within the ground electronic state. The excited molecules cascade to the lower levels by collisions and spontaneous decay. We adopt
\[
\begin{gathered}
 \Gamma_{\rm gr} = k_{\rm gr}\nh^2 \abn{HI}\braket{0.2 + \frac{4.2\eV}{1 + n_{\rm cr}(\ce{H2})/\nh}}
  \\
 \Gamma_{\rm \ce{H-}} = k_{\ce{H-}} \nh^2 \abn{\ce{H-}}\abn{HI} \frac{3.53\eV}{1 + n_{\rm cr}(\ce{H2})/\nh}   \\
 \Gamma_{\rm \ce{H2^+}} = k_{\ce{H2^+}} \nh^2 \abn{\ce{H2^+}}\abn{HI} \frac{1.83\eV}{1 + n_{\rm cr}(\ce{H2})/\nh}  \\
 \Gamma_{\rm three} = 
  (k_{19} \abn{H}^3 + k_{20} \abn{H}^2\abn{\ce{H2}})\nh^3 \frac{4.48\eV}{1 + n_{\rm cr}/\nh} 
 ,
 \end{gathered}
\]
respectively, for the chemical heating rates \citep{1979_HollenbachMcKee, 2000_Omukai}. 
Here, $k_{\rm gr}$, $k_{\ce{H-}}$, $k_{\ce{H2^+}}$, $k_{19}$, and $k_{20}$ are the corresponding reaction coefficients (reactions~k23, k8, k10, k19, k20 in \tref{tab:rates}),
and 
\[
 n_{\rm cr} (\ce{H2}) = \frac{10^6 T^{-1/2}}{1.6\abn{HI}e^{-(400/T)^2} + 1.4\abn{\ce{H2}}e^{-12000/(T+1200)} } \cm^{-3} 
\]
is the critical density.

As for chemical cooling, we take into account the collisional dissociation of \ce{H2} and the collisional ionization of atomic hydrogen. These processes reduce the gas energy by the binding energies of the species, namely, $4.48\eV$ and $13.6\eV$ for molecular and atomic hydrogen, respectively, per destruction.

\subsection{Carbon ionization heating}
\newcommand{\abnsimple}[1]{[{#1}]}
In Papers~I and II, we have constructed a simplified carbon thermochemistry based on \cite{1997_NelsonLanger} and \cite{2000_RichlingYorke}, where atomic carbon is not explicitly incorporated as a chemical species assuming that the photoionization/photodissociation fronts of \ce{C+}/C/CO are identical. In the present study, we update the thermochemistry model by explicitly including \ion{C}{1} as a chemical species, carbon photoionization due to FUV (\ce{C + {\gamma} -> C+ + e}), and the associated heating.
The reaction rates are taken from the UMIST RATE12 database \citep{2012_UMIST} as 
\gathering{
        k_{\rm C,ph} = 1.8\e{-10} \FFUV  e^{-3.3\Av} e^{-\tau_{\rm C}} \frac{e^{-\tau_{\ce{H2}}b/\pi v_1^2}}{1 + \tau_{\ce{H2}} b/ \pi v_1^2} \sec^{-1},   
        \label{eq:kcph}
}
where $k_{\rm C,ph}$ is the photoionization rate \citep{1988_vanDishoeck}, 
$e^{-\tau_{\rm C}} $ accounts for FUV attenuation due to the self-shielding of atomic carbon \citep{1970_Werner}, 
and $\exp[-\tau_{\ce{H2}}b/\pi v_1^2] (1 + \tau_{\ce{H2}} b/ \pi v_1^2)^{-1}$ is a mutual shielding factor due to \ce{H2} \citep{1980_deJong, 1985_TielensHollenbach}. 
The dimensionless parameters in the mutual shielding factor are given as 
\[
    \begin{array}{c c}
     \tau_{\rm C} = \frac{\col{C}}{10^{17} \cm^{-2}} &
     \tau_{\ce{H2}} = \braket{\frac{ \col{\ce{H2}}}{0.83\e{14}\cm^{-2}}} \delta \tilde{v}_{\rm d} ^{-1}\\
     b = 9.2\e{-3} \delta \tilde{v}_{\rm d}^{-1}&
     v_1 = 5\e{2} \delta \tilde{v}_{\rm d}^{-1},
    \end{array}
\]
where $\col{C}$ is the column density of atomic carbon, $\col{\ce{H2}}$ is that of \ce{H2}, and $\delta \tilde{v}_{\rm d}$ is the Doppler width in units of $\kms$ which we take to be $\delta \tilde{v}_{\rm d} = \cs/1\kms$ in this study. 

Regarding the carbon ionization heating, approximately $\Delta E_{\rm C, ph}\sim 1\eV$ of energy is deposited to the gas per \ion{C}{1} ionization \citep{1987_Black, 2004_Jonkheid}, and thus the heating rate is given by
\[
    \Gamma_{\rm C,ph}   =   \Delta E_{\rm C,ph} k_{\rm C, ph} n_{\rm C}, 
\]
where $n_{\rm C}$ is the number density of atomic carbon.


\section{Chemical reactions} \label{app:chemicalnetwork}
	In Paper~I and II, we have developed a reduced chemical network. 
	We update it by adding (photo)chemical reactions 
	relevant to C/CO formation to accurately derive their abundances.
	It is essential to calculate the heating/cooling rates in the evolved disk, 
	where \ce{H2} pumping and \ion{C}{1} ionization heating are possible to be dominant over photoelectric heating. 
	We have selected the (photo)chemical reactions in our chemical network
	so that it can largely reproduce a benchmark for the chemical structures 
	of the photodissociation region \citep{2007_Rollig}.
	The new chemical network incorporates 27 species:
	\ce{H, H+, H- , H2, H2+, e, C, C+, CO, O, He, He+,
	H3+, CH, CH+, CH2, CH2+, CH3+, CO+, HCO+, OH, OH+, H2O, H2O+, H3O+, O+, O2}.
	The spatial distribution of the chemical abundances is derived in a time-dependent and non-equilibrium manner
	by solving \eqnref{eq:chemevoeq}. Note that our model takes into account the advection terms for each species.
	We list the incorporated (photo)chemical reactions in \tref{tab:rates}, 
	and the reaction coefficients are taken from the papers shown in the third column of the table. 

We have tested our updated chemical network by running the PDR benchmark tests presented in \citet{2007_Rollig} to compare the resulting temperature and abundance distributions, as have been done in \citet{2020_Gressel}. 
\begin{figure*}
    \centering
    \includegraphics[clip, width = \linewidth]{\figdir/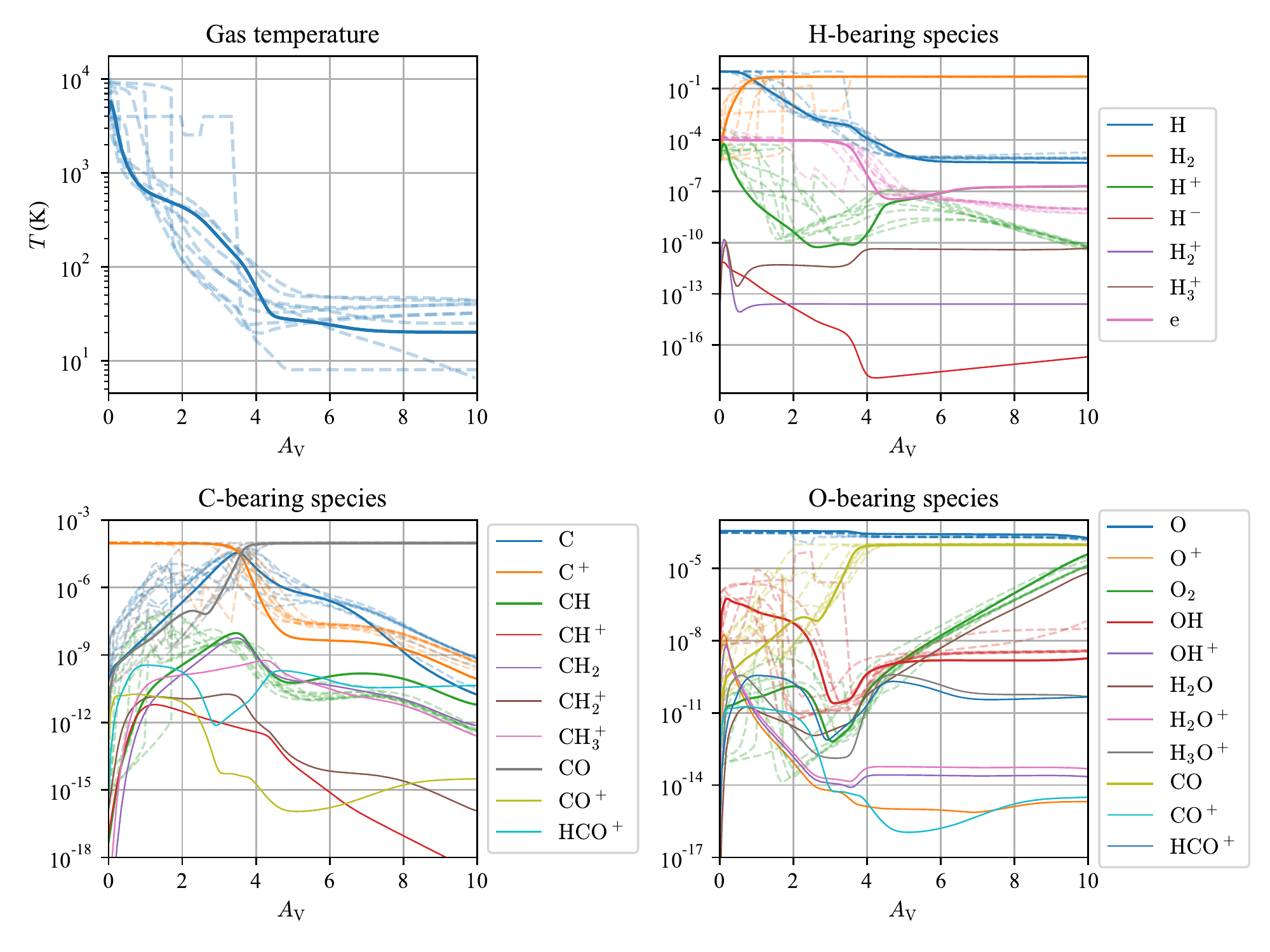}
    \caption{Distributions of temperature (top left), H-bearing species abundances (top right), C-bearing species abundances (bottom left), and O-bearing species abundances (bottom right) as functions of the visual extinction. The solid lines show our results, and they are compared with those derived by other various PDR codes \citep[dashed lines with the corresponding colors][]{2007_Rollig} on temperature and abundances of \ce{H, H2, H+, e, C, C+, CO, CH, O, O2, OH}. The data has been taken from {\tt http://zeus.ph1.uni-koeln.de/site/pdr-comparison/}.  }
    \label{fig:benchmark}
\end{figure*}
\fref{fig:benchmark} shows the results of our test where plane-parallel FUV with $G_0 = 1.7\e{5}$ is incident on a uniform-density gas slab ($\nh = 10^{5.5}\cm^{-3}$). We have given a uniform, fixed dust temperature of $T_{\rm dust} = 50 \Kelvin $ for simplicity. Despite our reduced chemical network (and other differences in computational methods), the physical quantities well agree with those derived by the more detailed chemical networks of other PDR codes \citep{2007_Rollig}.

\newcommand{\omukai}{1}
\newcommand{\umist}{2}
\newcommand{\paperone}{3}
\newcommand{\papertwo}{4}
\startlongtable
\begin{deluxetable*}{ c c c}
  \tablecaption{List of Chemical Reactions and Rate Coefficients \label{tab:rates}}
  \tablehead{
    {\rm Label}  & {\rm Reaction}   &   {\rm Reference}  
  }
  \startdata

\input{chem_reactions}

 \enddata
  \tablecomments{
  References: (\omukai) \citet{2000_Omukai}, (\umist) \citet{2012_UMIST} (UMIST RATE12), 
  (\paperone) \citet{2018_Nakatani}, (\papertwo) \citet{2018_Nakatanib}, 
  (5) \citet{1996_DraineBertoldi}, (6) \citet{1996_Lee},
  (7) \citet{1985_TielensHollenbach}
  }
\end{deluxetable*}

}}

\par 
\ \ \

\end{appendix}

\end{document}

%% file: newcommands.tex
\newcommand{\eV}{\,{\rm eV}}
\renewcommand{\sec}{{\,\rm s}}

\newcommand{\kms}{\,{\rm km\,s^{-1}}}

\newcommand{\AU}{ \, {\rm au}}
\newcommand{\au}{\AU}
\newcommand{\Msun}{{\rm \, M_{\odot}}}
\newcommand{\Mearth}{{\rm \, M_{\oplus}}}
\newcommand{\smetal}{Z_{\odot}}
\newcommand{\Rsun}{\,R_{\odot}}
\newcommand{\Lsun}{\,L_{\odot}}

\newcommand{\yr}{\, {\rm yr} }

\newcommand{\Myr}{\, {\rm Myr} }

\newcommand{\dd}{{\rm d}}

\newcommand{\cm}{{\, \rm cm}}

\newcommand{\K}{{\rm K}}
\newcommand{\erg}{{\rm \, erg}}

\ifx\arcsec\undefined
\newcommand{\arcsec}{\, {\rm arcsec}}
\fi

\newcommand{\rad}{{\,\rm rad}}
\newcommand{\gram}{{\,\rm g}}
\renewcommand{\arraystretch}{1.4}

\newcommand{\eq}[1]{\begin{equation}#1\end{equation}}

\newcommand{\gathering}[1]{\begin{gather}#1\end{gather}}
\newcommand{\splitting}[1]{\begin{split}#1\end{split}}
\newcommand{\braket}[1]{\left(#1\right)}

\newcommand{\e}[1]{\times 10^{#1}}

\newcommand{\beqas}{\begin{eqnarray*}}
\newcommand{\eeqas}{\end{eqnarray*}}

\ifx\del\undefined
\newcommand{\del}[2]{\frac{\dd #1}{\dd #2}}
\fi

\ifx\appref\undefined
\newcommand{\appref}[1]{Appendix~\ref{#1}}
\fi
\newcommand{\fref}[1]{Figure~\ref{#1}}
\newcommand{\tref}[1]{Table~\ref{#1}}

\ifx\secref\undefined
\newcommand{\secref}[1]{Section~\ref{#1}}
\fi
\newcommand{\eqnref}[1]{Eq.(\ref{#1})}

\newcommand{\mum}{\, {\rm \mu m}}
\newcommand{\unit}[2]{\, {\rm #1}^{#2}}

\newcommand{\cs}{c_{\rm s}}

\newcommand{\Av}{A_{\rm V}}

\ifx\dif\undefined
\newcommand{\dif}[2]{\frac{{\rm d}#1}{{\rm d}#2}}
\fi

\newcommand{\col}[1]{N_{\rm #1}}

\newcommand{\metal}{Z}
\makeatletter
\newcommand*{\rom}[1]{\expandafter\@slowromancap\romannumeral #1@}
\makeatother
\ifx\ion\undefined
\newcommand\ion[2]{#1$\;${\small\rmfamily\rom{#2}}\relax}
\fi

\newcommand{\HI}{\ion{H}{1}}

\newcommand{\HII}{\ion{H}{2}}

\newcommand{\CI}{\ion{C}{1}}

\newcommand{\CII}{\ion{C}{2}}

\newcommand{\OI}{\ion{O}{1}}

\newcommand{\dgratio}{\mathscr{DG}}     
\newcommand{\abn}[1]{y_{\text{\rm #1}}}
\newcommand{\nh}{n_{\text{\rm H}}}

\newcommand{\nspe}[1]{n_{\text{\rm #1}}}

\newcommand{\Kelvin}{{\rm \, K}}

\newcommand{\GFUV}{G_{\rm FUV}}
\newcommand{\LFUV}{L_{\rm FUV}}
\newcommand{\LEUV}{\Phi_{\rm EUV}}
\newcommand{\FFUV}{\GFUV}

\ifx\fh\undefined
\newcommand{\fh}{f_{\rm h}}
\fi

\ifx\rmxaa\undefined
\newcommand{\rmxaa}{}
\fi

%% file: blowout.tex
\replaced 
{
We have determined the minimum grain size by \eqnref{eq:maxs}, assuming $Q_{\rm pr} = 1$. The resulting minimum grain size has been $10\mum$ for the A-type star model and $1\mum$ for the solar-type star model. 
Technically, the transfer efficiency $Q_{\rm pr}$ decreases with $s$ for $s \lesssim \lambda_{\rm peak}$. The decrease of $Q_{\rm pr}$ mitigates increase of $\beta$ with lowering $s$. Therefore there are two roots for the equation, $\beta = 1$, depending on the bulk density and composition of grains. Generally, it is the case for stars with $1\Lsun \lesssim L_* \lesssim 10 \Lsun$ \citep{2015_Pawellek}. The blown-out grain size has a lower limit, $s_{\rm esc,min}$, as well as an upper limit, $s_{\rm esc,max}$, for those stars. 
Having the lower limit indicates that small grains lower than $s < s_{\rm esc,min}$ can remain in the evolved systems without being blown out. The lower limit is typically $s_{\rm sec, min} \sim 0.1 \mum$ at $L_* \approx 1\Lsun$ and decreases for higher stellar luminosity; it is $s_{\rm sec, min} \sim 0.001 \mum$ at $L_* \approx 10\Lsun$. 

We derive $Q_{\rm pr}$ and $\beta$ for our model, using the dust model and model parameters (\fref{fig:Qpr}). 
\begin{figure}
    \centering
    \includegraphics[clip, width = \linewidth]{figs/Debris_plotBetaQpr.pdf}
    \caption{Magnitude ratio of radiation pressure to the host star gravity $\beta$ (top), and the momentum transfer efficiency $Q_{\rm pr}$ (bottom) for our models. The red and blue lines represent these quantities for the A- and solar-type models, respectively. The cyan-shaded region shows $\beta > 1$, where radiation pressure dominates the gravity, leading to radial acceleration of grains. Note that $Q_{\rm pr}$ scales as $\propto s$ for $s \ll \lambda_{\rm peak}$; the black dashed line shows a reference.}
    \label{fig:Qpr}
\end{figure}
Indeed, $\beta > 1$ is satisfied for all $s$ with $s\lesssim10\mum$ with the A-type star model, while it is limited within $0.1\mum \gtrsim s \gtrsim 1\mum$ for the solar-type star model (see the cyan-shaded region in \fref{fig:Qpr}). 
Hence, not including all of small grains with $s\lesssim1\mum$ in our solar-type star model (\secref{sec:solartype}) is likely unrealistic. If small grains and PAHs are present, FUV can yield a mass-loss rate of $\approx 10^{-8}\Msun \yr^{-1}$, as has been shown by prior works \citep[e.g.,][]{2009_GortiHollenbach, 2018_Nakatani, 2018_Nakatanib}. 
In the A-type star model, our assumption on the minimum grain size is well justified. Thus, we confirm the results and conclusions in \secref{sec:results}.

Even for grains in a range of the blown-out size, the aerodynamical friction between gas and dust may affect to restrain the grains accelerated by the radiation pressure within the disk. It is also necessary to assess such possibilities to validate our assumptions on the minimum grain sizes for the A-type star model. 
The magnitude of the aerodynamical friction is characterized by the Stokes number,
\[
\splitting{
    {\rm St}  & = \Omega t_{\rm tstop}  
    =	 \sqrt{\frac{\pi}{8}}\frac{\rho_{\rm b} s }{\rho \cs} \Omega\\
    & \approx 0.64 \braket{\frac{\rho_{\rm b}}{1.4\gram \cm^{-3}}} \braket{\frac{s}{1\mum}}\braket{\frac{M_*}{2\Msun}}^{1/2}\\
 & \times \braket{\frac{\nh}{10^6\cm^{-3}}}^{-1} \braket{\frac{r}{30\au}}^{-3/2} \braket{\frac{\cs}{1\kms}}^{-1} . }
\]
The friction is strong enough to couple the motion of grain and gas for grains with ${\rm St} < 1$. 
This condition sets the lower limit of grain sizes that dynamically decouples from gas 
\eq{
\splitting{
    s & > s_{\rm stop, max} \equiv 1.56
    \braket{\frac{\rho_{\rm b}}{1.4\gram \cm^{-3}}}^{-1} \braket{\frac{M_*}{2\Msun}}^{-1/2}\\
    & \times \braket{\frac{\nh}{10^6\cm^{-3}}} \braket{\frac{r}{30\au}}^{3/2} \braket{\frac{\cs}{1\kms}} \mum . }
    \label{eq:mins}
}
Grains satisfying both of ${\rm St} > 1$ and $\beta > 1$ can escape from the optically-thin layers in a short time without being held back by gas. 
\footnote{Note that grain sizes satisfying both of ${\rm St} > 1$ and $\beta > 1$ necessarily fulfills $t_{\rm stop} > t_{\rm esc}$, i.e., the grains escape from the system before dynamically coupling with the gas.} 
\footnote{In cases of $t_{\rm stop} < t_{\rm esc}$, 
transferred radiation momentum is immediately transferred to the gas. If the transferred momentum is sufficiently large, it can drive an outward movement of the gas. \mycomment{this footnote will be removed soon. It is here at this moment just for my use.}}
Note that this is a sufficient condition; not only these grains are capable of escaping from the system (see below). 
For the existence of grains with ${\rm St} > 1$ and $\beta > 1$, 
$s_{\rm stop, max} < s_{\rm esc,max}$ is required. This necessity reduces to a restriction to gas density as
\eq{
\splitting{
        \nh & < n_{\rm H, crit}   \equiv  1.1\e{7}\braket{\frac{\cs}{1\kms}}^{-1} 
        \braket{\frac{Q_{\rm pr}}{1}}\\
        &\times \braket{\frac{L_*}{20\,L_\odot}}
        \braket{\frac{r}{30\au}}^{-3/2}\braket{\frac{M_*}{2\Msun}}^{-1/2} \cm^{-3}.   \label{eq:critdensity}
}
}
Hence, grains are able to escape by the effects of radiation pressure without being dragged by gas friction in the low-density layers where $\nh > n_{\rm crit}$. 
The critical gas density 
implicitly depends on $s$ and $\rho_{\rm b}$ in terms of $Q_{\rm pr}$. For $s < \lambda_{\rm peak}$, $n_{\rm H,crit}$ is small. Typical peak wavelength of stellar radiation is at $\sim 0.5\mum$ for A-type stars and at $\sim 1\mum$ for solar-type stars with the blackbody radiation.

We stress again that \eqnref{eq:maxs} is a necessary condition for escape driven by radiation pressure, while \eqnref{eq:critdensity} is a sufficient condition; that is, not only the grains satisfying \eqnref{eq:critdensity} but also small grains with ${\rm St} \ll 1$ are able to move outward to escape from the system under strong radiation. Acceleration due to radiation pressure yields nonzero radial velocity unless ${\rm St}  = 0$. Very small ${\rm St}$ affects to reduce the radiation pressure force. The radial distance of small grains accelerated by radiation pressure would increase at an approximate rate of 
\[
\splitting{
    v_{r, \rm drag} & \equiv r \Omega (\beta - 1) {\rm St} \\
    & \approx 1.04 \, \braket{\frac{\rho_{\rm b}}{1.4\gram \cm^{-3}}} \braket{\frac{s}{1\mum}}\\
 & \times   \braket{\frac{M_*}{2\Msun}}
  \braket{\frac{\nh}{10^6\cm^{-3}}}^{-1} 
  \braket{\frac{r}{30\au}}^{-2}\\ 
  & \times  \braket{\frac{\cs}{1\kms}}^{-1} (\beta - 1) \au \yr^{-1}.\\
    }
\]
The accelerated grains are blown out from a disk with a size of $R_{\rm disk}$ on a timescale of 
\eq{
\splitting{
   t_{\rm blow} & \equiv \frac{R_{\rm disk}}{v_{r,\rm drag}}\\
        & \approx 10^{-4}\,  \braket{\frac{\rho_{\rm b}}{1.4\gram \cm^{-3}}}^{-1} \braket{\frac{s}{1\mum}}^{-1}\\
 & \times   \braket{\frac{M_*}{2\Msun}}^{-1}
  \braket{\frac{\nh}{10^6\cm^{-3}}}
  \braket{\frac{r}{30\au}}^{2}\\ 
  & \times  \braket{\frac{\cs}{1\kms}} 
  \braket{\frac{R_{\rm disk}}{100\au}} (\beta - 1)^{-1}  \Myr.\\
  \label{eq:blow}
  }
}
All grains that have $\beta$ fairly larger than unity appear to be blown out within $\sim 1\Myr$ from the optically-thin layer of PPDs, even with ${\rm St} \ll 1$. 
Hence, radiation pressure is expected to blows out small grains with $s < s_{\rm esc, max}$ around A-type stars within a fairly short time regardless of the size (or Stokes number).

To summarize, the adopted minimum size of grains are well justified for the A-type star model. On the other hand, the blow-out effects of the radiation pressure is likely significantly overestimated in our solar-type star model in \secref{sec:solartype}. The solar-type star model corresponds to an extreme case where small grains have been depleted to an unrealistic extent. We conclude that EUV-driven photoevaporation is dominant in the evolved disks around A-type stars because of a reduced photoelectric heating, while for solar-type stars, small grains can remain in the disks so that FUV-driven photoevaporation is active.
This spectral-type dependence gives an interesting implication to the statistics of gas-rich debris disks, and we discuss it in the next section. 
}
{

\newcommand{\tdifver}{t_{\rm dif, \perp}}
We have set the minimum grain size in the evolved disks to $a_{\rm min, rem}$, which is determined by $\beta = 1$ (cf.~\eqnref{eq:maxs}). The adopted minimum grain size has been $a_{\rm min} = 4\mum$ for the A-type star model and $a_{\rm min} = 1\mum$ for the solar-type star model. 
However, small grains ($a<a_{\rm min, rem}$) can be continuously produced by collisional fragmentation of larger bodies near the midplane. The fragments are stirred up in the vertical direction by turbulent diffusion on a timescale of $\tdifver \sim \Omega^{-1}\alpha^{-1}(z/H)^2$, where $\alpha$ is the viscous alpha \citep{1973_ShakuraSunyaev}. These effects supply small grains into the upper layers of the evolved disks, and if sufficiently piled up, it would result in significantly increasing mass-loss rates by rising photoelectric heating rates there \citep{2015_Gorti}. 
In this section, we examine if the evolved disks can maintain a grain-depleted state with radiation forces, taking into account the effects of fragmentation, turbulent diffusion, gas drag, and grain production. 
We assume a hypothetical planetesimal disk where the collisional cascade of planetesimals continuously produces small grains and forms a debris belt, as is supposed for debris disks.

Given that small grains ($\lesssim 1 \mum$) are collisionally produced at the highest surface density region of the dust disk to the extent not to be radially optically thick, 
grains are susceptible to radiation force. 
The irradiated grains orbit {\it slower} than the gas in gas-rich disks because radiation pressure effectively reduces the gravity. The gas angular momentum is deposited to the grains to move them radially at a velocity of 
\eq{
    v_{R, \rm d} \approx 
    \frac{v_{R, \rm g} + \nondimts (\beta - \eta) R\Omega}{1 + \nondimts^2}
    \label{eq:terminalvelocity2}
}
\citep{2003_TakeuchiLin}, 
where $v_{R, \rm g}$ is the radial velocity of gas, $\nondimts$ is dimensionless stopping time
\eq{
\splitting{
    \nondimts  & 
    =   \sqrt{\frac{\pi}{8}}\frac{\rho_{\rm b} a }{\rho \cs} \Omega
    =   \frac{\pi}{2} \frac{\rho_{\rm b} a }{\Sigma \exp\braket{-z^2/2H^2}}\\
 &\approx 
 0.022  \braket{\frac{\rho_{\rm b}}{1.4\gram \cm^{-3}}} 
        \braket{\frac{a}{1\mum}}\\
 & \times 
        \braket{\frac{\Sigma}{0.01\gram\cm^{-2}}}^{-1} 
        \exp\braket{\frac{z^2}{2H^2}},
 } 
 \label{eq:nondimts}
}
and $\eta = -(\partial P/\partial R)/(R\Omega^2 \rho)$ is the ratio of pressure gradient to the stellar gravity in the radial direction and is of the order of $(H/R)^2$.
The nominal $\Sigma$ corresponds to the surface density of the disk with $\approx 10^{-4}\Msun$ at $30\au$. Gas density is the highest at around this radius in our evolved disk model. For reference, the last factor gives $\approx 7, 90, 3000$ for $z = 2H, 3H, 4H$, respectively.
We can assume $\beta \gg \eta$ (cf. \fref{fig:Qpr}) for well-coupled grains ($\nondimts \ll 1$).  \eqnref{eq:terminalvelocity2} reduces to 
\eq{
    v_{R, \rm d} \approx v_{R, \rm g} + \beta \nondimts R \Omega,   \label{eq:terminalvelocity}
}
in this case.

\eqnref{eq:terminalvelocity} shows that grains of gas-rich disks can obtain an outward velocity even for $\beta < 0.5$, as opposed to in gas-less disks where orbits of grains with $\beta < 0.5$ are always closed \citep[e.g.,][]{2006_Krivov}. 
The radial gas velocity $v_{R,\rm g}$ can be negative if the radial gas velocity is dominated by viscous accretion. The radial gas velocity is approximately $v_{R, \rm g} \sim - \alpha (H/R)^2 R\Omega$, and in this case, 
\eq{
\splitting{
     \exp\braket{\frac{z^2}{2H^2}} 
     & >  \frac{2 \alpha\Sigma (H/R)^2}{\pi \rho_{\rm b}a \beta} \\
     & \approx 4.6\e{-4} 
     \braket{\frac{\alpha}{10^{-3}}}
     \braket{\frac{a}{1\mum}}^{-1}
     \braket{\frac{H/R}{0.1}}^2\\
     &\times 
     \braket{\frac{\Sigma}{0.01\gram\cm^{-2}}}
     \braket{\frac{\rho_{\rm b}}{1.4\gram\cm^{-3}}}^{-1}
     \braket{\frac{\beta}{1}}^{-1}, \label{eq:zradial}
     }
}
is necessary for grains to have an outward velocity, $v_{R, \rm d} > 0$. 
The nominal aspect ratio $H/R$ is taken from our simulations. 
This condition is easily met for $\mum$- and nm-sized grains, and thus the radial velocity of small grains is likely dominated by the latter term of \eqnref{eq:terminalvelocity}, $\beta \nondimts R \Omega$. 
We define a outward-drifting timescale as $t_{\rm out}  \equiv {R}/{v_{R, \rm d}}  \approx \braket{\beta \nondimts \Omega}^{-1}$, and, in a dimensionless form, 
\eq{
\splitting{
\tilde{t}_{\rm out }  & = \braket{\beta \nondimts}^{-1}
    \\  &   
    \approx 45  
    \braket{\frac{\beta}{1}}^{-1}
    \braket{\frac{\rho_{\rm b}}{1.4\gram \cm^{-3}}} ^{-1}
    \braket{\frac{a}{1\mum}}^{-1}
    \\& 
    \times \braket{\frac{\Sigma}{0.01\gram\cm^{-2}}}
    \exp\braket{-\frac{z^2}{2H^2}}
.   \label{eq:tblow}
}
}
The radial profile of $\tilde{t}_{\rm out}$ essentially follows that of $\Sigma$, 
implying that the radial motion of grains gets faster compared to the orbital motion, as they go outward. 
The stopping time can gets $t_{\rm s} \gg 1$, as grains drift to the outer tenuous disk. 
There, grains are blown out on the Keplerian timescale if $\beta > 1$. 
On the other hand, grains with $\beta < 1$ can have a steady circular orbit at the radius where $\eta(R)\approx \beta$. This is the case for relatively large grains (e.g., $\approx 10\mum$ with $L_*\approx20\Lsun$).

In order to keep a grain-depleted state in the evolved disks, $t_{\rm out}$ needs to be shorter than the production timescale of small grains. As mentioned above, we consider a steady-state collisional cascade of a planetesimal disk for the grain production \citep[e.g.,][]{2007_Wyatta}. The production timescale is given by the depletion timescale of planetesimals \citep{2010_KobayashiTanaka}
\[
\splitting{
    t_{\rm p} &= 1.2\e{3} \Myr 
    \braket{\frac{M_{\rm tot}}{1\Mearth}}^{-1}
    \braket{\frac{a_{\rm p}}{30\au}}^{4.18}
    \\&\times
    \braket{\frac{\Delta a_{\rm p}/a_{\rm p}}{0.1}} 
    \braket{\frac{e}{0.1}}^{-1.4}
    \braket{\frac{\rho_{\rm b} }{1.4 \gram \cm^{-3}}}^{0.89}
    \\&\times
    \braket{\frac{m}{10^{21}\gram}}^{0.64} 
    \braket{\frac{Q_0}{9.5\e{8}\erg \gram^{-1}}}^{0.68}, 
}
\]
where $M_{\rm tot}$ is the total solid mass, $m$ is the mass of planetesimals,
$a_{\rm p}$ and $e$ are the mean semimajor axis and eccentricity of planetesimals, 
$\Delta a_{\rm p}$ is the width of the planetesimal disk, 
and $Q_0$ is the specific impact energy required for the catastrophic disruption, $Q_D^*$, for $m=10^{21} \gram$. We use the relation of $Q_{\rm D}^* = Q_0 (\rho_{\rm b} / 3\gram\cm^{-3})^{0.55} (m/10^{21}\gram)^{0.45}$ \citep{1999_BenzAsphaug}.
Note that the nominal $m$ approximately corresponds to mass of $\sim 10^2 {\rm \,km}$-sized planetesimals.
The ratio of the production timescale to the outward-drifting timescale is then
\eq{
\splitting{
    \frac{t_{\rm p}}{t_{\rm out}}
    &\approx 
    1\e{6}
    \braket{\frac{M_{\rm tot}}{1\Mearth}}^{-1}
    \braket{\frac{a_{\rm p}}{30\au}}^{4.18}
    \\&\times
    \braket{\frac{\Delta a_{\rm p}/a_{\rm p}}{0.1}} 
    \braket{\frac{e}{0.1}}^{-1.4}
    \braket{\frac{\rho_{\rm b} }{1.4 \gram \cm^{-3}}}^{1.89}
    \\&\times
    \braket{\frac{m}{10^{21}\gram}}^{0.64} 
    \braket{\frac{Q_0}{9.5\e{8}\erg \gram^{-1}}}^{0.68}
    \\&\times
    \braket{\frac{M_*}{2\Msun}}^{1/2}
    \braket{\frac{R}{30\au}}^{-3/2}
    \braket{\frac{\beta}{1}}
    \\&\times
    \braket{\frac{a}{1\mum}}
    \braket{\frac{\Sigma}{0.01\gram\cm^{-2}}}^{-1} 
    \exp\braket{\frac{z^2}{2H^2}}
    .
}   \label{eq:provsblow}
}
The production timescale is much longer than the outward-drifting timescale $\tilde{t}_{\rm out}$ with the nominal values. It means that produced small grains are immediately removed.

For $\tilde{t}_{\rm p} / \tilde{t}_{\rm out} < 1$, produced small grains would remain near the birthplace, preventing the evolved disks from remaining in a grain-depleted state. 
If those grains pile up to be (marginally) optically thick to FUV radiation and/or to produce an abundant \ce{H2}, it could result in a strong photoelectric heating and \ce{H2} pumping. 
This is possible only in the inner region of massive disks with a small truncated radius and small $a_{\rm p}$, according to \eqnref{eq:provsblow}. 
With strong radiation pressure ($\beta \sim 10$), a grain-depleted state is likely maintained even for disks similar to PPDs in mass and geometry.
Therefore, reproduction of small grains is safely neglected for the evolved disks investigated in the present study.

Aside from the comparison between $t_{\rm p}$ and $t_{\rm out}$, it is noteworthy that radiation force also affects the vertical motion of grains. 
For $t_{\rm s} \ll 1$, the terminal velocity is derived as 
\[
     v_{z, \rm d} = (\beta - 1) \nondimts z\Omega,
\]
where turbulence is subdominant. 
The velocity is directed upward with $\beta > 1 $. This may also work to decrease $\nondimts$ of grains by lifting them up to a low-density region as well as turbulent diffusion. 
On the other hand, the velocity is directed downward for $\beta < 1$, and $|z/v_{\rm z, \rm d}|\sim \tdifver$ sets an equilibrium height at which settling balances with turbulent stirring. This would limit the removal of grains only to the radial direction. It is disadvantageous (advantageous) to sweep out (retain) the produced small grains. Grain removal process is inferred to be physically different between $\beta > 1$ and $\beta < 1$; it can be more efficient for $\beta > 1$ than for $\beta < 1$ owing not only to stronger radiation force but also to more rapid reduction of $\nondimts$ by upward motion. Note that both of vertical and radial outward-drifting timescales are given by $\tilde{t}_{\rm out}$ for $\beta \gg 1$, and gas density decreases exponentially in the vertical direction while in a power-law manner to the radial direction. 

}


\deleted
{
The collisional velocity among small grains is estimated from the Brownian-motion induced random velocity as
\begin{equation}
 v_{\rm col} = 40 \left(\frac{a}{0.01\mum}\right)^{-3/2}
\left(
\frac{T}{100\Kelvin}
\right) 
\cm \sec^{-1} .
\end{equation}
Such low-velocity collisions result in sticking growth of dust grains. 
Since the collisional timescale is much shorter than the system's age even in optically-thin disks, 
collisional fragmentation of large bodies is capable of producing small grains continuously. 
The ejected fragments initially have specific orbital energies of 
\begin{equation}
 e_{\rm orb}  \approx (\beta -1/2) r^2 \Omega^2 . \label{eq:ene_frag}
\end{equation}
This equation indicates that $\beta > 1/2$ is a necessity for grains to be blown out from the system \citep[e.g.,][]{2006_Krivov}.
In gas-rich disks, the ejected small grains are dragged by gas prior to being blown out of the disks and lose the specific energy of the order of
\begin{equation}
 \Delta e_{\rm loss} = \frac{r^2 \Omega}{t_{\rm s}},
\end{equation}
where $t_{\rm s} = (\pi /8)^{1/2} (a/\cs) (\rho_{\rm b,d}/\rho)$ is the stopping time \citep{1976_Adachi}. 
For $1/2 < \beta < 1$, 
grains remain in the disks if $e_{\rm orb} \lesssim \Delta e_{\rm loss}$, 
which reduces to 
\eq{
\splitting{
 \nh &\ga 
  \left(\beta - \frac{1}{2}\right) \left(\frac{\pi}{8}\right)^{1/2}
  \frac{a \Omega \rho_{\rm b,d}}{\cs m}\\
  &\approx  6\e{3} \left(\beta - \frac{1}{2}\right)
   \braket{\frac{a}{10^{-2}\mum}} \braket{\frac{M_*}{2\Msun}}^{1/2}
   \\
  &~~ \times 
    \braket{\frac{\cs}{1\kms}} 
\left(\frac{r}{30 \au}
\right)^{-3/2} 
 \cm^{-3},
}
}
where $m$ is the gas mass per hydrogen nucleus. 
Therefore, grains with $\beta < 1$ survives in the disks with which we are concerned. 

For $\beta > 1$, grains are always unbound from the host star. 
Gas drag delay their blow-out. 
The terminal velocities of blow-out grains are 
given by the balance between gas drag, radiation pressure, and stellar gravity
\begin{equation}
 v_{\rm b} = (\beta -1) r \Omega^2 t_{\rm s}. 
\end{equation}
Therefore, the blow-out timescale is estimated to be
\begin{equation}
 t_{\rm b} = \frac{r}{v_{\rm b}} = 1 / (\beta -1) \Omega^2 t_{\rm s}. 
\end{equation}
To sustain small grains, $t_{\rm b}$ is longer than the production timescale of small grains or the collisional timescale of grains, $1/\tau \Omega$, where 
$\tau$ is the vertical optical depth of grains. 
Small grains are effectively maintained if
\begin{equation}
 t_{\rm s} \Omega \la \frac{\tau}{\beta-1}. 
\end{equation}
The condition is rewritten as
\begin{eqnarray}
 n_{\rm H} &\ga& 
  \left(\frac{\beta - 1}{\tau}\right) 
  \left(\frac{\pi}{8}\right)^{1/2}
  \frac{a \Omega \rho_{\rm b,d}}{\cs m_0},
  \nonumber
  \\
  &\approx& 
  5 \times 10^{6} \left(\frac{\beta - 1}{10}\right) \left(\frac{a}{0.01\,\mum}\right)
  \nonumber
  \\
 && \times 
  \left(\frac{\tau}{10^{-2}}\right)
  \left(\frac{T}{100\,K}\right)^{1/2} 
  \left(\frac{r}{30 \,{\rm AU}}
  \right)^{-3/2} \, {\rm cm}^{-3}. 
\end{eqnarray}
As the simulations show in \S.\ref{sec:structure}, 
photoevaporation effectively occurs at $n_{\rm H} \la 10^5\,{\rm cm}^{-3}$.  
Small grains with $\beta > 1$ are depleted in the photoevaporation effective layers. 
}

%% file: chem_reactions.tex
k1  & \ce{H + e -> H+ + 2e} 	& \omukai \\
k2  & \ce{H+ + e -> H + \gamma} & \omukai\\
k7  & \ce{H + e -> H- + \gamma  }  	& \omukai\\
k8  & \ce{H- + H -> H2 + e}         & \umist\\
k9  & \ce{H + H+ -> H2+  }          & \umist\\ 
k10 & \ce{H2+ + H -> H2 + H+}       & \omukai\\
k11 & \ce{H2 + H+  -> H + H2+}      & \omukai\\
k12 & \ce{H2 + e -> 2H + e }        & \omukai\\ 
k13 & \ce{H2 + H -> 3H}             & \omukai\\
k14 & \ce{H- + e -> H + 2e}         & \omukai \\
k15 & \ce{H- + H+ -> 2H}            & \umist \\
k16 & \ce{H- + H+ -> H2+ + e}       & \omukai\\
k17 & \ce{H2+  + e -> 2H}           & \umist\\
k18 & \ce{H2+ + H- -> H2 + H}       & \umist\\
k19 & \ce{3H -> H2 + H}             & \omukai\\
k20 & \ce{2H + H2 -> 2H2}           &\omukai\\
k21 &\ce{2H2 -> 2H + H2}            &\omukai \\
k22 &\ce{2H -> H+ + e + H}          &\omukai\\
k23 &\ce{2H  ->C[dust] H2}          &\omukai\\
kCRH    &\ce{H -> H+ + e}   &   \umist  \\
kCRH21  &\ce{H2 -> H+ + H-} &   \umist  \\
kCRH22  &\ce{H2 -> H + H+ + e}& \umist  \\
kCRH23  &\ce{H2 -> H2+ + e}  &  \umist  \\
kCRH24  &\ce{H2 -> H+ + e}  &   \umist  \\
kCRPHH  &\ce{H -> H+ + e}   &   \umist  \\
k1He    &\ce{He+ + e -> He}	&	\umist	\\
k2He    &\ce{He+ + CO -> O + C+ + He}	&	\umist	\\
k3He	&\ce{He+ + C -> C+ + He}		&	\umist	\\
k4He	&\ce{H + He+ -> He + H+}		&	\umist	\\
k5He	&\ce{H2 + He+ -> He + H2+}		&	\umist	\\
k6He	&\ce{H2 + He+ -> He + H+ + H}	&	\umist	\\
xi1He	&\ce{He + {CRP} -> He+ + e}	&	\umist	\\
xi2He	&\ce{He + {CRPHOT} -> He+ + e}	&	\umist	\\
k001    &\ce{C + O -> CO}   &   \umist  \\
k28     &\ce{C+ + e -> C + {\gamma}}    & \umist    \\ 
kCRC    &\ce{C + CRP -> C+ + e}   &   \umist  \\
kCRPHC  &\ce{C + CRPHOT -> C+ + e}   &   \umist  \\
kCRCPHCO&\ce{CO + CRPHOT -> C + O}   &   \umist  \\
k0      &\ce{C+ + H2 ->  CH2+}  &   \umist  \\
kea2pl  &\ce{CH2+ + e -> CH + H}&   \umist  \\
keb2pl  &\ce{CH2+ + e -> C + H2}&   \umist  \\
kec2pl  &\ce{CH2+ + e -> C +  H +  H}&  \umist  \\
kCO     &\ce{CH  + O ->  CO +  H}   &   \umist  \\
k2\_2pl &\ce{CH2+ + H2 ->  CH3+ + H}&   \umist  \\
ke\_pl  &\ce{CH+  + e -> C + H} &   \umist  \\
kH\_pl  &\ce{CH+  + H ->  C+ +  H2}&    \umist  \\
k2\_pl  &\ce{CH+  + H2 ->  CH2+ + H}    &   \umist  \\
kH\_a   &\ce{CH +   H ->  C +   H2}& \umist  \\
ke\_a3pl&\ce{CH3+ + e -> CH2 +  H}  &\umist   \\
ke\_b3pl&\ce{CH3+ + e -> CH +   H2} &\umist   \\
ke\_c3pl&\ce{CH3+ + e -> CH + H + H}&\umist   \\
p\_a2pl &\ce{CH2+ + CH -> H+}   &  \umist 	\\
p\_b2pl &\ce{CH2+   +   CH+ ->  H   }   &\umist    \\
p\_c2pl &\ce{CH2+   +   C+ ->  H2   }  &\umist\\
p\_a    &\ce{CH ->   CH+ +  e}& \umist\\
p\_b    &\ce{CH ->   C  +  H}&  \umist\\
p\_pl   &\ce{CH+    ->  C   +   H+}&    \umist\\
p\_a3pl &\ce{CH3+ ->   CH+  + H2}   &  \umist\\
p\_b3pl &\ce{CH3+ ->   CH2+ + H}    &  \umist\\
k1CH    &\ce{C + H  -> CH}          &    \umist\\
k1CH2   &\ce{C+ + CH2 -> CH2+ + C}  &    \umist\\
k2CH2   &\ce{H+ + CH2 -> CH2+ + H}  &    \umist\\
k3CH2   &\ce{H2+ + CH2 -> CH2+ + H2}&   \umist\\
kCRPHOTCH2\_a   &\ce{CH2 ->  CH2+ + e}& \umist\\
kCRPHOTCH2\_b   &\ce{CH2 ->  CH   + H}& \umist\\
k4CH2    &\ce{H+ + CH2 -> CH+ +  H2 }   &\umist   \\
k6CH2    &\ce{CH2 + O -> CO + H + H}    &\umist   \\
k7CH2    &\ce{CH2 + O -> CO + H2}       &  \umist \\
k8CH2    &\ce{H +  CH2 -> CH + H2}      & \umist  \\
k9CH2    &\ce{CH2 -> CH2+ + e}          &   \umist\\
k10CH2   &\ce{CH2 -> CH + H}            &   \umist\\
k11CH2   &\ce{H- + CH -> CH2 + e}       &   \umist\\
k12CH2   &\ce{CH3+ + e -> CH2 + H}      &   \umist\\
k13CH2   &\ce{H2 + C -> CH2}            &   \umist\\
k1COp    &\ce{H2+ + CO -> CO+ + H2}     &   \umist\\
k2COp    &\ce{CO -> CO+ + e}            &   \umist\\
k3COp    &\ce{C+ + O2 -> CO+ + O}       &   \umist\\
k4COp    &\ce{C+ + OH -> CO+ + H}       &   \umist\\
k5COp    &\ce{CH+ + O -> CO+ + H}       &   \umist\\
k6COp    &\ce{HCO+ -> CO+ + H}          &   \umist\\ 
k7COp    &\ce{C+ + O -> CO+}            &   \umist\\
k8COp    &\ce{C + CO+ -> CO + C+}       &   \umist\\
k9COp    &\ce{H + CO+ -> CO + H+}       &   \umist\\
k10COp   &\ce{C + O+ -> CO+}            &   \umist\\
k11COp   &\ce{CO+ + e -> O + C}         &   \umist\\
k13COp   &\ce{H2 + CO+ -> HCO+ + H}     &   \umist\\ 
k15COp   &\ce{CO+ -> C+ + O}            &   \umist\\
k16COp   &\ce{O + CO+ -> CO + O+}       &   \umist\\
k1HCOp   &\ce{HCO+ + e -> CO + H}       &   \umist\\
k3HCOp   &\ce{CH + O -> HCO+ + e}       &   \umist\\
k4HCOp   &\ce{H2+ + CO -> HCO+ + H}     &   \umist\\
k6HCOp   &\ce{H3+ + CO -> HCO+ + H2}    &   \umist\\
k9HCOp   &\ce{H2O + C+ -> HCO+ + H}     &   \umist\\
k1OH     &\ce{H3+ + O -> OH+ + H2}      &   \umist\\
k2OH'    &\ce{H3+ + O -> H2O+ + H}      &   \umist\\
k2OH     &\ce{H2 + OH+ -> H2O+ + H}     &   \umist\\
k3OH     &\ce{H2O+ + e -> OH + H}       &   \umist\\
k4OH     &\ce{H2O+ + e -> O + H2}       &   \umist\\
k5OH    &\ce{H2O+ +   e -> O +  H +   H}&    \umist\\
k6OH    &\ce{H2O+ +   H2 ->  H3O+ + H}  &    \umist\\
k7OH    &\ce{H3O+ +   e -> O + H2 + H}  &    \umist\\
k8OH    &\ce{H3O+ +   e -> OH + H2}     &    \umist\\
k9OH    &\ce{H3O+ +   e -> OH + H + H}  &    \umist\\
k10OH   &\ce{H3O+ +   e -> H2O + H}     &    \umist\\
p11OH   &\ce{OH -> O + H}               &    \umist\\
k12OH   &\ce{C + O2 -> CO + O}          &    \umist\\
k13OH   &\ce{C + OH -> CO + H}          &    \umist\\
k14OH   &\ce{O + OH -> O2 + H}          &    \umist\\
k15OH   &\ce{C+ + O2 -> CO + O+}        &    \umist\\
k16OH   &\ce{H + O+ -> O + H+}          &    \umist\\
k18OH   &\ce{O+ + e -> O}               &    \umist\\
k19OH   &\ce{H+ + O -> O+ + H}          &    \umist\\
k20OH   &\ce{O+ + H2 -> OH+ + H}        &    \umist\\
k21OH   &\ce{H + O -> OH}               &    \umist\\
k22OH   &\ce{H2 + O -> OH + H}          &    \umist\\
k1O2    &\ce{O + O -> O2}               &    \umist\\
kO2gamma    &\ce{O2 -> O + O}           &    \umist\\
kOplCRP &\ce{O -> O+ +  e}              &    \umist\\
kOplCRPHOT  &\ce{O -> O+ +  e}          &    \umist\\
p1H2O   &\ce{H2O -> OH + H}             &    \umist\\
pO2     &\ce{O2 -> O + O}               &    \umist\\
pOHpl   &\ce{OH+ -> O+ + H}             &    \umist\\
k0H3p   &\ce{H2+ + H2 -> H3+ +  H}      &    \umist\\
k01H3p  &\ce{H3+ + e -> H2 + H}         &    \umist\\
k02H3p  &\ce{H3+ + e -> H +   H +   H}  &    \umist\\
p1      &\ce{H +  {\gamma}_{EUV}   -> H+ + e}   & \paperone     \\
p2      &\ce{H +  {\gamma}_{X-ray} -> H+ + e}   & \papertwo     \\
p3      &\ce{H2 + {\gamma}_{X-ray} -> H2+ + e}  & \papertwo     \\
p4      &\ce{H2 + {\gamma}_{FUV}  -> 2 H}       & \paperone, 5     \\
p5      &\ce{CO + {\gamma}_{FUV} -> C + O}      & \paperone, 6 \\
p6      &\ce{C + {\gamma}_{FUV} -> C+ + e}      & \umist, 7 (cf. \eqnref{eq:kcph})  \\
p7      &\ce{H- + {\gamma} -> H + e }           & \umist\\
